\documentstyle[12pt,tgrind,vi]{mitthesis}
\begin{document}

\input epsf
\def\edth{\;\raise1.0pt\hbox{$'$}\hskip-6pt\partial\;}
\def\baredth{\;\overline{\raise1.0pt\hbox{$'$}\hskip-6pt
\partial}\;}
\def\bi#1{\hbox{\boldmath{$#1$}}}
\def\gsim{\raise2.90pt\hbox{$\scriptstyle
>$} \hspace{-6.4pt}
\lower.5pt\hbox{$\scriptscriptstyle
\sim$}\; }
\def\lsim{\raise2.90pt\hbox{$\scriptstyle
<$} \hspace{-6pt}\lower.5pt\hbox{$\scriptscriptstyle\sim$}\; }

%
%
%
%
%
%
%
\title{Fluctuations in the Cosmic Microwave Background}

\author{Matias Zaldarriaga}
\department{Department of Physics}
\degree{Doctor of Philosophy}
\degreemonth{May}
\degreeyear{1998}
\thesisdate{May 1, 1998}


\supervisor{Edmund Bertschinger}{Professor}
\cosupervisor{Uro\v s Seljak}{Harvard Smithsonian
Center for Astrophysics Fellow}

\chairman{Thomas J. Greytak}{Associate Department Head for Education}

\maketitle



\newpage
\setcounter{savepage}{\thepage}
\begin{abstractpage}
%
%
%
%
%
%
In this thesis we investigate several aspects related to the theory of
fluctuations in the Cosmic Microwave Background. We develop a new
algorithm to calculate the angular power spectrum of the
anisotropies which is two orders of magnitude faster than the
standard Boltzmann hierarchy approach (Chapter \ref{chaplos}). 
The new algorithm will become
essential when comparing the observational results of the next
generation of CMB experiments with theoretical predictions. The
parameter space of the models is so large that an exhaustive 
exploration to find the best fit model will only be feasible with this
new type of algorithm. We also investigate the polarization properties
of the CMB field. We develop a new formalism to describe the statistics
of the polarization variables that takes into account their spin two
nature (Chapter \ref{chapstatcmb}). 
In Chapter \ref{chapinfpol} 
we explore several physical effects that create distinct
features in the polarization power spectrum. We study the signature of
the reionization of the universe and a stochastic background of
gravitational waves. We also describe how the polarization correlation
functions can be used to test the causal structure of the universe.  
Finally in Chapter \ref{chapconstraints} 
we quantify the amount of information the next generation of
satellites can obtain by measuring both temperature and polarization
anisotropies. We calculate the expected error bars on the cosmological
parameters for the specifications of the MAP and Planck satellite missions.

\end{abstractpage}


\newpage

\section*{Acknowledgments}

The time has come to thank all the people that made this work 
possible. I would like to thank Ed for his support and for giving me
the freedom to pursue my own interests and collaborate with other
people. When I arrived at MIT three years ago he suggested I should
talk to Uro\v s.
This proved to be an excellent suggestion. We started working on
CMBFAST, not to enthusiastically at times, ``well
it is faster, so what?'' we thought.  Since then we have done 
several other things together and I think we
had and are having a lot of fun. This is in my mind
the most important thing.
I have benefited immensely from talking and working with him, and
hopefully this will continue into the future (if he has time to spend
with the rest of us mortals now that he will become a distinguished
professor). 

I want to thank the members of my thesis committee for their
suggestions, in particular Ed and Uro\v s for going carefully through
the thesis to help improve my poor english. 
At MIT I enjoyed the company other graduate students and
postdocs, Greg, Jim, Lam and Rennan among them.

I also want to thank David Spergel, several parts of this thesis were
done in collaboration with him. It was a real pleasure to be able to
work with him both during my visits to Princeton 
and over e-mail. I am very grateful for his support. 
I also want to thank John Bahcall for inviting me to
Princeton for extended periods of time giving me the chance to 
work more closely with Dave and talk with the 
people at the IAS.
 
Several people have made these years much more fun. My ``roommates''
Alejandro, Fernando, Lucas and Susana share among them extremely important
qualities. Formidable cooking abilities, exquisite taste for movies,
obsession for cleanness, passion for sports and all other necessary 
characteristic of the perfect roommate. Patricio and our  ``ex-social
chair'' Pablo contributed to my well-being providing up to date party
information and many fun weekday lunches. I also enjoyed keeping
e-mail and phone contact with my old friends from the university and high
school. Special thanks go to 
Fernando for his periodic updates on the absurd twists of argentinian
politics and his regular one line messages to learn how things were
going,  ``JJ the great'' for
explaining me much more that I ever wanted to know about X-ray lasers and
other perhaps more useful thoughts about life and Arturo, 
my friend from the ``real
world'' with whom I have shared lots of things since our highschool
days. 

My parents and brothers have been a constant source of support. I
enjoyed greatly our regular phone calls and visits which remind me
that there are people who care about me back home. Hopefully the
large ``spatial distance'' that separates us will not get in our way.


\pagestyle{plain}
\include{contents}
\chapter{Introduction}

Since the initial discovery of anisotropies in the temperature of the
microwave background radiation (CMB) by COBE in 1992 \cite{1.smoot92}, 
the field has been transformed from a theoretical exercise to an
active experimental area of research. The large amount of
data from CMB experiments and other observations
is beginning to
revolutionize cosmology and has the potential of changing
dramatically our view of the universe. In the next decade we might obtain
answers to very fundamental questions such as the age, size and fate
of the universe we live in. A
few percent accuracy on most cosmological parameters
seems achievable in the very near
future, when the next CMB satellite missions fly. 
Should the 
inflationary paradigm prove to be correct, detailed measurements of the
anisotropies coming from a number of ground based, 
balloon borne and satellite 
experiments could be used to determine a large set of cosmological
parameters such as
the density of
the universe ($\Omega_0$) or the Hubble constant ($H_0$)
with unprecedented accuracy \cite{1.jungman,1.bet,1.zss}.

At the time
the anisotropies in the microwave background were produced the
universe was almost perfectly homogeneous. The fluctuations were
at the $10^{-5}$ level and a linear
approximation to the evolution equations suffices. This allows us
to make very
accurate predictions which are to be compared with  experiments
\cite{1.be87,1.hu95,1.bond96,1.los}. 
Both the physics and the equations to be solved are relatively simple.
CMB anisotropies are very  clean from the theoretical point of
view. This distinguishes them from other more local probes of
cosmology where nonlinear effects complicate the theoretical
interpretation of the data.  

\begin{figure*}
\begin{center}
\leavevmode
\end{center}
\caption{Compilation of measurements of the CMB anisotropy power
spectra (from Tegmark http://www.sns.ias.edu/\~{}max/cmb/experiments.html). The theoretical
prediction for COBE normalized SCDM is shown for comparison.} 
\label{1.figexper}
\end{figure*}

A large number of other detections have followed the initial discovery
by COBE 
\cite{1.firs,1.tenerife,1.sp,
1.bam,1.argo,1.max,1.python,1.iac,1.iab,1.msam, 1.sask,1.cat,1.suzie}. Figure
\ref{1.figexper} shows all the results in the literature at the time
this thesis was written (April 1998) in terms of the angular power
spectrum of the temperature 
fluctuations, $C_l$. The easiest way to understand the meaning of the
angular power spectrum is to consider a small region of the sky (for
example the simulation in Figure \ref{tqups}). In that case the
temperature map can be expanded in Fourier modes and the $C_l$ is
the  power (mean square amplitude) as a function of the 
spatial frequency denoted with $l$.
The rapid evolution of the field 
will make the figure outdated soon. 
We can expect new results from several 
ground and balloon borne
experiments in the near future. The  MAP satellite from NASA 
will be launched  by the year 2000 
and a few years later Planck from ESA should be launched too. 
The sensitivity and angular
resolution of these
satellites will be outstanding; MAP is expected to have a noise level
of $20 \mu K$ per resolution element (of approximately
$0.3^o \times 0.3^o$). This translates into an uncertainty for
each multipole due to noise in the detectors of 
$\Delta C_l\approx \sqrt{2/(2l+1)}\times (0.11\mu K)^2$.
The Planck satellite will be
equipped with very sensitive bolometers in the highest
frequency channels (above 100 GHz).
These channels will also have the best angular resolution,
$\sim 0.1^o$. For Planck the noise in the multipoles is expected to
be $\Delta
C_l\approx \sqrt{2/(2l+1)}\times (0.01\mu K)^2$. 
The high sensitivity of 
the experiments coupled to the accurate theoretical predictions are
the primary  motivation for using
CMB anisotropies as a probe for cosmology. 

The impressive angular resolution and noise levels of the future
satellite missions create certain technical problems for the data
analysis. The first step in the analysis is to go from the 
time ordered data (temperature or temperature differences measured by the
different detectors on the satellite at each time) 
to maps of the CMB  sky. A computationally feasible 
algorithm for doing this has recently been developed \cite{1.maps}. 
In the next step one wants 
to optimally measure the power
spectrum of the anisotropies from the maps which have  
$10^5-10^6$ pixels. Direct
evaluation of the likelihood function involves the inversion of a huge
correlation matrix and is computationally too expensive. A practical
algorithm solving this problem has recently been developed for MAP 
\cite{1.david}. 

Once 
the power spectrum of the anisotropies has been measured 
we would like to find the model that best fits the data and the confidence
intervals on the model parameters. The power spectrum of the CMB
anisotropies depends on a large number of parameters. We would like to
learn about the total density of the universe $\Omega_0$, Hubble
constant $H_0$, baryon density $\Omega_b$, the 
cosmological constant $\Lambda$ and the amplitude and scale
dependence of the initial
fluctuations. The exploration of this vast
parameter space that characterizes the models
will require many calculations of the theoretical power
spectrum. This is computationally too expensive
with the usual Boltzmann codes which need to solve a large system of
differential equations at a large number of wavenumbers.
This computational barrier was overcome by the
development of a new algorithm based on a line of sight 
integral \cite{1.los}
which we present in Chapter \ref{chaplos}.  
This algorithm has proved very useful for the rapid exploration of 
the phenomenology of CMB anisotropies. The power spectrum for any
cosmological model can be obtained in only a few minutes on a normal
workstation. It has also been extended to include active sources 
(as opposed to the primeval potential fluctuations of standard
models), so 
that it
can be used to make accurate predictions of the anisotropies produced
by topological defect models \cite{1.urostur}. 
The development of this new line
of sight algorithm is one of the most important results presented in
this thesis.

There is still a need to develop ways of extracting
additional information from the CMB maps  not encoded in the
temperature power
spectrum, which  could prove important if topological defects or other 
non-gaussian sources are shown to be the cause 
of the anisotropies. If
inflationary models are correct, the power spectrum fully describes the
statistical properties of the primary anisotropies, but 
secondary effects like gravitational
lensing can leave interesting information in higher moments. These
effects tend to be small, at least on large angular scales.

The CMB radiation field 
is also expected to be linearly polarized due to Thomson
scattering of the photons with free electrons 
\cite{1.bond87,1.crittenden93,1.fr94,1.coul94,1.critt95,1.zh95}.
The polarization of the CMB can provide in principle 
the same information about our universe 
that will become available from the temperature maps.
The main
disadvantage of polarization 
is that the CMB is predicted to be only $1-10\%$
polarized depending on the angular scale. 
Future satellite missions will have the required
sensitivity and  the most likely source of problems will be 
contamination from galactic foregrounds. The nature and level
of the foregrounds for polarization remain uncertain
and we may have to wait until the satellites fly for a
definitive answer. There will also be several ground based experiments
before the satellites, eg. \cite{1.wisconsin,1.polatron}. They may help 
determine the level of foreground contamination to be expected for the
satellite missions and may also detect cosmological polarization. 

The detection of polarization is
challenging but nevertheless the rewards can be great.
We have shown that polarization may be able to
constrain the ionization history of the universe 
very accurately \cite{1.zalrei}. 
We know that the light from the first generation of
objects reionized the universe after
recombination, but the time when this occurs remains very uncertain. The
scatterings of the CMB photons with the electrons after reionization
will leave a signature in the CMB polarization that could 
be our first direct detection of the epoch of reionization.
We will also show that large angular scale correlations in the
polarization can be used to test  
the causal structure of the universe and could provide a direct test
of the inflationary paradigm \cite{1.sperzal}. Finally the pattern of
the polarization vectors can be used to detect
the presence of a stochastic background of gravitational waves
\cite{1.selzal,1.kks}. We
will discuss all of these issues in Chapter \ref{chapinfpol}. 

The 
polarization of the CMB has the potential of providing very
interesting tests for our theories of the universe.  
The study of the 
properties of the polarization is a unifying theme of this
thesis. In Chapter \ref{chapstatcmb} we introduce the physics of CMB
anisotropy and polarization and 
we develop a statistical
description for the CMB polarization field. 
Chapter \ref{chaplos} describes the algorithm for computing both
temperature and polarization spectra. In Chapter
\ref{chapinfpol} we look at several ways in which polarization can
provide extra information not available in the temperature power spectrum. 
In Chapter \ref{chapconstraints} we quantify the benefits of
using polarization for the determination of cosmological parameters by
the MAP and Planck satellites. 

The anisotropies in the CMB radiation field encode  an enormous
amount of information about the universe we live in. In the next years
the main driver of the field will be the rapid
progress of experimental research. The results of these
experiments  will be used in
conjunction with the results of other observations to test our
theories of the universe with an
accuracy which was unimaginable just  a decade ago.
It is a great time for cosmology, the availability of large
quantities of data is making all the difference and hopefully will
provide several interesting surprises.

\def\edth{\;\raise1.0pt\hbox{$'$}\hskip-6pt\partial\;}
\def\baredth{\;\overline{\raise1.0pt\hbox{$'$}\hskip-6pt
\partial}\;}
\def\bi#1{\hbox{\boldmath{$#1$}}}
\def\gsim{\raise2.90pt\hbox{$\scriptstyle
>$} \hspace{-6.4pt}
\lower.5pt\hbox{$\scriptscriptstyle
\sim$}\; }
\def\lsim{\raise2.90pt\hbox{$\scriptstyle
<$} \hspace{-6pt}\lower.5pt\hbox{$\scriptscriptstyle\sim$}\; }

\chapter{Statistical Treatment of CMB Fluctuations$^1$}\label{chapstatcmb}

\setcounter{footnote}{1}

In this chapter \footnotetext{Partially based on M. Zaldarriaga \&
U. Seljak, Phys. Rev. D {\bf 55}, 1830 (1997).} 
we will develop the statistical description of
the CMB radiation field. We will enphasize the 
spin 2 nature of polarization and the mathematical tools needed to
analyze it. We will present a physical interpretation of the
new variables introduced by making a connection with the theory
of gravitational lensing and a brief summary of the mechanisms that
can produce anisotropies and polarization in our universe.

Figure \ref{tqups} shows a simulated map of CMB anisotropies (both
temperature and polarization) for the standard cold dark matter model 
(SCDM)\footnote{ This model has $\Omega_0=1$,
$\Omega_{\Lambda}=0$, $\Omega_b=0.05$, $H_0=50$ km/sec/Mpc, a COBE
normalized power spectrum of initial fluctuations with a spectral
index $n=1$ and no stochastic background of  gravitational 
waves.}. The anisotropies can tell us about
the state of the universe at a very early time, 
only 300.000 years after the big bang when hydrogen recombined and the
universe became transparent to the CMB photons. Since then the
photons have been  able to travel an enormous distance, approximately
$10^{10}$ light years. Thus the CMB anisotropies map the most
distant parts of our universe at a very early stage.

The main cause for the 
temperature anisotropies are fluctuations in the photon density at
recombination. We see more photons coming from what were denser
regions, producing a hot spot in the map. Photons coming from
different directions also come from regions which were moving at
different velocities relative to us; the consequent 
difference in the Doppler shifts 
they suffered also cause anisotropies.
Gravitational redshift is the last source of anisotropies.
Fluctuations in the gravitational potential will lead to different
amounts of gravitational redshift for photons coming from different
parts of the universe. We can also consider the presence of a 
stochastic background of
gravitational waves.  They will also create temperature anisotropies as
photons get redshifted or blueshifted by different amounts 
according to their direction of
propagation relative to that of the gravity wave.

Polarization is produced by Thomson scattering of the CMB photons and 
electrons. Thus it can only be produced when hydrogen is
ionized. Furthermore only radiation that is anisotropic can lead to
non zero polarization because by symmetry if the radiation incident on an
electron is isotropic there is no preferred direction for the
polarization of the scattered light. The sources of anisotropies described in 
the previous paragraph will create some polarization when Thomson
scattering is present.  

\section{Characterization of the Radiation Field}

The aim of this section is to develop the mathematical tools needed to
describe the CMB anisotropies. We will emphasize
the description of the polarization which is new to this thesis.
The CMB anisotropy field is characterized by a $2\times 2$ intensity 
tensor $I_{ij}$. For convinience we normalize this tensor so it represents 
the fluctuations in
units of the mean intensity ($I_{ij}=\delta I /I_0$), 
it is dimensionless. The intensity tensor is a
function of direction on the sky $\hat{\bi{n}}$ and  two directions
perpendicular to $\hat{\bi{n}}$ that are  used to define its components
(${\bf \hat e}_1$,${\bf \hat e}_2$).
The Stokes parameters $Q$ and $U$ are defined as
$Q=(I_{11}-I_{22})/4$ and $U=I_{12}/2$, while the temperature 
anisotropy is
given by $T=(I_{11}+I_{22})/4$ (the factor of $4$ relates fluctuations
in the intensity with those in the temperature, $I\propto T^4$).  
In Figure \ref{tqups} we represent the
polarization using vectors with magnitude $P=\sqrt{Q^2+U^2}$ that form
an angle $\alpha={1\over 2}\arctan({U/Q})$ with ${\bf \hat e}_1$.
In principle the fourth  
Stokes parameter $V$ that describes circular polarization would also 
be needed, but in cosmology it can be ignored because it cannot
be generated through Thomson scattering\footnote{The presence of a
large magnetic field at recombination could generate a small component
of circular polarization.}.  
While the temperature is invariant
under a right handed rotation in the plane perpendicular to direction
$\hat{\bi{n}}$,
$Q$ and $U$ transform under rotation by an angle $\psi$ as
\begin{eqnarray}
Q^{\prime}&=&Q\cos 2\psi  + U\sin 2\psi  \nonumber \\  
U^{\prime}&=&-Q\sin 2\psi  + U\cos 2\psi 
\label{QUtrans} 
\end{eqnarray}
where ${\bf \hat e}_1^{\prime}=\cos \psi\ {\bf \hat e}_1+\sin\psi\ 
{\bf \hat e}_2$ 
and ${\bf \hat e}_2^{\prime}=-\sin \psi\ {\bf \hat e}_1+\cos\psi\ 
{\bf \hat e}_2$. 
We  can construct two quantities from the Stokes $Q$ 
and $U$ parameters that have 
a definite value of spin 
(see Appendix
\ref{appa}
for a review of spin-weighted functions and their properties),  
\begin{equation}
(Q\pm iU)'(\hat{\bi{n}})=e^{\mp 2i\psi}(Q\pm iU)(\hat{\bi{n}}).
\end{equation}
We may therefore expand each of the quantities 
in the appropriate
spin-weighted basis 
\begin{eqnarray}
T(\hat{\bi{n}})&=&\sum_{lm} a_{T,lm} Y_{lm}(\hat{\bi{n}}) \nonumber \\
(Q+iU)(\hat{\bi{n}})&=&\sum_{lm} 
a_{2,lm}\;_2Y_{lm}(\hat{\bi{n}}) \nonumber \\
(Q-iU)(\hat{\bi{n}})&=&\sum_{lm}
a_{-2,lm}\;_{-2}Y_{lm}(\hat{\bi{n}}).
\label{Pexpansion}
\end{eqnarray}
$Q$ and $U$ are defined at each direction $\hat {\bi{n}}$
with respect to the spherical coordinate system $(\hat{{\bf e}}_\theta,
\hat{{\bf e}}_\phi)$.
Using the first equation in (\ref{propYs}) one can show that
the expansion coefficients for the polarization variables
satisfy $a_{-2,lm}^*=a_{2,l-m}$. For the temperature the relation is
$a_{T,lm}^*=a_{T,l-m}$.

\begin{figure*}
\begin{center}
\leavevmode
\end{center}
\caption{Simulated temperature and polarization 
map for SCDM ($2.5^o\times 2.5^o$
field). The
polarization vectors are shown together with the map of the
temperature. The temperature ranges from $-1690 \mu K < T < 1810 \mu K$
while the maximum amplitude of the polarization vectors is $P=128 \mu K$.} 
\label{tqups}
\end{figure*}

\cleardoublepage

\vspace*{10cm}

\cleardoublepage

The main difficulty when computing the power spectrum of 
polarization in the
past originated in the fact that the Stokes parameters
are not invariant under
rotations in the plane
perpendicular to $\hat{\bi{n}}$ (equation \ref{QUtrans}). The usual
approach starts by expanding all perturbations in Fourier modes with
wavevector ${\bi {k}}$.
While $Q$ and $U$ are easily  calculated in a 
coordinate system where the wavevector ${\bi{k}}$ is 
parallel to $\hat{\bi{z}}$,
the superposition 
of the different modes is complicated by the behavior of $Q$ and $U$
under rotations. For each wavevector
$\bi k$ and direction on the 
sky $\hat{\bi{n}}$ one has to rotate the $Q$ and $U$ parameters from this
$\bi{k}$ and $\hat{\bi{n}}$
dependent basis where the calculation is done
into a fixed basis on the sky before adding them. 
Only in the 
small scale limit is this process well defined. This 
approximation had always been assumed in previous work 
\cite{2.crittenden93,2.frewin,2.coulson,2.coutur,2.uros,2.kosowsky96}. 
However, one can use the spin raising and lowering operators 
$\edth$ and $\baredth$ defined in Appendix \ref{appa}
to obtain spin zero quantities. These  
have the advantage of being rotationally invariant
like the temperature and no ambiguities connected with the 
rotation of coordinate system arise. Acting twice with 
$\edth$, $\baredth$ on $Q\pm iU$ in equation (\ref{Pexpansion}) leads to 
\begin{eqnarray}
\baredth^2(Q+iU)(\hat{\bi{n}})&=&
\sum_{lm} \left[{(l+2)! \over (l-2)!}\right]^{1/2}
a_{2,lm}Y_{lm}(\hat{\bi{n}})
\nonumber \\
\edth^2(Q-iU)(\hat{\bi{n}})&=&\sum_{lm} \left[{(l+2)! \over (l-2)!}\right]^{1/2}
a_{-2,lm}Y_{lm}(\hat{\bi{n}}).
\end{eqnarray}

The expressions for the expansion coefficients are
\begin{eqnarray}
a_{T,lm}&=&\int d\Omega\; Y_{lm}^{*}(\hat{\bi{n}}) T(\hat{\bi{n}})
\nonumber  \\  
a_{2,lm}&=&\int d\Omega \;_2Y_{lm}^{*}(\hat{\bi{n}}) (Q+iU)(\hat{\bi{n}})
\nonumber  \\
&=&\left[{(l+2)! \over (l-2)!}\right]^{-1/2}
\int d\Omega\; Y_{lm}^{*}(\hat{\bi{n}}) 
\baredth^2 (Q+iU)(\hat{\bi{n}})
\nonumber \\  
a_{-2,lm}&=&\int d\Omega \;_{-2}Y_{lm}^{*}(\hat{\bi{n}}) (Q-iU)(\hat{\bi{n}}) 
\nonumber  \\  
&=&\left[{(l+2)! \over (l-2)!}\right]^{-1/2}
\int d\Omega\; Y_{lm}^{*}(\hat{\bi{n}})\edth^2 (Q-iU)(\hat{\bi{n}}).\
\label{alm}
\end{eqnarray}

Instead of $a_{\pm 2,lm}$ it is convenient to introduce their
linear combinations \cite{2.np},
\begin{eqnarray}
a_{E,lm}&=&-(a_{2,lm}+a_{-2,lm})/2 \nonumber \\ 
a_{B,lm}&=&i(a_{2,lm}-a_{-2,lm})/2.
\label{aeb}
\end{eqnarray}

We can define two quantities in real space,
\begin{eqnarray}
E(\hat{\bi{n}})&=&\sum_{l,m}a_{E,lm} \ Y_{lm}(\hat{\bi{n}}) \nonumber
\\
B(\hat{\bi{n}})&=&\sum_{l,m}a_{B,lm} \ Y_{lm}(\hat{\bi{n}}). 
\label{EBreal}
\end{eqnarray}
The temperature is a
scalar quantity
under a rotation of the coordinate system,
$T^{\prime}(\hat{\bi{n}}^{\prime}={\bf \cal R} \hat{\bi{n}})
=T(\hat{\bi{n}})$, where  $\bf {\cal
R}$ is the rotation matrix. We denote with a
prime the quantities in the transformed coordinate system. Equation 
(\ref{alm}) shows that up to an $l$-dependent factor $a_{\pm 2,lm}$ are
the expansion coefficients of two spin zero quantities 
$\baredth^2 (Q+iU)$ and $\edth^2 (Q-iU)$. As a result
$E(\hat{\bi{n}})$ and $B(\hat{\bi{n}})$ are also 
invariant under rotations.

It is interesting to analyze the behavior of $E$ and $B$ under a parity
transformation. We will consider the case where we reverse the sign of
the $x$ coordinate but leave the others unchanged. In spherical
coordinates this amounts to changing the sign of $\phi$ while 
$\theta$ and
$r$ remain the same. Under this transformation  
$\hat{{\bf e}}_\phi^\prime=-\hat{{\bf e}}_\phi$ and 
$\hat{{\bf e}}_\theta^\prime=\hat{{\bf e}}_\theta$ so the Stokes
parameters transform as $Q^{\prime}(\hat{\bi{n}}^{\prime})=Q(\hat{\bi{n}})$
and $U^{\prime}(\hat{\bi{n}}^{\prime})=-U(\hat{\bi{n}})$.
With the aid of equation (\ref{edth}) we can show that
\begin{eqnarray}
\baredth^2 (Q+iU)^{\prime}(\hat{\bi{n}}^{\prime}) &=& 
\edth^2 (Q-iU)(\hat{\bi{n}}) \nonumber \\
\edth^2 (Q-iU)^{\prime}(\hat{\bi{n}}^{\prime}) &=&
\baredth^2 (Q+iU)(\hat{\bi{n}}).
\end{eqnarray}
This implies that (equations \ref{alm}
\ref{aeb} and \ref{EBreal}), 
\begin{eqnarray}
E^{\prime}(\hat{\bi{n}}^{\prime})&=&E(\hat{\bi{n}}) \nonumber \\
B^{\prime}(\hat{\bi{n}}^{\prime})&=&-B(\hat{\bi{n}}).
\end{eqnarray}
These two new variables
behave differently under parity:
while $E$ remains unchanged $B$ changes  sign \cite{2.np}, in 
analogy  
with electric and magnetic fields.
The sign convention in equation (\ref{aeb}) makes
these expressions consistent with those defined 
previously in the small scale limit \cite{2.uros}.

To characterize the statistics of the CMB perturbations
only four power spectra are needed, 
those for $T$, $E$, $B$ and the cross correlation between $T$ and $E$.
The cross correlation between $B$ and $E$ or $B$ and 
$T$ vanishes because 
$B$ has the opposite parity to $T$ or $E$.
The power spectra are defined as the rotationally invariant quantities
\begin{eqnarray}
C_{Tl}&=&{1\over 2l+1}\sum_m \langle a_{T,lm}^{*} a_{T,lm}\rangle 
\nonumber \\
C_{El}&=&{1\over 2l+1}\sum_m \langle a_{E,lm}^{*} a_{E,lm}\rangle 
\nonumber \\
C_{Bl}&=&{1\over 2l+1}\sum_m \langle a_{B,lm}^{*} a_{B,lm}\rangle 
\nonumber \\
C_{Cl}&=&{1\over 2l+1}\sum_m \langle a_{T,lm}^{*}a_{E,lm}\rangle 
\label{Cls}
\end{eqnarray}
in terms of which,
\begin{eqnarray}
\langle a_{T,l^\prime m^\prime}^{*} a_{T,lm}\rangle&=&
C_{Tl} \delta_{l^\prime l} \delta_{m^\prime m} \nonumber \\
\langle a_{E,l^\prime m^\prime}^{*} a_{E,lm}\rangle&=&
C_{El} \delta_{l^\prime l} \delta_{m^\prime m} \nonumber \\
\langle a_{B,l^\prime m^\prime}^{*} a_{B,lm}\rangle&=&
C_{Bl} \delta_{l^\prime l} \delta_{m^\prime m} \nonumber \\
\langle a_{T,l^\prime m^\prime}^{*} a_{E,lm}\rangle&=&
C_{Cl} \delta_{l^\prime l} \delta_{m^\prime m} \nonumber \\
\langle a_{B,l^\prime m^\prime}^{*} a_{E,lm}\rangle&=&
\langle a_{B,l^\prime m^\prime}^{*} a_{T,lm}\rangle=
0. 
\label{stat}
\end{eqnarray}
The brackets $\langle \cdots \rangle$ are ensemble averages.

\subsection{Correlators in Real Space}

Sometimes it is useful to investigate the correlations 
in real space. These expressions
are needed, for example, to write the likelihood function 
of a measured CMB map. 
We rewrite 
equation (\ref{Pexpansion}) as
\begin{eqnarray}
T({\bf \hat n})&=&\sum_{lm} a_{T,lm}\ Y_{lm}({ \bf \hat n}) \nonumber \\
Q({\bf \hat n})&=&-\sum_{lm} 
a_{E,lm}\ X_{1,lm}({\bf \hat n}) 
+i a_{B,lm}\ X_{2,lm}({\bf \hat n}) \nonumber \\
U({\bf \hat n})&=&-\sum_{lm} 
a_{B,lm}\ X_{1,lm}({\bf \hat n})-i a_{E,lm}\ X_{2,lm}({\bf \hat n}).
\label{Pexpansion2}
\end{eqnarray}
We have introduced
$X_{1,lm}({\bf \hat n})=(\;_2Y_{lm}+\;_{-2}Y_{lm})/2$
and $X_{2,lm}({\bf \hat n})=(\;_2Y_{lm}-\;_{-2}Y_{lm})/ 2$.
They satisfy $X^{*}_{1,lm}=X_{1,l-m}$  and $X^*_{2,lm}=-X_{2,l-m}$ which
together with $a_{E,lm}=a_{E,l-m}^*$ and
$a_{B,lm}=a_{B,l-m}^*$ make  $Q$ and $U$ real quantities.

In fact, $X_{1,lm}({\bf \hat n})$ and $X_{2,lm}({\bf \hat n})$ 
can be written in the form
$X_{1,lm}({\bf \hat n})=\sqrt{(2l+1) / 4\pi}$ $F_{1,lm}(\theta)\ e^{im\phi}$
and $X_{2,lm}({\bf \hat n})
=\sqrt{(2l+1) / 4\pi}$ $F_{2,lm}(\theta)\ e^{im\phi}$, where 
$F_{(1,2),lm}(\theta)$ can be expressed in terms of Legendre
polynomials \cite{2.kks} \footnote{A subroutine that calculates these
functions is available at http://arcturus.mit.edu/\~{}matiasz/CMBFAST .}:
\begin{eqnarray}
F_{1,lm}(\theta)&=&2 \sqrt{(l-2)!(l-m)! \over (l+2)!(l+m)!}
\Biggl\{ -\biggl[{l-m^2 \over \sin^2\theta} 
+{1 \over 2}l(l-1)\Big]P_l^m(\cos \theta)
\biggr] \Biggr.\nonumber \\
\Biggl.
&&+(l+m) {\cos \theta \over \sin^2 \theta} 
P_{l-1}^m(\cos\theta) \Biggr\} \nonumber \\
F_{2,lm}(\theta)&=&2 \sqrt{(l-2)!(l-m)! \over (l+2)!(l+m)!}\ {m \over
\sin^2 \theta}
\biggl[ -(l-1)\cos \theta P_l^m(\cos \theta)\nonumber \\
&&+(l+m) P_{l-1}^m(\cos\theta)\biggr]. 
\label{spinuseful}
\end{eqnarray}
Note that $F_{2,lm}(\theta)=0$ if $m=0$, a necessary condition 
given that the
Stokes parameters are real.

The correlation functions can be calculated using equations (\ref{stat}) 
and (\ref{Pexpansion2}), 
\begin{eqnarray}
\langle T(1)T(2) \rangle&=&\sum_l C_{Tl} \biggl[\sum_m \;_0Y_{lm}^*(1) 
\;_0Y_{lm}(2)\biggr] \nonumber \\ 
\langle Q(1)Q(2) \rangle&=&\sum_l C_{El} \biggl[\sum_m X_{1,lm}^*(1) 
X_{1,lm}(2)\biggr]+C_{Bl}\biggl[\sum_m X_{2,lm}^*(1) 
X_{2,lm}(2)\biggr] \nonumber \\ 
\langle U(1)U(2) \rangle&=&\sum_l C_{El} \biggl[\sum_m X_{2,lm}^*(1) 
X_{2,lm}(2)\biggr]+C_{Bl}\biggl[\sum_m X_{1,lm}^*(1) 
X_{1,lm}(2)\biggr] \nonumber \\ 
\langle T(1)Q(2) \rangle&=&\sum_l C_{Cl} \biggl[\sum_m \;_0Y_{lm}^*(1) 
X_{1,lm}(2)\biggr] \nonumber \\
\langle T(1)U(2) \rangle&=&i\sum_l C_{Cl} \biggl[\sum_m \;_0Y_{lm}^*(1) 
X_{2,lm}(2)\biggr] 
\label{correl}
\end{eqnarray}
where $1$ and $2$ stand for the two directions in the sky
${\bf \hat n}_1$ and  ${\bf \hat n}_2$.
These expressions can be further simplified
using the addition theorem for the spin harmonics,
\begin{equation}
\sum_m \;_{s_1} \bar Y_{lm}^*({\bf \hat n}_1) 
\;_{s_2} Y_{lm}({\bf \hat n}_2)=\sqrt{2l+1 \over 4 \pi} 
\;_{s_2} Y_{l-s_1}(\beta,\psi_1)e^{-is_2\psi_2},
\label{addtheo}
\end{equation}
where $\beta$ is the angle between ${\bf \hat n}_1$ and
${\bf \hat n}_2$, and $(\psi_1$,$\psi_2)$ are the angles 
$(\hat e_\theta, \hat e_\phi)$ at ${\bf \hat n}_1$ and
${\bf \hat n}_2$ need to be rotated to become aligned with
the great circle going through both points.
In the case of the temperature, 
equation (\ref{addtheo}) gives the usual
relation, 
\begin{equation}
\langle T_1T_2 \rangle=\sum_l {2l+1 \over 4 \pi}
C_{Tl} P_l(\cos \beta).
\end{equation}

For polarization the addition
relations for $X_{1,lm}$ and $X_{2,lm}$ are calculated from equation 
(\ref{addtheo}),
\begin{eqnarray}
\sum_m X_{1,lm}^*(1) 
X_{1,lm}(2)&=&{2l+1 \over 4 \pi} 
\biggl[F_{1,l2}(\beta)  \cos 2\psi_1 \cos 2\psi_2 - F_{2,l2}(\beta)
\sin 2\psi_1 \sin 2\psi_2\biggr] \nonumber \\
\sum_m X_{2,lm}^*(1) 
X_{2,lm}(2)&=&{2l+1 \over 4 \pi} 
\biggl[F_{1,l2}(\beta)  \sin 2\psi_1 \sin 2\psi_2 - F_{2,l2}(\beta)
\cos 2\psi_1 \cos 2\psi_2\biggr] \nonumber \\
\sum_m X_{1,lm}^*(1) 
X_{2,lm}(2)&=&i{2l+1 \over 4 \pi} 
\biggl[F_{1,l2}(\beta)  \sin 2\psi_1 \cos 2\psi_2 + F_{2,l2}(\beta)
\cos 2\psi_1 \sin 2\psi_2\biggr] \nonumber \\ 
\sum_m \;_0Y_{lm}^*(1) 
X_{1,lm}(2)&=&{2l+1 \over 4 \pi} 
F_{1,l0}(\beta)  \cos 2\psi_2 
\nonumber \\
\sum_m \;_0Y_{lm}^*(1) 
X_{2,lm}(2)&=&-i{2l+1 \over 4 \pi} 
F_{1,l0}(\beta)  \sin 2\psi_2 .
\label{addrelpol}
\end{eqnarray}
We can equivalently write $F_{1,l0}(\beta)=\sqrt{(l-2)!/(l+2)!}
\ P_l^2(\beta)$.

The correlations in 
equation (\ref{correl}) with the sums given by equation 
(\ref{addrelpol}) are all that is 
needed to analyze any given experiment. These
relations are simple to understand \cite{2.kks}; the
natural coordinate system to express the correlations is one in which
the  ${\bf \hat e}_1$ vector for each point (1 and 2)
is aligned with the great circle
connecting the two directions, the ${\bf \hat e}_2$ vectors 
perpendicular to the ${\bf \hat e}_1$. With this choice of reference frame 
we have \cite{2.kks},
\begin{eqnarray}
\langle Q_{r}(1)Q_{r}(2) \rangle&=&\sum_l {2l+1 \over 4 \pi} \biggl[C_{El}
F_{1,l2}(\beta)-C_{Bl} F_{2,l2}(\beta)\biggr]  \nonumber \\ 
\langle U_{r}(1)U_{r}(2) \rangle&=&\sum_l {2l+1 \over 4 \pi}
\biggl[C_{Bl} F_{1,l2}(\beta)-C_{El} F_{2,l2}(\beta) \biggr] \nonumber \\ 
\langle T(1)Q_{r}(2) 
\rangle&=&-\sum_l {2l+1 \over 4 \pi} C_{Cl} F_{1,l0}(\beta)\nonumber \\
\langle T(1)U_{r}(2) \rangle&=&0 .
\label{QUr}
\end{eqnarray}
The subscript $r$ 
here indicates that the Stokes parameters are measured in this
particular coordinate system.
We can use
the transformation laws in equation (\ref{QUtrans})
to write $(Q,U)$ in terms of $(Q_r,U_r)$. When we use
equation (\ref{QUr}) for their correlations  we recover
our final result given by equations
(\ref{correl}) and (\ref{addrelpol}).

\subsection{Small Scale Limit}

There are several temperature and 
polarization experiments being planned or built. Many of them
will attempt to measure the anisotropies in small patches of
the sky. In this limit the sky can be approximated as flat and
spherical harmonics can be replaced by Fourier modes. This has
important practical benefits because it allows the use of the 
fast Fourier transforms to reduce computational time. 
In this
section we will study how polarization is described in this limit. 
In the small scale limit one considers only directions in the sky 
$\hat{\bi{n}}$ which are close to
$\hat{\bi{z}} $, in which case instead of spherical decomposition
one may use a plane wave expansion.
For temperature anisotropies
we replace 
\begin{equation}
\sum_{lm}a_{T,lm}Y_{lm}(\hat{\bi{n}}) \longrightarrow
\int d^2\bi{l}\ T(\bi{l})e^{i\bi{l}\cdot\bi{\theta}}, 
\end{equation}
so that
\begin{equation}
T(\hat{\bi{n}})=(2\pi)^{-2}\int d^2 \bi{l}\;\;
T(\bi{l})e^{i\bi{l} \cdot \bi{\theta}}. 
\label{eq220}
\end{equation}
The Fourier coefficients satisfy,
\begin{equation}
\langle T(\bi{l})T(\bi{l}^\prime)\rangle=(2\pi)^2\ C_{Tl}\ 
\delta^D(\bi{l}-\bi{l}^\prime).
\end{equation}

To expand $s=\pm 2$ weighted
functions we use
\begin{eqnarray}
_2Y_{lm}=
 \left[{(l-2)!\over (l+2)!}\right]^{1\over 2}\edth^2 Y_{lm}
&\longrightarrow&(2\pi)^{-2}{1\over l^2}\edth^2 e^{i\bi{l} \cdot \bi{\theta}}
\nonumber \\ 
_{-2}Y_{lm}=
 \left[{(l-2)!\over (l+2)!}\right]^{1\over 2}
\baredth^{2} Y_{lm}
&\longrightarrow&(2\pi)^{-2}{1\over l^2}\baredth^2 
e^{i\bi{l} \cdot \bi{\theta}}.
\end{eqnarray}
They lead to the following expression
\begin{eqnarray}
(Q+iU)(\hat{\bi{n}})&=&-(2\pi)^2\int d^2 \bi{l}\;\;
[E(\bi{l})+iB(\bi{l})] {1\over l^2}\edth^2 
e^{i\bi{l} \cdot \bi{\theta}} \nonumber \\
(Q-iU)(\hat{\bi{n}})&=&-(2\pi)^2\int d^2 \bi{l}\;\;
[E(\bi{l})-iB(\bi{l})]
{1\over l^2}\baredth^2 
e^{i\bi{l} \cdot \bi{\theta}}.
\label{SSL1}
\end{eqnarray} 
From equation (\ref{edth}) we obtain in the small scale limit
\begin{eqnarray}
{1\over l^2}\edth^2 
e^{i\bi{l} \cdot \bi{\theta}}&=& - e^{-2i(\phi-\phi_{l})}
e^{i\bi{l} \cdot \bi{\theta}} \nonumber \\
{1\over l^2}\baredth^2 
e^{i \bi{l} \cdot \bi{\theta}}&=& - e^{2i(\phi-\phi_{l})}
e^{i \bi{l} \cdot \bi{\theta}} \nonumber \\
\label{SSL2}
\end{eqnarray}
where $(l_x+il_y)=le^{i\phi_{l}}$.

The above expression was derived in the spherical basis where
$\hat{{\bi e}}_1=\hat{{\bi e}}_{\theta}$ and $\hat{{\bi e}}_2
=\hat{{\bi e}}_{\phi}$,
but in the small scale limit one can define a fixed basis in the sky
perpendicular to $\hat{\bi{z}}$,
$\hat{{\bi e}}_1'=\hat{{\bi e}}_{x}$ and $\hat{{\bi e}}_2'
=\hat{{\bi e}}_{y}$.
The Stokes parameters in the two coordinate
systems are related by
\begin{eqnarray}
(Q+iU)'&=&e^{-2i\phi}(Q+iU)\nonumber \\
(Q-iU)'&=&e^{2i\phi}(Q-iU).
\label{SSL3}
\end{eqnarray}
Combining equations (\ref{SSL1})-(\ref{SSL3}) we find
\begin{eqnarray}
Q'(\bi{\theta})&=&(2\pi)^{-2}\int d^2 \bi{l}\;\;
[E(\bi{l}) \cos(2\phi_{l})
-B(\bi{l}) \sin(2\phi_{l})]
e^{i\bi{l} \cdot \bi{\theta}} 
\nonumber \\
U'(\bi{\theta})&=&(2\pi)^{-2}\int d^2 \bi{l}\;\;
[E(\bi{l}) \sin(2\phi_{l})
+B(\bi{l}) \cos(2\phi_{l})]
e^{i\bi{l} \cdot \bi{\theta}}. 
\label{QUreal}
\end{eqnarray}
These relations agree with those given in \cite{2.uros}, which were 
derived in the small scale approximation. As already shown there,
power spectra and correlation functions for $Q$ and $U$ used in 
previous work on the subject
\cite{2.crittenden93,2.frewin,2.coulson,2.coutur,2.kosowsky96} 
can be simply derived from these
expressions. 

In the small scale limit the correlation functions are (in 
their  natural coordinate system
denoted with an $r$ in equation \ref{QUr}), 
\begin{eqnarray}
C_T({\theta})&=&\int {d^2{\bi l}\over (2\pi)^2}\   
e^{il\theta \cos \phi_{l}}
\ \ C_{ Tl} \nonumber \\
C_Q({\theta})&=&\int {d^2{\bi l}\over (2\pi)^2}\  
e^{il\theta \cos \phi_{l}}
 \ [C_{El} \cos^2(2\phi_{\bi l})
+C_{Bl} \sin^2(2\phi_{\bi l})]\nonumber \\
C_U({\theta})&=&\int {d^2{\bi l}\over (2\pi)^2}\  
e^{il\theta \cos \phi_{l}}
 \ [C_{El} \sin^2(2\phi_{\bi l})
+C_{Bl} \cos^2(2\phi_{\bi l})] \nonumber \\
C_C({\theta})&=&\int {d^2{\bi l}\over (2\pi)^2}\
  e^{il\theta \cos \phi_{l}}
\ C_{Cl} \cos(2\phi_{\bi l}),
\end{eqnarray}
or equivalently,
\begin{eqnarray}
C_T(\theta)&=&\int {l dl\over 2\pi}
\ C_{Tl}\ J_0(l\theta) \nonumber \\ 
C_{Q}(\theta)+C_{U}(\theta)&=&\int {l dl\over 2\pi}
\ (C_{El}+C_{Bl})\ 
J_0(l\theta) \nonumber \\ 
C_{Q}(\theta)-C_{U}(\theta)&=&\int {l dl\over 2\pi}
\ (C_{El}-C_{Bl})\ 
J_4(l\theta) \nonumber \\ 
C_C(\theta)&=&-\int {l dl\over 2\pi}
\ C_{Cl}\ J_2(l\theta), 
\label{fullcorr}
\end{eqnarray}
$J_\nu$ are cylindrical Bessel functions.

These relations can be inverted to obtain the power spectra in terms
of the correlations,
\begin{eqnarray}
C_{Tl}&=&2\pi \int_0^\pi \theta d\theta \ C_T(\theta) \ J_0(l\theta) \nonumber \\
C_{El}&=&2\pi \int_0^\pi \theta d\theta \
\{ [C_Q(\theta)+C_U(\theta)]\ J_0(l\theta) 
+ [C_Q(\theta)-C_U(\theta)]\ J_4(l\theta) \} \nonumber \\
C_{Bl}&=&2\pi \int_0^\pi \theta d\theta \
\{ [C_Q(\theta)+C_U(\theta)]\ J_0(l\theta) 
- [C_Q(\theta)-C_U(\theta)]\ J_4(l\theta) \} \nonumber \\
C_{Cl}&=&-2\pi \int_0^\pi \theta d\theta \
C_C(\theta)\ J_2(l\theta). 
\label{defcl}
\end{eqnarray}

\subsection{Analysis of All-Sky Maps}

In this section we discuss issues related to simulating and
analyzing all-sky polarization and temperature
maps. This should be specially useful
for future satellite missions \cite{2.map,2.cobra}, which will measure 
temperature anisotropies and polarization over 
the whole sky
with  high angular resolution. 
Such an all-sky analysis will be of particular importance 
if reionization and tensor fluctuations
are important, in which case polarization will have
useful information on large angular scales (Chapter \ref{chapinfpol}),
where Fourier analysis 
(i.e. division of
the sky into locally flat patches) is not possible. In addition, it is
important
to know how to simulate an all-sky map which preserves proper correlations
between neighboring patches of the sky and with which small scale
analysis can be tested for possible biases.

To make an all-sky map we need to generate the multipole moments
$a_{T,lm}$, $a_{E,lm}$ and $a_{B,lm}$.
This can be done by a generalization of the method given
in \cite{2.uros}. For each $l$ one diagonalizes
the correlation matrix $M_{11}=C_{Tl}$,
$M_{22}=C_{El}$, $M_{12}=M_{21}=C_{Cl}$ and generates 
from a normalized gaussian distribution two pairs of
random numbers (for real and imaginary components of $a_{l\pm m}$).
Each pair is multiplied with the square root of
eigenvalues of $M$ and rotated back to the original frame.
This gives a realization of $a_{T,l\pm m}$ and $a_{E,l\pm m}$ with correct
cross-correlation properties. For $a_{B,l\pm m}$ the procedure is simpler,
because it does not cross-correlate with either $T$ or $E$, so a pair
of gaussian random variables is multiplied with $C_{Bl}^{1/2}$ to
make a realization of $a_{B,l\pm m}$. 

Once $a_{E,lm}$ and $a_{B,lm}$ are generated we can form their linear
combinations $a_{2,lm}$ and $a_{-2,lm}$.
Finally, to make a map of $Q(\hat{\bi n})$ and $U(\hat{\bi n})$ in the
sky we perform the sum in equation (\ref{Pexpansion}).
To reconstruct the polarization
power spectrum from a map of $Q(\hat{\bi n})$ and
$U(\hat{\bi n})$ one first combines them in $Q+iU$ and $Q-iU$ to obtain
spin $\pm 2$ quantities. Performing the integral
over ${}_{\pm 2}Y_{lm}$ (equation \ref{alm})
projects out ${}_{\pm 2s}a_{lm}$, from which $a_{E,lm}$ and $a_{B,lm}$
can be obtained. It is important to remember that in this treatment
$Q$ and $U$ are defined using the spherical coordinate system.

Once we have the multipole moments we can 
construct various power spectrum estimators and analyze their
variances. In the case of full sky coverage one may generalize the
approach in \cite{2.knox95} to estimate the variance in the power spectrum 
estimator in the presence of noise. We will assume that we are given a
map of temperature
and polarization with $N_{pix}$ pixels and 
that the noise is uncorrelated from pixel to pixel
and also between $T$, $Q$ and $U$. 
The rms noise in the temperature is
$\sigma_T$ and that in $Q$ and $U$ is $\sigma_P$. If temperature and
polarization are obtained from the same experiment by adding and
subtracting the intensities between two orthogonal polarizations then
the rms noise in temperature and polarization
are related by $\sigma_T^2=\sigma_P^2/2$.

Under these conditions and using the orthogonality
of the $\;_sY_{lm}$ we obtain the statistical property of noise,
\begin{eqnarray}
\langle (a_{T,lm}^{{\rm noise}})^{*}a^{{\rm noise}}_{T,l^{\prime}m^{\prime}}\rangle 
&=& {4\pi \sigma_T^2 \over N_{pix}} 
\delta_{l l^{\prime}} \delta_{m m^{\prime}}
\nonumber \\
\langle (a^{{\rm noise}}_{2,lm})^{*}a^{{\rm noise}}_{2,l^{\prime}m^{\prime}}\rangle 
&=& {8\pi \sigma_P^2 \over N_{pix}} 
\delta_{l l^{\prime}} \delta_{m m^{\prime}}
\nonumber \\
\langle (a^{{\rm noise}}_{-2,lm})^{*}a^{{\rm noise}}_{-2,l^{\prime}m^{\prime}}\rangle 
&=& {8\pi \sigma_P^2 \over N_{pix}} 
\delta_{l l^{\prime}} \delta_{m m^{\prime}}
\nonumber \\
\langle (a^{{\rm noise}}_{-2,lm})^{*}
a^{{\rm noise}}_{2,l^{\prime}m^{\prime}}\rangle 
&=& 0.
\end{eqnarray}
By assumption there are no correlations between the noise in 
temperature and polarization. 
With these and equations
(\ref{aeb}) and (\ref{stat}) we find 
\begin{eqnarray}
\langle a_{T,lm}^{*}a_{T,l^{\prime}m^{\prime}}\rangle 
&=& (C_{Tl} e^{-l^2 \sigma_b^2} + w_T^{-1})
\delta_{l l^{\prime}} \delta_{m m^{\prime}}
\nonumber \\
\langle a_{E,lm}^{*}a_{E,l^{\prime}m^{\prime}}\rangle 
&=& (C_{El} e^{-l^2 \sigma_b^2} + w_P^{-1}) 
\delta_{l l^{\prime}} \delta_{m m^{\prime}}
\nonumber \\
\langle a_{B,lm}^{*}a_{B,l^{\prime}m^{\prime}}\rangle 
&=& (C_{Bl} e^{-l^2 \sigma_b^2} +w_P^{-1}) 
\delta_{l l^{\prime}} \delta_{m m^{\prime}}
\nonumber \\
\langle a_{E,lm}^{*}a_{T,l^{\prime}m^{\prime}}\rangle 
&=& C_{Cl} e^{-l^2 \sigma_b^2} 
\delta_{l l^{\prime}} \delta_{m m^{\prime}}
\nonumber \\
\langle a_{B,l^\prime m^\prime}^{*} a_{E,lm}\rangle&=&
\langle a_{B,l^\prime m^\prime}^{*} a_{T,lm}\rangle=
0. 
\label{almvar}
\end{eqnarray}
For simplicity we characterized the beam smearing by 
$e^{l^2 \sigma_b /2}$ where $\sigma_b$ is the gaussian size of the beam
and we defined $w_{T,P}^{-1}=4\pi\sigma_{T,P}^2/N_{pix}$
\cite{2.uros,2.knox95}.

The estimator for the temperature power spectrum is \cite{2.knox95},
\begin{eqnarray}
\hat{C}_{Tl}&=&\left[\sum_m{ |a_{T,lm}|^2 \over 2l+1} -  w^{-1}_T 
\right]e^{l^2\sigma_b^2}
\end{eqnarray}
Similarly for polarization and cross correlation the optimal
estimators are given by,
\begin{eqnarray}
\hat{C}_{El}&=&\left[\sum_m{ |a_{E,lm}|^2 \over 2l+1} -  w^{-1}_P 
\right]e^{l^2\sigma_b^2}\nonumber \\
\hat{C}_{Bl}&=&\left[\sum_m{ |a_{B,lm}|^2 \over 2l+1}-  w^{-1}_P 
\right]e^{l^2\sigma_b^2}\nonumber \\
\hat{C}_{Cl}&=&\left[\sum_m{ (a_{E,lm}^{*}a_{T,lm}+a_{E,lm}a_{T,lm}^{*}) 
\over 2(2l+1)}\right]e^{l^2 \sigma_b^2}.
\end{eqnarray}

The covariance matrix between the different estimators, 
${\rm Cov }(XX^{\prime})=\langle (\hat X - \langle \hat X \rangle)
(\hat X^{\prime} - \langle \hat X^{\prime} \rangle)\rangle$
is easily calculated using equation (\ref{almvar}). 
The diagonal terms are given by
\begin{eqnarray}
{\rm Cov }(\hat{C}_{Tl}^2)&=&{2\over 2l+1}({C}_{Tl}+
w_T^{-1}e^{l^2 \sigma_b^2})^2
\nonumber \\
{\rm Cov }(\hat{C}_{El}^2)&=&{2\over 2l+1}({C}_{El}+
w_P^{-1}e^{l^2 \sigma_b^2})^2
\nonumber \\
{\rm Cov }(\hat{C}_{Bl}^2)&=&{2\over 2l+1}({C}_{Bl}+
w_P^{-1}e^{l^2 \sigma_b^2})^2
\nonumber \\
{\rm Cov }(\hat{C}_{Cl}^2)&=&{1\over 2l+1}\left[{C}_{Cl}^2+
({C}_{Tl}+w_T^{-1}e^{l^2 \sigma_b^2})
({C}_{El}+w_P^{-1}e^{l^2 \sigma_b^2})\right].
\label{covdiag}
\end{eqnarray}
The non-zero off diagonal terms are
\begin{eqnarray}
{\rm Cov }(\hat{C}_{Tl}\hat{C}_{El})&=&{2\over 2l+1}{C}_{Cl}^2
\nonumber \\
{\rm Cov }(\hat{C}_{Tl}\hat{C}_{Cl})&=&{2\over 2l+1}{C}_{Cl}
(C_{Tl}+w_T^{-1}e^{l^2 \sigma_b^2})
\nonumber \\
{\rm Cov }(\hat{C}_{El}\hat{C}_{Cl})&=&{2\over 2l+1}{C}_{Cl}
(C_{El}+w_P^{-1}e^{l^2 \sigma_b^2}).
\end{eqnarray}
Note that the theoretical analysis is significantly more
complicated if all four power spectrum estimators are used to deduce
the underlying cosmological model. For example, to test the sensitivity of 
the spectrum to the underlying parameters one uses the Fisher information
matrix approach \cite{2.jungman,2.zss}. If only temperature information is
given then for each $l$ a derivative of the temperature
spectrum with respect to the parameter under investigation is computed
and this information is then summed over all $l$ weighted  
by ${\rm Cov }^{-1}(\hat{C}_{Tl}^2)$. 
In the more general case discussed here instead of a single derivative 
we have a vector of four derivatives and the
weighting is given by the inverse of the covariance matrix,
\begin{equation}
\alpha_{ij}=\sum_l \sum_{X,Y}{\partial C_{Xl} \over \partial s_i}
{\rm Cov}^{-1}(C_{Xl}C_{Yl}){\partial C_{Yl} \over \partial s_j},
\label{intfisher}
\end{equation}
where $\alpha_{ij}$ is the Fisher information or curvature 
matrix, ${\rm Cov}^{-1}$ is the inverse of the covariance matrix,
$s_i$ are the cosmological parameters one would like to 
estimate and $X,Y$ stands for $T,E,B,C$. For each $l$ one has to
invert the covariance matrix and sum over $X$ and $Y$,
which makes the numerical evaluation of this expression significantly
more involved.

\section{Understanding E and B Polarization}

In the previous sections
we have discussed in detail the mathematical
formalism needed to characterize the anisotropies in the CMB. In this 
subsection we want to take another look at the new polarization
variables we introduced, $E$ and $B$. By doing so we will develop 
further intuition as to what $E$ and $B$ physically mean.
We will also study the power spectra and correlation functions for
SCDM to point out the similarities 
and differences between the temperature and polarization anisotropies.

\subsection{E and B in Real Space}

The main characteristic of the polarization field is that it is a
spin-2 field. The new variables $E$ and $B$ allow us to describe it more
conveniently in terms of two spin zero quantities. So far we have done
this in Fourier space (more generally in  $l$-space when not working in the
small scale limit). In this section
we want to explore $E$ and $B$-type polarizations
directly in real space. We will work in the small scale limit to
make the notation simpler; the generalization to all sky modes is
straightforward.

As in equation (\ref{eq220}) we can define
\begin{eqnarray}
E(\bi{\theta})&=&(2\pi)^{-2}\int d^2{\bi l}\ 
e^{i{\bi l}\cdot{\bi \theta}}\ E({\bi l}) \nonumber \\
B(\bi{\theta})&=&(2\pi)^{-2}\int d^2{\bi l}\ 
e^{i{\bi l}\cdot{\bi \theta}}\ B({\bi l}).
\label{ebreal1}
\end{eqnarray}
These two quantities  describe completely the polarization field.
Figure \ref{equps} shows the $E$ polarization field and the
polarization vectors for the same simulation shown in Figure \ref{tqups}.
There is only $E$-type polarization associated with SCDM because
this is the only pattern that is produced by density perturbations and
we did not include a stochastic background of gravity waves.
We will discuss this point further in the next section and in Chapter
\ref{chapinfpol}.

In Figure \ref{equps} we can see that hot spots of
the $E$ map correspond to points with tangential polarization patterns
(negative $Q_r$). We find radial polarization patterns around the 
cold spots of $E$. The polarization pattern in our simulation did not
have any $B$-type polarization so to get a
better intuition as to what $B$-type polarization means we can take 
the same $E$ field in Figure \ref{equps} and pretend that it was
actually $B$-type polarization. 
The polarization vectors are shown
in Figure \ref{bqups}, they  correspond to the ones in 
Figure \ref{equps} rotated by $45^o$.  Hot and cold spots of the
$B$ field correspond to places where the polarization vectors
circulate around in opposite directions. It is clear from this Figure
that the polarization pattern is not invariant under reflections. 
The distinction between $E$ and $B$-type of polarization is not in the
size or orientation of the polarization vector at a given point but in
the relative orientations of the vectors, namely the pattern.

We can stress the characteristics of $E$ and $B$-type of polarization 
by looking at Figure
\ref{2.ebpatt} which summarize what we found in the simulated
maps. Places on the sky where the polarization pattern is tangential
correspond to hot spots of $E$ while a radial pattern will
produce a cold spot. The patterns that give rise to $B$ are
different in nature; while the radial and tangential patterns are
symmetric under reflections those that create $B$ are not. 
One pattern transforms into the other under a parity
transformation. This is precisely what we mean when we say $B$ is a 
pseudoscalar, if the pattern of polarization vectors on the sky was
such that we had a hot spot of $B$ when we looked at the mirror
image of that pattern we would have a cold spot of $B$.

We can use Figure \ref{2.ebpatt} to argue why density perturbations
cannot create $B$-type polarization while gravity waves can, 
although the detailed mathematical proof is left for Chapter \ref{chaplos}.
We can consider only one Fourier mode of density perturbations because
an arbitrary density field can always be expanded in Fourier modes.
In the linear regime the total $B$ generated by all the
modes will be the sum of the contributions by each mode. 
We will prove the total $B$-type polarization is  
zero because the $B$ polarization generated every mode is zero.

Figure \ref{2.densp2} illustrates our point.
In the upper half we consider the perturbations induced by a single
density mode.  
The problem has two important symmetries, symmetry under rotation
around $\hat{\bi k}$ and symmetry under a reflection about
any plane containing $\hat{\bi k}$. It is clear that these two symmetries
do not allow for a non zero $U$ in the coordinate system 
shown in the figure. The polarization vectors must be either along
$\hat {\bi e}_\theta$ or $\hat {\bi e}_\phi$. The two possible
orientations of the polarization consistent with the symmetries are
shown in the upper right hand side. This can be shown to imply
that $B$-type polarization is zero. We first focus on 
the particular case when 
the direction of observation coincides $\hat{\bi k}$ because 
it is easier to visualize.
We can use Figure \ref{2.ebpatt} with 
to recognize the polarization patterns that are
allowed around $\hat{\bi k}$. The two $B$-type polarization patterns 
cannot be produced because they are not invariant under parity.
We leave the generalization of the argument to directions of
observation different than $\hat{\bi k}$ for later, when we introduce
a way of calculating $E$ and $B$ directly in real space. 

\begin{figure*}
\begin{center}
\leavevmode
\end{center}
\caption{Simulated polarization map for SCDM($2.5^o\times 2.5^o$
field). The polarization vectors are shown  together with the $E$-type
polarization. The $E$ field varies between $-140\mu K < E < 158 \mu
K$, the largest polarization vector has an amplitude $P=128 \mu K$.}
\label{equps}
\end{figure*}

\cleardoublepage

\vspace*{10cm}

\cleardoublepage

Gravitational waves do not possess either rotational or reflection
symmetry. We illustrate this in Figure \ref{2.densp2} using the
deformation of a ring of test particles in the $\hat{\bi x}-\hat{\bi
y}$ plane when a gravitational wave traveling along the $\hat{\bi z}$
passes by. The influence of a gravitational wave does not have the symmetry
under reflections that made zero the $B$-type polarization for density
perturbations and will thus produce some $B$.

We can 
combine equation (\ref{ebreal1}) 
with the definition of $E({\bi l})$ and
$B({\bi l})$  in terms of $Q$ and $U$,
\begin{eqnarray}
E(\bi{l})&=&\int d^2{\bi \theta}\
[Q(\bi{\theta})\cos(2\phi_l)+U(\bi{\theta})\sin(2\phi_l)] \ e^{-i{\bi l}\cdot
{\bi \theta}} \nonumber \\
B(\bi{l})&=&\int d^2{\bi{\theta}}\
[U(\bi{\theta})\cos(2\phi_l)-Q(\bi{\theta})\sin(2\phi_l)] \ e^{-i{\bi l}\cdot
{\bi \theta}}, 
\end{eqnarray}
to obtain an expression relating $E$ and $B$ in real space directly to
$Q$ and $U$, 
\begin{eqnarray}
E(\bi{\theta})&=&-\int d^2{\bi \theta}^{\prime}\ 
\omega(\tilde \theta)\ [Q({\bi \theta}^{\prime})
\cos(2\tilde\phi) - U({\bi \theta}^{\prime})
\sin(2\tilde\phi)]\nonumber \\
&=&-\int d^2{\bi \theta}^{\prime}\ 
\omega(\tilde \theta)\ Q_r({\bi \theta}^{\prime}) \nonumber \\
B(\bi{\theta})&=&-\int d^2{\bi \theta}^{\prime}\ 
\omega(\tilde \theta)\ [U({\bi \theta}^{\prime})
\cos(2\tilde\phi) + Q({\bi \theta}^{\prime})
\sin(2\tilde\phi)]\nonumber \\
&=&-\int d^2{\bi \theta}^{\prime}\ 
\omega(\tilde \theta)\ U_r({\bi \theta}^{\prime}).
\label{ebreal2}
\end{eqnarray}
The variables $(\tilde \theta,\tilde \phi)$ are the polar coordinates
of the vector $\bi{\theta}-\bi{\theta}^{\prime}$.
In equation (\ref{ebreal2})
$Q_r$ and $U_r$ are the Stokes parameters in the
polar coordinate system centered at $\bi{\theta}$. For example
if $\bi{\theta}$ is zero, $Q_r=\cos 2\phi^{\prime}\
Q(\bi{\theta}^{\prime}) - \sin 2\phi^{\prime}\
U(\bi{\theta}^{\prime})$ and $U_r=\cos 2\phi^{\prime}\
U(\bi{\theta}^{\prime}) + \sin 2\phi^{\prime}\
Q(\bi{\theta}^{\prime})$. The window can be shown to be 
 $\omega(\theta)=1/\pi \theta^2\;
(\theta\neq 0)$, $\omega(\theta)=0 \;
(\theta= 0)$. We can read directly from equation (\ref{ebreal2}) the
relations between the pattern of polarization and the sign of $E$ and
$B$, for example a tangential pattern of polarization which has
negative $Q_r$ will produce positive $E$ (hot spot of $E$). 

From equation (\ref{ebreal2}) we can understand 
how we achieve two rotationally invariant quantities out
of the spin 2 field: to get $E({\bi \theta})$ and $B({\bi \theta})$ 
we average the
values of $Q_r$ and $U_r$ respectively, 
over circles centered at ${\bi
\theta}$. Each circle is weighted by $\omega({\tilde \theta})$. By
construction these two quantities are rotationally invariant: the
Stokes parameters $Q_r$ and $U_r$ do not depend on the coordinate
system, they are defined relative to the 
$\bi{\theta}-\bi{\theta}^{\prime}$
vector and the weight function $\omega$ is also rotationally invariant.
We are
giving the same weight to all the points in each circle and we are
using the Stokes parameters defined in their natural coordinate
system. The variable $B$ is clearly a pseudoscalar because it is the average
of $U_r$ and $U_r$ changes sign under parity.

We can use Equation (\ref{ebreal2}) to prove that $B$-type
polarization is zero in any direction for a single density mode. 
We had argued that this was the case when the direction of observation
and $\hat{\bi k}$ where the same. In general the two directions 
form an angle and $\hat{\bi k}$ will intersect the plane perpendicular
to $\hat{\bi n}$ at a point different from $\hat{\bi n}$. We
illustrate this in Figure \ref{2.densp2} where we considered the case
in which  the polarization vectors are radial around $\hat{\bi k}$, but the
argument is the same if they were tangential (the only allowed
patterns at a given distance away from $\hat{\bi k}$). To compute $B$ we
integrate $U_r$ along circles centered at $\hat{\bi n}$, 
but the polarization pattern has
reflection symmetry across the line connecting $\hat{\bi k}$ and 
$\hat{\bi n}$ which implies that on opposite sides of the circle
$U_r$ has different signs and the integral cancels. 

There is nothing special about the weight $\omega(\theta)$ as far as
constructing a scalar and a pseudo-scalar, but this choice of weight
has several important properties. It makes
the correlation functions of $E$ and $B$ the closest to that of $Q$ and
$U$. For example with this choice
$C_Q(\theta)+C_U(\theta)=C_E(\theta)+C_B(\theta)$. Another property is
that this choice of weight preserves the nature of white noise. White
noise in $Q$ and $U$ becomes white noise in $E$ and $B$. On the other
hand this choice of window requires integration over all $\theta$ and
is non-local. 

\begin{figure*}
\begin{center}
\leavevmode
\end{center}
\caption{Simulated polarization map for SCDM($2.5^o\times 2.5^o$
field) where the $E$-type polarization has been
changed into $B$-type. The polarization vectors of the this figure are
rotated $45^o$ respect to the ones in previous ones.} 
\label{bqups}
\end{figure*}

\cleardoublepage

\vspace*{10cm}

\cleardoublepage

Another expression for $E({\bi \theta})$  and $B({\bi \theta})$ can be
obtained in terms of derivatives,
\begin{eqnarray}
E(\bi{\theta})&=&\nabla^{-2}(\partial^2_x-\partial_y^2)\ Q({\bi \theta})
+ \nabla^{-2} (2\partial_{x}\partial_y)\ U({\bi \theta}) \nonumber \\
B(\bi{\theta})&=&\nabla^{-2}(\partial^2_x-\partial_y^2)\ U({\bi \theta})
- \nabla^{-2} (2\partial_{x}\partial_y)\ Q({\bi \theta}). 
\label{ebrealder}
\end{eqnarray} 
This equation makes the point that the extra $l$ factors that would be
introduced in the power spectra by the derivatives are compensated by the
inverse Laplacian.  This is what we mean when we say that the spectrum
of $E$ and $B$ are the closest to that of $Q$ and
$U$. The inverse Laplacian is responsible for the
non-local nature of $E(\bi{\theta})$ and $B(\bi{\theta})$ evident also
in equation (\ref{ebreal2}).

\subsection{Analogy with Weak Lensing}\label{weaklens}

There is a very close analogy between the $E$ and $B$ formalism we are
using to describe CMB polarization and the mathematical framework used
to study weak lensing. As photons travel
across the universe their trajectories are deflected. As a result the
observed intensity is related to the true one by
$I_{obs}({\bi \theta})=I_{true}({\bi \theta}+\delta{\bi \theta})$. This
mapping will induce ellipticities in the images of the observed
galaxies. The deformation tensor ${\bf \Phi}_{ij}$ ($i,j=x,y$ in the small
scale limit) can be calculated in
terms of a projected gravitational potential
$\phi_{proj}$\footnote{The interested reader may find detailed
expressions for the projected potential and other quantities in
\cite{2.uroslens}.}, 
\begin{equation}
{\bf \Phi}_{ij}=\partial_i\partial_j \phi_{proj}.
\label{deftens}
\end{equation} 

The deformation tensor is usually decomposed into trace $\kappa$ and
the two components of the shear $\gamma_1$ and $\gamma_2$, 
${\bf \Phi}_{xx}=-\kappa-\gamma_1$, ${\bf
\Phi}_{yy}=-\kappa+\gamma_1$, ${\bf \Phi}_{xy}={\bf \Phi}_{yx}
=-\gamma_2$. The trace $\kappa$ is also the dimensionless 
projected mass density.
The shear part of the tensor can be 
obtained from the observed ellipticities of
the galaxies \cite{2.kaisqui}. 

\begin{figure*}
\begin{center}
\leavevmode
\end{center}
\caption{Polarization patterns that lead to positive and negative
values of the $E$ and $B$ polarization 
fields. The Stokes parameters are measured
in the polar coordinate system centered at the cross. All four
patterns are invariant under rotation but the two patterns that
generate $B$ are not invariant under reflections.} 
\label{2.ebpatt}
\end{figure*}

\begin{figure*}
\begin{center}
\leavevmode
\end{center}
\caption{Upper half:
A single mode of density perturbations has symmetries under
rotation around the $\hat{\bf k}$ axis and reflection about any plane
containing it. Only $I$ and $Q$ Stokes parameters can be present in this
reference frame and thus no $B$ polarization is created. The integral
along a circle of $U_r$, used to calculate $B$ directly in real space,
is zero because the pattern of polarization 
is symmetric under reflections across the line containing 
$\hat{\bf k}$ and $\hat{\bf n}$. Gravitational
waves do not have these symmetries as illustrated by the deformation
suffered by a ring of test particles as wave traveling along
$\hat{\bf z}$ passes by.} 
\label{2.densp2}
\end{figure*}

The two quantities $\gamma_{(1,2)}$ are the analogues of $(Q,U)$ for
the CMB polarization field. They also form spin two variables $\gamma_1 \pm i
\gamma_2$. Equations (\ref{ebrealder}) and (\ref{deftens}) can be used
to show that the $E$ variable in this case is nothing but the
dimensionless projected mass density $\kappa$,  
\begin{equation}
E^{WL}({\bi \theta})=\nabla^{-2}(\partial^2_x-\partial_y^2)\ 
\gamma_1({\bi \theta})
+ \nabla^{-2} (2\partial_{xy})\ \gamma_2({\bi \theta})
= \nabla^2 \phi_{proj}({\bi \theta})=\kappa({\bi \theta})
\end{equation}
Thus in the case of weak lensing $E$ has an important physical:
interpretation it is the projected mass density. This  
is clear in Figure \ref{equps}: hot
spots of the $E$ field are the analog of high mass concentrations,
so we see in the polarization vectors the typical orientation of the
shear that we see in the deformation of galaxies around clusters.  
The analogous calculation of $B$ for weak lensing 
is identically zero,
as it is the $\hat {\bi z }$ component 
of the curl of the gradient of the projected
potential. 

\section{Correlators for SCDM}\label{corrcdm}

\begin{figure*}
\begin{center}
\leavevmode
\end{center}
\caption{Intensity patterns incident on an electron and the resulting
polarization of the scattered light in the direction perpendicular to
the page. The dashed line indicates a smaller intensity while the
full lines represent an excess in the number of photons. 
A uniform intensity field or a
dipole pattern will produce no polarization. The rods around each
diagram represent the direction of the polarization of the scattered
light which come from that particular direction before the scattering. The
total scattered light can be obtained by ``adding'' the four rods
whose length is proportional to the amount of photons incident on the
electron from each direction. Only the quadrupolar pattern
will induce some polarization, because the summed lengths of the top
and bottom rods differs from that of the left and right rods.} 
\label{trthom}
\end{figure*}

In this section we present the correlators for COBE normalized
SCDM to point out some of the properties of CMB anisotropies
and to understand their physical origin. Hydrogen recombination plays
a crucial role in the generation of CMB anisotropies. As the universe
expands it cools and around $3\times 10^5$ years after the big bang
hydrogen atoms are able to recombine to form neutral hydrogen 
and the universe becames
transparent to the CMB photons. Recombination occurs very quickly. Before
recombination photons and electrons scatter very efficiently forming a
single fluid, while after recombination the photons are free to travel to
the observer. The photons we detect coming from a particular direction
come from a very small region of our universe (of size $\sim 50\
h^{-1}$ Mpc) at a distance $D \sim 6000\ h^{-1}$ Mpc away from us. The
collection of all this regions in the universe is called the 
last scattering surface. 

In order to understand the
main features in Figure \ref{tqups} one can work in the   
thin scattering surface approximation, where one assumes recombination
occurs instantaneously. In this approximation 
the final temperature fractional 
anisotropy in direction $\hat {{\bf n}}$ on the sky is 
\begin{eqnarray}
T(\hat {{\bf n}})={\frac{\delta _R}4}|_{\tau _{*}}-\hat {{\bf n}}\cdot {\bf {%
v_R}}|_{\tau _{*}}-{\frac 12}\int_{\tau _{*}}^{\tau _0}d\tau \dot h_{ij}\hat
{{\bf n}}^i\hat {{\bf
n}}^j,  
\label{tempapprox}
\end{eqnarray}
$\delta_R$ and ${\bf { v_R}}$ stand for the relative density
perturbation and velocity of
the photon-baryon plasma while $h_{ij}$ describes the perturbation in
the metric. Conformal time is denoted with $\tau$; $\tau_0$ and
$\tau_{*}$ correspond to the times today and at recombination, respectively.
The first two terms are evaluated at the last scattering surface and the
third term is an integral along the line of sight. The anisotropies
have three distinct origins: an overdensity of the photon baryon fluid
somewhere on the last scattering translates 
into an increase in the measured 
temperature when we observe in that particular direction 
(first term). Furthermore, if the fluid is
moving there will be an additional Doppler shift contribution 
(second term). Finally, gravitational redshift will change the photon 
temperature as photons travel toward us (third term). 
Note that this effect can take place anywhere along the path. 

In order to generate polarization 
we need Thomson scattering between photons and electrons, which means
that polarization cannot be generated after recombination (if there is
no reionization). But Thomson scattering is not enough,
the radiation 
incident on the electrons must also
be anisotropic. Figure \ref{trthom} 
illustrates this point. If the radiation incident on the
electron is isotropic, by symmetry there can be no net
polarization after the scattering. We consider a dipole
anisotropy next: in the figure we illustrate the case where the
intensity of the incident radiation is higher from the top and lower from the
bottom, with the average intensity incident from the sides. 
The scattered radiation from photons incident from either  the
bottom or top will be polarized in the horizontal direction, while
that coming from the sides will vertically polarized. 
The total degree of polarization is obtained by looking at all 
the scattered light, which in this case
is unpolarized because the excess of radiation coming from above
(relative to the sides) is
compensated by the smaller amount coming from the bottom. Since a
dipole pattern of anisotropies will not create any
polarization,  
we need a quadrupolar pattern of
intensity: an excess in photons must come from both top and bottom.

In the context of CMB, velocity gradients in the photon-baryon 
fluid will create the
quadrupole that generates polarization. 
The photons that are
scattered from a given electron come 
from places where the fluid has velocity ${\bf {v}
}$. Because of the tight coupling between photons and
electrons the photon distribution function
has a dipole term 
$T_1=\hat {{\bf n}}\cdot {\bf {v}}$. Furthermore, gradients in
the velocity field across the mean free path of the photons ($\lambda _p$)
create a quadrupole $T_2=\lambda _pn^in^j\partial _iv_{Rj}$ in the photon
distribution as seen in the rest frame of the electron. 
The velocity
of the fluid will only create a dipole; one needs at least a velocity
gradient to create a quadrupole. 

The
scattered radiation field is given by $(Q+iU)=-3/4\sigma _T\int d\Omega
^{\prime }/4\pi ({\bf m}\cdot \hat {{\bf n}}^{\prime })T_2(\hat {{\bf n}%
}^{\prime })\propto \lambda _p{\bf m}^i{\bf m}^j\partial
_iv_j|_{\tau_*}$, 
where $\sigma _T$ is the Thomson scattering cross section and we have
written the
scattering matrix as $P({\bf m},\hat {{\bf n}}^{\prime })=-3/4\sigma _T|{\bf %
m}\cdot \hat {{\bf n}}^{\prime }|^2$ with ${\bf m}=\hat {{\bf e}}_1+i\hat {%
{\bf e}}_2$ . In the last step we integrated over all directions of the
incident photons $\hat {{\bf n}}^{\prime }$.
As photons decouple from the baryons their mean free path grows very
rapidly, so a more careful analysis is needed to obtain the final
polarization\footnote{The velocity in this
equation is in the conformal gauge.}\cite{2.zalhar}, 
\begin{eqnarray}
(Q+iU)(\hat {{\bf n}})\approx 0.17\Delta \tau _{*}{\bf m}^i{\bf m}^j\partial
_iv_j|_{\tau _{*}}  
\label{polapprox}
\end{eqnarray}
where $\Delta \tau _{*}$ is the width of the last scattering surface and is
giving a measure of the distance photons travel between their last two
scatterings. The appearance of $m^i m^j$ in equation (\ref{polapprox}) assures
that $(Q+iU)$ transforms correctly under rotations of $(\hat {{\bf e}%
}_1,\hat {{\bf e}}_2)$. 

We can  combine equations (\ref{ebrealder}) and (\ref{polapprox}) 
to understand why $B$-type
polarization is zero for anisotropies produced by density
perturbations.  
We take
$\hat {{\bf n}}=\hat {{\bf z}}$ and ${\bf m}=\hat {{\bf x}}+i\hat {
{\bf y}}$. We then have $\nabla ^2B\propto(\partial
_y^2-\partial _x^2)U+2\partial _x\partial _yQ$ which gives $
\nabla ^2B\propto \nabla ^2(\hat {{\bf z}}\cdot 
{\bf {\nabla }\times {v})}$ and
is zero because the velocity field produced by density perturbations is
irrotational, ${\bf v}=\nabla \psi$ and $\nabla \times \nabla \psi=0$. 
On the other hand $\nabla ^2E \propto \nabla
^2(\partial_xv_x+\partial_y v_y)$.

\begin{figure*}
\begin{center}
\leavevmode
\end{center}
\caption{Two dimensional universe analogy. A particular density mode
is oscillating in time before decoupling. At decoupling the photons
are free to escape and have traveled a distance D to reach the
observer (at the center of the circle). 
Although the density field has continued to evolve 
after recombination, the photons carry the
information of the state of the plasma at the time the last
scattered. We would see a pattern of hot and cold regions in this one
dimensional sky because photons arriving from different directions
come from places with varying photon energy density. This pattern is
shown as a dotted line on the right side of the circle.} 
\label{trlss}
\end{figure*}

Before recombination Thomson scattering keeps the photons and baryons
tightly coupled. They form a fluid with pressure provided by the
photons and inertia by the baryons. This fluid supports the
analog of acoustic oscillations where both
the density  and velocity are  oscillating
functions of time \cite{2.uroscmb,2.wayneth}. 
The density  is proportional to  $\cos (c_s
k\tau)$ while the velocity to $\sin(c_s k\tau)$, 
$c_s$ is the sound speed.  After hydrogen recombines the photons
are free to travel to the observer without further scatterings.

We illustrate the important ingredients in the generation of the
anisotropies with a two dimensional analogy in Figure \ref{trlss}. 
We show only one density mode which is oscillating as a function of
time before recombination. Suddenly the hydrogen recombines when this
mode was in a particular phase in the oscillation. The photons travel
freely a distance $D$ to the observer who sees more photons coming
from the what were denser regions at recombination. The pattern of hot
and cold temperature produced by this mode is not a perfect sinusoid
because we are intersecting a plane wave with a circle, but it has a
typical period. The contribution peaks at $l\sim k D$.

In Figure \ref{figclcdm} we show the three power spectra needed to
characterize the CMB anisotropies.
The oscillations in the CMB spectra can be analytically understood
within this picture. Waves of different wavenumber $k$ are at different
phases in their oscillations at this particular time, which
translates into an oscillating amplitude for the modes as a function
of $k$. Physical size and angle in the sky are related by the angular
diameter distance to recombination, so the oscillating amplitude of the
modes also produces acoustic peaks in $l$ space. On small
scales ($l \geq  1000$), when the wavelength of the perturbation becomes
comparable to the mean free path of the photons prior to
recombination, the anisotropies are suppressed
by photon diffusion. 
Photons can diffuse out of density peaks, thereby erasing the
anisotropies. 

The curves in Figure  \ref{figclcdm} illustrate the differences between
temperature and polarization anisotropies. The large angular scale
polarization is greatly suppressed. Correlations over large angles can
only be created by the long
wavelength perturbations, but these cannot produce a large
polarization signal because of the tight coupling between photons and
electron prior to recombination. Multiple scatterings make the
plasma very homogeneous; only wavelengths that are small enough to
produce anisotropies over the mean free path of the photons will give
rise to a significant quadrupole in the temperature distribution, and
thus to polarization. 

\begin{figure*}
\begin{center}
\leavevmode
\end{center}
\caption{ Power spectra for COBE normalized SCDM. The $E$ and the
$T-E$ spectra have been rescaled for convenience.} 
\label{figclcdm}
\end{figure*}

On subdegree  angular 
scales temperature anisotropy, polarization and their cross correlation show
acoustic oscillations (Figure \ref{figclcdm}), 
but in the polarization and cross
correlation spectra the peaks are much sharper. The polarization is produced
by  velocity gradients of the photon-baryon fluid
at the last scattering surface (equation \ref{polapprox}). 
The temperature receives
contributions from density and velocity perturbations and  these partially
cancel each other making the features in the temperature spectrum less
sharp. The dominant contribution to the temperature comes from the 
oscillations in the density, which are out of phase with the velocity.
This explains the difference in location between temperature and
polarization peaks. The extra gradient in the polarization signal,
equation 
(\ref{polapprox}), explains why its overall amplitude peaks at a smaller
angular scale. The fact that the polarization field has relatively
more small scale power is evident when we compare the the $T$ and $E$
fields in Figures \ref{tqups} and \ref{equps}.

\begin{figure*}
\begin{center}
\leavevmode
\end{center}
\caption{Correlation functions in real space for COBE normalized SCDM
model. The spectra have been smoothed with a $\theta_{fwhm}=0.2^o$
corresponding to the beam size of MAP.} 
\label{figcorrcdm}
\end{figure*}

Figure \ref{figcorrcdm} show the correlation functions in real
space. The spectrum has been smoothed with a $\theta_{fwhm}=0.2^o$
gaussian, similar to the MAP beam. An interesting point is that both
polarization auto-correlation functions are negative for some range of
angles, which does not happen for the temperature. 
To interpret the cross correlation we can
consider the polarization pattern around a hot spot ($T>0$). The cross
correlation starts positive, implying a radial  pattern of
polarization. Not all the polarization is correlated with the
temperature so it is hard to see this trend in Figure \ref{tqups}. In
Figure \ref{tqucps} we only plot the correlated part of the
polarization, here it is clear that the vectors are preferentially
radial around hot spots.

As we move out to larger angles the cross correlation changes sign. 
When it is
negative the pattern becomes tangential. For large separations the
polarization around a hot (cold) spot is tangential (radial).
A point worth noting is that
the cross correlation goes to zero as $\theta$ goes to zero, in
contrast to what happens for the other correlation functions. 
Symmetry arguments dictate that the $Q$ Stokes parameter  
at a given point cannot be correlated with the temperature at that
same point. What sign would this correlation have? Equivalently, 
in what direction
would the polarization be? Only when we consider two points separated
by some distance is  the symmetry broken. The vector joining the
two points becomes the privileged direction and the polarization can be
preferentially parallel or perpendicular to this direction.

\begin{figure*}
\begin{center}
\leavevmode
\end{center}
\caption{Simulated temperature and polarization 
map for SCDM ($2.5^o\times 2.5^o$
field). Only the correlated part of the
polarization vectors are shown together with the map of the
temperature. The temperature ranges from $-1690 \mu K < T < 1810 \mu K$
while the maximum amplitude of the polarization vectors is $P=45 \mu K$.} 
\label{tqucps}
\end{figure*}

\cleardoublepage

\vspace*{10cm}

\cleardoublepage

\def\bi#1{\hbox{\boldmath{$#1$}}}

\chapter{The Line of Sight Integration$^{1}$}\label{chaplos} 

\setcounter{footnote}{1}

The field of cosmic microwave background (CMB) anisotropies has seen 
a rapid development since its first detection by the COBE satellite
only a few years ago\footnotetext{ Based on U. Seljak \&
M. Zaldarriaga, Astrophys. J. {\bf 469}, 437 (1996) and
M. Zaldarriaga, U. Seljak \& E. Bertschinger, Astrophys. J. {\bf 494},
491(1998).}. There are now several reported experimental 
results that are detecting anisotropies
on degree angular scales (see \cite{3.Scott95} and
\cite{3.Bond96} for a recent review), 
which together with a few upper limits on smaller 
angular scales already give interesting limits on 
cosmological models. 
With 
the development of the new generation of experiments now being 
proposed one hopes to accurately map the CMB sky
from arcminute scales to several degree scales. 
The amount of data thus provided would 
allow for an unprecedented accuracy in the determination of cosmological 
parameters. Theoretical modeling shows 
that CMB anisotropies are 
sensitive to most of the cosmological parameters and have a distinctive
advantage over other 
cosmological observations in that they probe the universe
in the linear regime. This avoids the complications caused by physical 
processes in the nonlinear regime and allows to use powerful statistical
techniques to search over the parameter space for the best cosmological 
model (see e.g. \cite{3.Jungman95,3.bet,3.zss}). 

A large stumbling block in this program has been the speed of 
theoretical model calculations, which are still too slow to allow for
a rapid search over the parameter space.
The development of a fast and accurate algorithm to calculate the
anisotropies becomes essential to analyze the high quality data that
the future promises to deliver. The parameter space describing the
possible models 
is so large that  a search over  parameter space would be
impossible without a fast algorithm. Very fast approximate methods have
been developed but can only provide a $10 \%$ accuracy. In this
chapter we present an
algorithm that is both very fast 
(more than two orders of magnitude faster that
the usual Boltzmann approach) and exact within linear perturbation
theory. 

\section{Einstein and Fluid Equations}

In this section we present the Einstein and fluid differential equations 
for the metric, cold dark matter (CDM)  and baryons that must be  solved
to calculate the CMB anisotropy spectra produced by density
perturbations in our Universe. We will also present the Einstein
equations for gravity waves. These equations are the 
basis of the traditional methods and are also used in the integral 
method, discussed this chapter.
The derivation of the Einstein and  fluid 
equations 
can be found in the literature (e.g. \cite{3.bert95}), 
so we only present the final results. We restrict the treatment to
spatially flat universes. The interested reader can find the
generalization to arbitrary Robertson-Walker background in
\cite{3.zsb,3.otamm}. 

The metric is written as  
\begin{eqnarray}
ds^2&=&-dt^2+a^2(\delta_{ij}+h_{ij})dx^i dx^j \nonumber \\
 &=& a^2 [-d\tau^2+(\delta_{ij}+h_{ij})dx^i dx^j],
\end{eqnarray}
where $a$ is the expansion factor, $x_i$ the comoving coordinates and
$\tau=\int dt /a$ the conformal time. We are using units in which
$c=1$.  The space part of the unperturbed metric is a Kronecker delta,
$\delta_{ij}$  and $h_{ij}$ is the metric
perturbation in synchronous gauge \cite{3.lifshitz}. The metric
perturbations have contributions from scalar (density) and tensor
(gravity waves) modes, $h_{ij}=h^S_{ij}+h^T_{ij}$.
Although all 
observable quantities are identical in different gauges the   
computational efficiency to obtain them within a given accuracy 
is not. This criterion lead us to 
work in synchronous gauge. In comparison to  
the longitudinal gauge \cite{3.bardeen} it is about 20\% more efficient with 
isentropic initial conditions and even more so with isocurvature
initial conditions, which are difficult to set up in 
the longitudinal gauge. 

We start by considering perturbations produced by density modes.
When working with linear theory in a flat universe it is convenient to use  
Fourier modes because
 they evolve independently. These modes 
are the eigenfunctions of the Laplacian operator
that we shall call $G(\bi{k},\bi{x})$,
\begin{equation}
\nabla^2 G(\bi{k},\bi{x})= -k^2 G(\bi{k},\bi{x}).
\label{Laplace}
\end{equation}  

We expand all the perturbations in terms of $G$ and its spatial 
covariant
derivatives. For example, the metric perturbations for a single mode are 
given by 
\begin{equation}
h^S_{ij}={h\over 3}\delta_{ij}G-
(h+6\eta)(k^{-2}G_{\mid ij}+{1\over 3}\delta_{ij}
G),
\label{metricscalar}
\end{equation}
where 
$h$ and $(h+6\eta)$ are the trace and traceless part of the metric 
perturbation.
The perturbed Einstein's equations result in the following equations for $h$
and $\eta$ (Bertschinger 1996),
\begin{eqnarray}
k^2 \eta-{1\over 2}{\dot a \over a} \dot{h}&=&-8\pi Ga^2\delta \rho 
\nonumber \\
k^2 \dot{\eta} &=&4\pi Ga^2 
({\bar\rho}+{\bar p})kv.  
\end{eqnarray}
($G$ here stands for the gravitational constant and should not be
confused with the mode functions $G(\bi{k},\bi{x})$);
$\delta \rho$ and $v$ characterize the density and velocity 
perturbations 
($v=i \hat{ \bi{k}} \cdot \bi{v}$), 
$\delta \rho = \sum_j {\bar \rho_j} \delta_j$, 
$({\bar\rho}+{\bar p})v = \sum_j ({\bar\rho_j}+{\bar p_j})v_j$,
where
${\bar\rho_j}$ and ${\bar p_j}$ are the mean density and pressure of 
the $j$-th species
and the sum is carried out over all the different species in the universe.   

The 
equation for the cold dark matter density perturbation $\delta_c$ is,
\begin{equation}
\label{cdm2}
        \dot{\delta_c} = -{\dot h \over 2},
\end{equation}
where by definition in this gauge the cold dark matter particles have
 zero peculiar velocities. The Euler equation for the baryons
has additional terms 
caused by Thomson scattering and  pressure. Baryons  
have velocities relative to the dark matter,
\begin{eqnarray}
\label{baryon2}
\dot{\delta}_b &=& -kv_b-{\dot{h}\over 2} \,, \nonumber\\
\dot{v}_b &=& -{\dot{a}\over a}v_b
+ c_s^2 k\delta_b
+ {4\bar\rho_\gamma \over 3\bar\rho_b}
 an_ex_e\sigma_T(v_{\gamma}-v_b) \,.
 \label{cdmb}
 \end{eqnarray}
Here $c_s$ is the baryon sound speed, $v_b$ is the baryon velocity,
$v_{\gamma}$ is given by the
temperature dipole $v_{\gamma}=3\Delta_{T1}$ and
$\bar\rho_\gamma$, $\bar\rho_b$
are the mean photon and baryon densities respectively.
The Thomson scattering cross section is $\sigma_T$, 
$n_e$ is the electron density and  $x_e$ is the ionization fraction.

There are two independent degrees of freedom or 
polarizations for gravity waves. The purturbed metric
for each is,
\begin{equation}
h^{T(\pm)}_{ij}=(\hat {\bi{e}}_1 \pm i \hat{ \bi{e}}_2)_i \ 
(\hat {\bi{e}}_1 \pm i \hat {\bi{e}}_2)_j \ h_t(\tau) \ G(\bi{k},\bi{x}),
\end{equation}
the $\pm$ labels the polarization mode of the gravity wave and $(\hat
{\bi{e}}_1 ,\hat {\bi{e}}_2,\hat {\bi{k}})$ form an orthogonal basis. The
Einstein equations lead to
\begin{equation}
\ddot h_t+2 {\dot a \over a} \dot h_t + k^2 h_t=0.
\label{einstens}
\end{equation}
We have neglected the source terms on the right hand side produced by
the neutrino and photon anisotropic stress. It is straightforward to
add this source,   but it has a negligible effect in practice because
for modes outside the horizon free streaming cannot create a shear
while for modes inside the horizon the $k^2 h_t$ term in
(\ref{einstens}) becomes dominant over the source terms.

\section{Boltzmann equation}

The photons are described using their distribution function which
depends on time, spatial position and direction of propagation of the
photons ($\hat{\bi{n}}$).  In fact polarization depends also on the
two axes perpendicular to $\hat{\bi{n}}$ used to define the Stokes
parameters. The evolution of the distribution function is described
using the Boltzmann equation which states that when one follows a
light ray the temperature or polarization of the radiation can change
for two independent reasons: gravitational redshifts or blueshifts 
and Thomson scatterings. We can then write,
\begin{equation}
{d X \over d\tau}={\partial X \over \partial \tau}+\hat{n}^i{\partial X
\over \partial x_i} = {\dot X}_{gravity}+{\dot X}_{thomson},
\label{bolgen}
\end{equation}
$X=T or (Q\pm i U)$ which we introduced in Chapter \ref{chapstatcmb}
to characterize the fractional fluctuations
in the CMB field (ie. $T={\delta T\over
T_0}={\delta I \over 4 I}$). 
The temperature of the CMB blackbody can be used
instead of the intensity at a particular wavelength
because both the gravitational redshift and Thomson scattering
preserve the blackbody nature of the spectrum. The gravity
source only acts on the temperature equation and not on the
polarization,
\begin{equation}
{\dot T}_{gravity}=-{1\over 2}\ \hat{n}^i \hat{n}^j {\dot h}_{ij}.
\label{gravgen}
\end{equation}
This equation is valid for both density modes and gravity waves.
We will derive the Thomson scattering source in \ref{secthomson}.

\subsection{Thomson scattering}\label{secthomson}

The Thomson scattering cross section is,
\begin{equation}
{ d \sigma \over d \Omega}={3 \sigma_T \over 8 \pi} |\tilde 
{\bi{\epsilon}} \cdot \tilde {\bi{\epsilon}}^{\prime}|^2,
\label{thcs}
\end{equation} 
where $\sigma_T$ is the Thomson scattering 
cross section and
 $\tilde{{\bi \epsilon}}$ and 
 $\tilde{{\bi \epsilon}}^\prime$
are the unit vectors that describe the polarization of the electric
field of the scattered and incoming radiation respectively. 
The scattering 
terms in equations (\ref{bolgen}) are 
most easily computed in 
the coordinate system where  the incident photons travel
along the $\hat{\bi{z}}$ axes and the electrons are at rest.
If  $\hat{\bi{n}}^\prime$ is the direction of the incident
photon and $\hat{\bi{n}}$ that of the scattered one then
$\hat{\bi{n}}^\prime=\hat z=(\theta=0,\phi=0)$ and 
$(\theta,\phi)$ describe $\hat{\bi{n}}$. 
For a given scattering event, 
the Thomson scattering matrix is the simplest when
expressed in terms of the
intensities of  radiation parallel ($\tilde T_{\parallel}$) 
and perpendicular ($\tilde T_{\perp}$) to the plane  
containing both $\hat{\bi{n}}$ and $\hat{\bi{n}}^{\prime}$. 
Equation (\ref{thcs}) leads to
the following relation
between incoming and scattered radiation,
\begin{eqnarray}
\tilde T_{\parallel}
&=&{3\over 16\pi} \cos^2\theta \ \tilde T_{\parallel}^\prime
\nonumber \\  
\tilde T_{\perp}&=&{3\over 16\pi} \ \tilde T_{\perp}^\prime
\nonumber \\  
\tilde U&=&{3\over 16\pi} \cos \theta 
\ \tilde U^\prime.
\label{eqn2}
\end{eqnarray}
We have normalized the equation (\ref{eqn2}) so that the number of
photons is conserved by the scattering. In this way we only need to
introduce the scattering rate to obtain the terms needed in the
Boltzmann eqaution.
The total intensity is the sum of the two components, 
$\tilde T=\tilde 
T_{\parallel}+\tilde T_{\perp}$, while the difference gives polarization
$\tilde Q=
\tilde T_{\parallel}-\tilde T_{\perp}$. Because 
the components are measured
using this coordinate system, defined by  the plane that contains both
$\hat{\bi{n}}$ and $\hat{\bi{n}}^\prime$, the Stokes parameters
of the incoming radiation  $\tilde Q^\prime$ 
and 
$\tilde U^\prime$ depend on the angle $\phi$ of the scattered
photon, while $\tilde Q $ and $\tilde U$
are already measured relative to the correct frame.
It is more useful to refer the Stokes parameters of the incoming 
radiation relative to a fixed frame.
To achieve this we  construct the
scattering matrix in terms of  $T^{\prime}$, 
$Q^\prime+iU^\prime
=\exp(2i\phi)(\tilde Q^\prime+i\tilde U^\prime)$
and $Q^\prime-iU^\prime=\exp(-2i\phi)(\tilde Q^\prime-
i\tilde U^\prime)$, where we have used the transformation law 
(equation \ref{QUtrans}) to
relate the two sets of Stokes parameters.

Equation (\ref{eqn2}) implies that 
the scattered radiation in
direction $\hat{\bi{n}}$ which initially came in direction
$\hat{\bi{n}}^{\prime}$ is
\begin{eqnarray}
\delta T(\hat{\bi{n}}^{\prime},\hat{\bi{n}})&=&{1 \over 4\pi}
\left[{3\over 4}(1+\cos^2 \theta) T^\prime +
{3\over 8}(\cos^2 \theta -1)e^{-2i\phi} (Q^\prime+ iU^\prime)+\right.  
\nonumber \\
&& \left.
{3\over 8}(\cos^2 \theta -1)e^{2i\phi} 
(Q^\prime- iU^\prime)\right] \nonumber \\
\delta (Q\pm iU)(\hat{\bi{n}}^{\prime},\hat{\bi{n}})&=&{1 
\over 4\pi}\left[{3\over 4}(\cos^2 \theta-1) T^\prime +
{3\over 8}(1 \pm \cos \theta )^2 e^{-2i\phi}(Q^\prime+ iU^\prime)+ \right. 
\nonumber \\
&& \left.
{3\over 8}(1 \mp \cos \theta )^2e^{2i\phi}(Q^\prime- iU^\prime)
\right]. \nonumber \\
\label{eqna}
\end{eqnarray}
We introduced a $\delta$ to indicate that 
the final expression
for the scattered field is an integral over all directions 
$\hat{\bi{n}}^\prime$, 
\begin{equation}
\dot X(\hat{\bi{n}})|_{Thomson}
= -a \sigma_T n_e \ x_e  \left[X(\hat{\bi{n}})+
\int d\Omega^{\prime} \delta X(\hat{\bi{n}}^{\prime},\hat{\bi{n}})\right],
\label{expfinal}
\end{equation}
$X$ stands for $T$ and $(Q\pm iU)$. We introduced the scattering
rate $a \sigma_T n_e \ x_e$ where $n_e\ x_e$ is the density of ionized
electrons. 
The first term 
accounts for the photons that are scattered away from the line of sight and
the expansion factor $a$ is introduced because we are calculating
the derivative with respect to conformal time. 

Equation (\ref{eqna}) for the scattering matrix is written in the
frame where $\hat{\bi{n}}^{\prime}=(\theta^{\prime}=0,\phi^{\prime}=0)$.
The first spherical harmonics are explicitly  
$ \ _{\pm 2} Y_2^2(\hat{\bi{n}})=\sqrt{1\over
16\pi}(1\mp\cos\theta)^2 e^{2i\phi}$,
$ \ _{\pm 2} Y_2^{-2}(\hat{\bi{n}})=\sqrt{1\over
16\pi}(1\pm\cos\theta)^2e^{-2i\phi}$ and
$ \ _{\pm 2} Y_2^{0}(\hat{\bi{n}})=\sqrt{15\over
64\pi}(1-\cos^2\theta)$ . These 
together with 
$ \ _0 Y_0^m(\hat{\bi{n}}^{\prime})=\sqrt{1\over 4\pi} \delta_{m0}$,
$ \ _0 Y_2^m(\hat{\bi{n}}^{\prime})=\sqrt{5\over 4\pi} \delta_{m0}$,
and $_{\pm 2} Y_2^m(\hat{\bi{n}}^{\prime})=\sqrt{5\over 4\pi} \delta_{m\mp
2}$ enable us to rewrite (\ref{eqna}) in a more useful form
($ \delta_{ij}$ is the Kronecker delta),
\begin{eqnarray}
\delta T(\hat{\bi{n}}^\prime,\hat{\bi{n}})&=&\sigma_T\sum_m\Big[\Big({1\over
10}\ _0Y_2^m(\hat{\bi{n}})\ _0\bar Y_2^{m}(\hat{\bi{n}}^\prime)
+\ _0Y_0^m(\hat{\bi{n}}) \ _0\bar
Y_0^{m}(\hat{\bi{n}}^\prime)  
\Big) T^\prime  
\nonumber \\
& &  -
{3\over 20} \sqrt{2\over 3}\ _0Y_2^m(\hat{\bi{n}})\ _2\bar Y_2^{m}
(\hat{\bi{n}}^\prime)\
(Q^\prime+ iU^\prime)\nonumber \\
& & -
{3\over 20} \sqrt{2\over 3}\ _0Y_2^m(\hat{\bi{n}})\ _{-2}\bar Y_2^{m}
(\hat{\bi{n}}^\prime)
(Q^\prime- iU^\prime)
\Big]  \nonumber \\
\delta (Q\pm i U)(\hat{\bi{n}}^\prime,\hat{\bi{n}})&=&
\sigma_T\sum_m\Big[-{6\over
20} \sqrt{2\over 3}\ _{\pm 2}Y_2^m(\hat{\bi{n}})\ _0\bar 
Y_2^{m}(\hat{\bi{n}}^\prime)   T^\prime \nonumber \\
& & +
{6\over 20} \ _{\pm 2}Y_2^m(\hat{\bi{n}})\ 
_2\bar Y_2^{m}(\hat{\bi{n}}^\prime)
(Q^\prime+ iU^\prime)+ \nonumber \\
&& {6\over 20} 
\ _{\pm 2}Y_2^m(\hat{\bi{n}})\ _{-2} \bar 
Y_2^{m}(\hat{\bi{n}}^\prime)(Q^\prime- iU^\prime)
\Big] . 
\label{scm}  
\end{eqnarray}
This form has the advantage of being independent of the coordinate system.
We will use the scattering matrix in the frame where 
$\bi{k} \parallel \hat{\bi{z}}$ and not $\hat{\bi{n}}^{\prime}=\hat{\bi{z}}$. 
Here $\bi{k}$ is the wavevector of the Fourier mode under
consideration.
The addition theorem for the spin harmonics (\ref{addtheo})
can be used 
to show that
the sum
$ \sum_m\ _sY_l^m(\hat{\bi{n}})\ _{s^\prime}\bar Y_l^{m}(\hat{\bi{n}}^\prime)$ 
acquires a phase change under rotation of the coordinate system 
that exactly cancels the phase 
change in the transformation of $(Q\pm i U)$ in equation 
(\ref{scm}). 
We may therefore use this equation 
in any coordinate system. 

\subsection{Density fluctuations}

As we are dealing with a linear problem with a spatially constant
background we may
consider only one eigenmode of the Laplacian at a time. 
We may choose without loss of generality that
$ \bi{k} \parallel \hat {\bi z}$. To define the 
Stokes parameters we use the spherical coordinate unit vectors
$ (\hat{\bi{ e}}_{\theta},\hat{\bi{e}}_{\phi})$. 

The density field produced by a single mode has two important
symmetries. It is invariant under rotations around $\bi{k}$ and parity
operations where the $x$ or $y$ axes change sign ($\bi{k} \parallel
\hat z$). By invariant we mean
that the transformed density field
$ \rho^{\prime}(\bi{x})=\rho(\bi{x})$, there is no prime in the
argument of $\rho^{\prime}$. What is directly relevant to the
calculation of the anisotropies is that the gravity source term 
$ {\hat n}^i{\hat n}^j {\dot h}^S_{ij}/2$ satisfies these symmetries.
The rotational symmetry implies that the
neither the temperature nor the Stokes parameters (in the spherical
basis) can depend on $\phi$.
The parity symmetry has interesting consequences on the allowed
polarization patterns as well. 
For simplicity we focus on the Stokes parameters at the
origin. Symmetry implies that $(Q\pm iU)^\prime(\hat{\bi{n}})=(Q\pm
iU)(\hat{\bi{n}})$ for both rotation and parity operations. We 
consider a parity operation that changes the sign of the $y$ axes.
If we consider 
the direction $ \hat{\bi{n}}=(\theta,0)$ when we apply the
transformation its location 
remains unchanged,  $\hat{\bi{n}}^\prime=\hat{\bi{n}}$. 
On the other hand $\hat{\bi{e}}_{\phi}$, used to define the Stokes
parameters changes sign. As a
consequence $U$ changes sign, 
$U^\prime(\hat{\bi{n}}^\prime)=-U(\hat{\bi{n}})$.  This is
inconsistent with the symmetry statement unless $U$ is zero. If $U$ is
zero in this particular direction it has to be zero in all directions
because of the rotational symmetry. In 
this particular coordinate system
only $Q$ is different from zero and we denote it by
$\Delta_P^S$, so that $\Delta_P^S=Q=Q\pm i U$ $(U=0)$. 
The temperature anisotropy for the single eigenmode is denoted
by $\Delta^S_T$. We introduced a $\Delta$ in
the notation to enphasize that these are the
contributions of one mode. The total
field $T$, $Q$ and $U$ are obtained by adding
the contribution of all modes.

For a plane wave, rotational symmetry implies that 
both $\Delta^S_T$ and $\Delta^S_P$ depend only on the angle between
$\hat{\bi{n}}$ and $\hat z$ ($\bi{k} \parallel \hat z$),
so only harmonics with $m=0$ (which do not depend on $\phi$), 
are needed in the expansion. To calculate 
 the evolution of these two
quantities we expand them as,
\begin{eqnarray}
\Delta^S_T(\bi{k}, \hat{\bi{n}})&=&\ G(\bi{k},\bi{x})
\sum_l (-i)^l \sqrt{4\pi(2l+1)}\Delta^S_{Tl} \; Y_l^0(\hat{\bi{n}}) \nonumber \\
&=&\ G(\bi{k},\bi{x})\sum_l (-i)^l (2l+1)\Delta^S_{Tl} P_l(\mu)
\nonumber \\
\Delta^S_P(\bi{k}, \hat{\bi{n}})
&=& \ G(\bi{k},\bi{x})\sum_l (-i)^l \sqrt{4\pi(2l+1)
(l+2)!/(l-2)!} \; _2\Delta^S_{Pl}
\; _2Y_l^0(\hat{\bi{n}})\nonumber \\
&=& \ G(\bi{k},\bi{x})\sum_l (-i)^l \sqrt{4\pi(2l+1)
(l+2)!/(l-2)!} \; _{-2}\Delta^S_{Pl}
\; _{-2}Y_l^0(\hat{\bi{n}})  \nonumber \\
&=&\ G(\bi{k},\bi{x})\sum_l (-i)^l (2l+1)
 \; _{\pm 2}\Delta^S_{Pl}
P_l^2(\mu) 
\label{expleg}
\end{eqnarray}
where $G (\bi{k},\bi{x}) =\exp(i\bi{k} \cdot
\bi{x})$ and $\mu=\hat{\bi{k}} \cdot \hat{\bi{n}}$. 
We added  a subindex $\pm 2$ to $_{\pm 2}\Delta^S_{Pl}$ to  denote that they
are the expansion coefficients in spin $\pm 2$ harmonics \footnote{The
relation between these coefficient and those used in Zaldarriaga \&
Seljak (1997)
is $_{\pm 2}\Delta^S_{Pl}= - \sqrt{(l-2)!/(l+2)!} \Delta^S_{El}$.} and we
used the explicit expression for spin $s$ harmonics with $m=0$
to write them  in terms of associated Legendre polynomials 
\cite{3.goldberg},
\begin{eqnarray}
Y_l^0(\theta,\phi)&=& \sqrt{(2l+1)\over 4\pi} P_l(\cos \theta)
\nonumber \\
\ _{\pm 2}Y_l^0(\theta,\phi)&=& \sqrt{{(2l+1)\over 4\pi}{(l-2)! 
\over (l+2)!}} 
P_l^2(\cos \theta).
\label{eqylm}
\end{eqnarray}

As stated above,
scalar modes in this reference frame have $U=0$, so $\Delta^S_P$
describes both spin $\pm 2$ quantities. 
For $m=0$ one has $\, _2Y_l^0=\,_{-2}Y_l^0$ and so
$ \; _{2}\Delta^S_{Pl}=\; _{-2}\Delta^S_{Pl}$. This is a very
important result, because 
the fact that both coefficients are equal implies
that this single Fourier mode does not produce any $B$ mode
polarization (equation \ref{aeb}). 
The observed polarization field is the superposition of
that produced by each Fourier mode, and thus we have shown that the
pattern of polarization produced by density perturbations has no $B$
component. Note that $B$ is invariant under rotations so the fact that
we have proved this result in a particular reference frame that
depends on $\bi{k}$ is not
important. This conclusion does not apply to $U$: although it is zero
in this particular frame it is not in others. When we superimpose the
perturbations of all modes $B$ remains zero while $U$ does not. 

We can replace equation (\ref{expleg}) into (\ref{bolgen}) and use the
expresion for the gravitational redshift, equations
(\ref{metricscalar}) and (\ref{gravgen}), to obtain
the Boltzmann equation for the CMB photons;
\begin{eqnarray} 
\dot\Delta^S_T +ik\mu \Delta^S_T 
&=&-{1\over 6}\dot h-{1\over 3}(\dot h+6\dot\eta)
P_2(\mu) + \dot \Delta^S_{T|Thomson}\nonumber \\   
\dot\Delta^S_P +ik\mu \Delta^S_P
&=&\dot \Delta^S_{P|Thomson}.
\label{boltzmann1}
\end{eqnarray}
The first term in the temperature equation represents
the effect of gravitational redshift, while $\dot{\Delta^S}_{T|Thomson}$
and $\dot{\Delta^S}_{P|Thomson}$ are the changes in the photon
distribution function produced by Thomson scattering. 
After inserting equation (\ref{expleg}) into equation (\ref{boltzmann1})
one obtains a system of two coupled hierarchies, one for the
temperature and the other for polarization, 
\begin{eqnarray}
\dot\Delta^S_{T0}
 &=& -k\Delta^S_{T1}-{\dot{h}\over 6}+\dot \Delta^S_{T0|Thomson}
 \, \nonumber \\
\dot \Delta^S_{T1} &=&
{k \over 3}\left[\Delta^S_{T0}-
2\Delta^S_{T2}
\right] +\dot \Delta^S_{T1|Thomson}
\,\nonumber\\
\dot \Delta^S_{T2} &=&{k \over 5}\left[2\Delta^S_{T1}-3
\Delta^S_{T3}\right] +{2  \over 15} k^2  \alpha+
\dot \Delta^S_{T2|Thomson} \nonumber \\
\dot \Delta^S_{Tl}&=&{k \over
2l+1}\left[l\Delta^S_{T(l-1)}-(l+1)
\Delta^S_{T(l+1)}\right]
+\dot \Delta^S_{Tl|Thomson} \,, l>2 \nonumber \\
\; _2\dot\Delta^S_{Pl}&=&{k\over 2l+1}\left[(l-2)\; _2\Delta^S_{Pl-1}-
(l+3)\; _2\Delta^S_{Pl+1}\right]+\; _2\dot \Delta^S_{Pl|Thomson},
\label{eqn1}
\end{eqnarray}
where $\alpha=(\dot{h}+6\dot{\eta})/ 2 k^2$, and we wrote sepatarely
the first 3 equations of the temperature hierarchy that contain the
gravity sources.
We also used the recurrence relations for the Legendre functions,
\begin{eqnarray}
\mu P_l(\mu)={1\over 2l+1}\left[ l\ P_{l-1}+
(l+1) P_{l+1}\right] \nonumber \\
\mu P_l^2(\mu)={1\over 2l+1}\left[(l+2)P_{l-1}^2+
(l-1)P_{l+1}^2\right].
\label{recurr}
\end{eqnarray}

Substituting the expansion for the Stokes parameters from equation
(\ref{expleg}) into equation (\ref{scm}) and using equation 
(\ref{expfinal})
we find
\begin{eqnarray}
\dot\Delta^S_{Tl}|_{Thomson}& \equiv & -a \sigma_T n_e x_e \left[\Delta^S_{Tl}
+ \int d \Omega 
\,_0Y_l^m(\hat{\bi{n}}) \delta T(\hat{\bi{n}})\right]
\nonumber \\
&=&\dot \kappa(- \Delta^S_{Tl}
+\Delta^S_{T0}\delta_{l0}+{\Pi\over
10}\delta_{l2}) \nonumber \\
\ _{\pm 2}\dot\Delta^S_{Pl}|_{Thomson}
&\equiv&-a \sigma_T x_en_e \left[\ _{\pm 2}\Delta^S_{Pl}
+ \int d \Omega \,_2Y_l^m(\hat{\bi{n}})\ \delta (Q\pm iU)(\hat{\bi{n}})\right]
\nonumber \\
&=&\dot \kappa (-\ _{\pm 2}\Delta^S_{Pl}-{\Pi\over
20}\delta_{l2} )
\label{eqnv}
\end{eqnarray}
with 
\begin{equation}
\Pi=\Delta^S_{T2}-6 (\ _2\Delta^S_{P2}+\ _{-2}\Delta^S_{P2}).
\end{equation} 
The differential optical depth for Thomson scattering is denoted
$\dot\kappa=an_ex_e\sigma_{T}$. Note that the polarization
has sources  only  at $l=2$. 
Equation (\ref{eqnv}) is valid in the rest frame of the electrons,
so in the reference frame where the baryon velocity is $v_b$
the distribution of scattered radiation has an additional 
dipole. 
The final expression for the Boltzmann hierarchy is
\begin{eqnarray}
\dot \Delta^S_{T0}
 &=& -k\Delta^S_{T1}-{\dot{h}\over 6}
 \, \nonumber \\
\dot \Delta^S_{T1} &=&
{k \over 3}\left[\Delta^S_{T0}-
2\Delta^S_{T2}
\right] + \dot{\kappa} \left(
{v_b \over 3}-\Delta^S_{T1}\right)\,\nonumber\\
\dot \Delta^S_{T2} &=&{k \over 5}\left[2\Delta^S_{T1}-3
\Delta^S_{T3}\right] +{2  \over 15} k^2  \alpha+
\dot{\kappa} \left[{\Pi \over 10}-\Delta^S_{T2}\right]\, \nonumber\\
\dot \Delta^S_{Tl}&=&{k \over
2l+1}\left[l\Delta^S_{T(l-1)}-(l+1)
\Delta^S_{T(l+1)}\right]
-\dot{\kappa}\Delta^S_{Tl} \,, l>2 \nonumber \\
\; _2\dot\Delta^S_{Pl}&=&{k\over 2l+1}\left[(l-2)\; _2\Delta^S_{Pl-1}-
(l+3)\; _2\Delta^S_{Pl+1}\right]-\dot\kappa \; _2\Delta^S_{Pl}-{1\over 20}
\dot\kappa \Pi \delta_{l2}.
\label{photon}
\end{eqnarray}

The last important ingredient is the relation between
$\Delta^S_{(T,P)}$ for each Fourier mode and the observed CMB power
spectra. For a single mode we can combine equations (\ref{alm}) and
(\ref{expleg}) to show
\begin{eqnarray}
a_{T,lm}&=&(-i)^l\ \sqrt{4\pi(2l+1)}\ \Delta^S_{Tl}\ G(\bi{k},\bi{x})\
\delta_{m,0} \nonumber \\
a_{E,lm}&=&(-i)^l\ \sqrt{4\pi(2l+1)}\ \Delta^S_{El}\ G(\bi{k},\bi{x})\
\delta_{m,0} \nonumber \\
a_{B,lm}&=&0
\end{eqnarray}
with $\Delta^S_{E,l}=-\sqrt{(l+2)!/(l-2)!}\; _{\pm 2}\Delta^S_{Pl}$.

The different Fourier modes are statistically independent so the 
observed total power in the integral of the powers produced by all modes, 
\begin{eqnarray}
C_{(T,E)l}&=&(4\pi)^2\int k^2dk
P(k)|\Delta^S_{(T,E)l}(k,\tau=\tau_0)|^2 \nonumber \\
C_{Cl}&=&(4\pi)^2\int k^2dk
P(k)\Delta^S_{Tl}(k,\tau=\tau_0)\Delta^S_{El}(k,\tau=\tau_0)
\label{cl}
\end{eqnarray}
with $P(k)$ the primordial power
spectrum. The primordial power spectrum gives
the power in each mode at the initial time.
For example if we choose to normalize the
initial conditions so that $\delta_c=1$ then
$P(k)$ gives the initial power in the cold
dark matter component.

\subsection{Gravity waves}

For the gravity wave modes 
the source term in the temperature equation is
$ {1 \over 2}\ \hat {n}^i\hat {n}^j {\dot h}^{T(\pm)}_{ij}= {1 \over 2}\
(1-\cos^2 \theta)e^{\pm i 2\phi}\ \ h_t\ G(\bi {k},\bi{x})$. 
There is an explicit $\phi$ dependence of the source 
so the symmetries  of  rotation and 
parity satisfied by the density modes discussed in the previous
section no longer apply. This implies that $U$ is no longer zero and
that gravity waves will produce a non-zero $B$. This is a crucial
point and may allow in the future to  detect  the
stochastic background of gravity waves if it is present in our
universe. We will discuss this further in section \ref{secnonscal}.   

For gravity waves we can expand,
\begin{eqnarray}
\Delta^{T(\pm)}_T(\bi{k}, \hat{\bi{n}})&=&\ G(\bi{k},\bi{x})
\sum_l (-i)^l \sqrt{4\pi(2l+1)\ (l+2)!/(l-2)!}\nonumber \\
&\times& \Delta^{T(\pm)}_{Tl} \;
Y_l^{\pm 2}(\hat{\bi{n}}) \nonumber \\
(\Delta^{T(\pm)}_Q+ i \Delta^{T(\pm)}_U)(\bi{k}, \hat{\bi{n}})
&=& \ G(\bi{k},\bi{x})\sum_l (-i)^l \sqrt{4\pi(2l+1)}
(l+2)!/(l-2)! \nonumber \\
&\times&\; _{+2}\Delta^{T(\pm)}_{Pl}
\; _2Y_l^{\pm 2}(\hat{\bi{n}})\nonumber \\
(\Delta^{T(\pm)}_Q- i \Delta^{T(\pm)}_U)(\bi{k}, \hat{\bi{n}})
&=& \ G(\bi{k},\bi{x})\sum_l (-i)^l \sqrt{4\pi(2l+1)}
(l+2)!/(l-2)! \nonumber \\
&\times& \; _{-2}\Delta^{T(\pm)}_{Pl}
\; _{-2}Y_l^{\pm 2}(\hat{\bi{n}}).
\end{eqnarray}
Because of the $\phi$ dependence of the source term we had to use
harmonics with $m=\pm 2$ for the two different gravity wave
polarizations.

Now that $ \; _{+2}\Delta^{T(\pm)}_{Pl}\neq \; _{-2}\Delta^{T(\pm)}_{Pl}$, 
we  need two different hierarchies to solve for both of
them separately.
The derivation of the hierarchies is analogous to that for the
density modes yielding, 
\begin{eqnarray}
\dot{\Delta}^{T(\pm)}_{Tl}&=&{k\over 2l+1}\biggl[(l-2)\Delta^{T(\pm)}_{Tl-1}-
(l+3) \Delta^{T(\pm)}_{Tl+1}\biggr]+{\dot{h}_t \over 30} \delta_{l2} 
+ \dot{\Delta^{T(\pm)}}_{Tl|Thomson} \nonumber \\
\;_{\pm 2}\dot{\Delta}^{T(\pm)}_{Pl}&=&{k\over 2l+1}
\biggl[{(l-2)^2\over l}\; _{\pm
2} \Delta^{T(\pm)}_{Pl-1} \pm {20 i \over l(l+1)} \; _{\pm
2} \Delta^{T(\pm)}_{Pl} \nonumber \\
&-& {(l+3)^2\over l+1}\; _{\pm
2} \Delta^{T(\pm)}_{Pl+1} \biggr]+\dot \Delta^{T(\pm)}_{Pl|Thomson},
\label{tensorl}
\end{eqnarray}
with
\begin{eqnarray}
\dot \Delta^{T(\pm)}_{Tl|Thomson}&=&
-\dot{\kappa}({\Delta^{T(\pm)}}_{Tl}-{1\over 10} \Pi\ \delta_{l2})
\nonumber \\  
\dot \Delta^{T(\pm)}_{Pl|Thomson}&=&
-\dot{\kappa}({\Delta^{T(\pm)}}_{Pl}+{1 \over 20}\Pi\ \delta_{l2})
\nonumber \\
\Pi&=&\Delta_{T2}^{T(\pm)}-6 \biggl[_{+2}\Delta^{T(\pm)}_{P2}
+  \; _{-2}\Delta^{T(\pm)}_{P2}\biggr].
\end{eqnarray}

We can change variables to obtain hierarchies for $E$ and $B$
directly,
\begin{eqnarray}
\dot{\Delta}^{T(\pm)}_{\tilde El}&=&{k\over 2l+1}\left[{(l-2)(l+2)\over l}\
\Delta^{T(\pm)}_{\tilde El-1} - {20 \over l(l+1)} \ 
\Delta^{T(\pm)}_{\tilde Bl}\nonumber \right. \\
&+& \left. {(l+3)(l-1)\over l+1}\
\Delta^{T(\pm)}_{\tilde El+1} \right]
-\dot{\kappa}\Delta^{T(\pm)}_{\tilde El}+\dot{\kappa}/20\Pi\
\delta_{l2}
\nonumber \\
&& \\
\dot{\Delta}^{T(\pm)}_{\tilde Bl}&=&{k\over 2l+1}\left[{(l-2)(l+2)\over l}\
\Delta^{T(\pm)}_{\tilde Bl-1} + {20 \over l(l+1)} \ 
\Delta^{T(\pm)}_{\tilde El}\nonumber \right. \\
&+& \left. {(l+3)(l-1)\over l+1}\
\Delta^{T(\pm)}_{\tilde El+1} \right]
-\dot{\kappa}\Delta^{T(\pm)}_{\tilde Bl}
\nonumber \\
& & \\
\Delta^{T(\pm)}_{\tilde T,l}&=&\sqrt{(l+2)!\over (l-2)!}\; 
\Delta^{T(\pm)}_{Tl}
\nonumber \\
\Delta^{T(\pm)}_{\tilde E,l}&=&-{(l+2)!\over (l-2)!}
\biggl(\; _{+2}\Delta^{T(\pm)}_{Pl}+\; _{-
2}\Delta^{T(\pm)}_{Pl}\biggr)/2 \nonumber \\
\Delta^{T(\pm)}_{\tilde B,l}&=&-{(l+2)!\over (l-2)!}
\biggl(\; _{+ 2}\Delta^{T(\pm)}_{Pl}-\; _{-
2}\Delta^{T(\pm)}_{Pl}\biggr)/2i, 
\end{eqnarray}
with $\Pi=\Delta^{T(\pm)}_{T2}-6(_{+2}\Delta^{T(\pm)}_{P2}
+  \;
_{-2}\Delta^{T(\pm)}_{P2})=
(\Delta^{T(\pm)}_{\tilde T2}
+\sqrt{6}\Delta^{T(\pm)}_{\tilde E2})/\sqrt{4!}$.
We introduced tildes in the notation to remind the reader that we
introduced extra $l$ dependent factors for later convenience. 
Free streaming couples $E$ and $B$ modes, but
the Thomson scattering only creates $E$ 
(the $\dot{\kappa} \Pi/20$ term is only present in the $E$ hierarchy). 
It is through freestreaming and not scattering 
that $B$ is created. 

Finally the relation between the observed CMB power spectra and the
perturbations for each mode is,
\begin{eqnarray}
C_{(T,E,B)l}&=&(4\pi)^2\int k^2dk
P_t(k)|\Delta^{T(\pm)}_{(\tilde T,\tilde E, \tilde B)l}
(k,\tau=\tau_0)|^2 \nonumber \\
C_{Cl}&=&(4\pi)^2\int k^2dk
P_t(k)\Delta^S_{\tilde Tl}(k,\tau=\tau_0)\Delta^{T(\pm)}_{\tilde El}
(k,\tau=\tau_0),
\label{clt}
\end{eqnarray}
where $P_h(k)$ denotes the initial power
spectrum for the gravity waves and we
normalized the initial conditions to $h_t=1$ 

The results for scalars $(m=0)$ and tensors $(m=2)$ 
can all be combined (the
following formulas  
also applies to vector modes with $m=1$). We expand all variables as,
\begin{eqnarray}
X&=&G(\bi{k},\bi{x})
\sum_l (-i)^l\ \sqrt{4\pi (2l+1) {(l+|s|)! \over (l-|s|)!}
{(l+|m|)! \over (l-|m|)!} } \;_{s}\Delta_{Xl} \; _{s}Y_l^m.
\end{eqnarray}
The expansion coefficients will satisfy the hierarchy,
\begin{eqnarray}
\dot {\Delta}_{Xl}&=&{k \over (2l+1)} \left[ {(l-|m|)(l-|s|)\over l}
{\Delta}_{Xl-1}-{(l+|m|+1)(l+|s|+1)\over (l+1)} \right.\nonumber \\
&\times& \left.
{\Delta}_{Xl+1} + {s m (2l+1) \over l(l+1)}{\Delta}_{Xl}\right]-\dot{\kappa}
{\Delta}_{Xl}+{1 \over 10}P_X \delta_{l2} \nonumber \\
P_T&=&\Pi \nonumber \\
P_{Q\pm iU}&=&\Pi / 2, 
\label{hiersvt}
\end{eqnarray}
with $\Pi=\Delta^{T(\pm)}_{T2}-6(_{+2}\Delta^{T(\pm)}_{P2}
+  \; _{-2}\Delta^{T(\pm)}_{P2})$.
Similar  expressions can be written for the $E$ and $B$ hierarchies. 
Equation (\ref{hiersvt}) 
does not include the gravitational redshift term or the Doppler shift 
caused by the the baryon
velocity which must be must be added to the temperature equation.
In the general case we can define 
\begin{eqnarray}
\Delta_{\tilde T,l}&=&\sqrt{
{(l+|m|)!\over (l-|m|)!}}\; 
\Delta_{Tl}
\nonumber \\
\Delta_{\tilde E,l}&=&-\sqrt{{(l+|m|)!\over (l-|m|)!}
{(l+2)!\over (l-2)!}}
\left(\Delta_{(Q+iU)l}+
\Delta_{(Q-iU)l}\right)/2 \nonumber \\
\Delta_{\tilde B,l}&=&-\sqrt{{(l+|m|)!\over (l-|m|)!}
{(l+2)!\over (l-2)!}}
\left(\Delta_{(Q+iU)l}-
\Delta_{(Q-iU)l}\right)/2i 
\end{eqnarray}
and use equation (\ref{clt}) to calculate the observer power spectra.

\section{The Line of Sight Method}

Instead of solving the coupled system of differential equations (\ref{photon})
one may formally integrate equations (\ref{boltzmann1}) 
along the photon past light cone to obtain (e.g. \cite{3.zal95}),
\begin{eqnarray}
\Delta_T^{S}(k,\tau_0) &=&\int_0^{\tau_0}d\tau e^{i k \mu (\tau -\tau_0)}
e^{-\kappa} 
\{ \dot\kappa e^{-\kappa} [\Delta^{S}_{T0}+i\mu v_b + {1\over 2} P_2(\mu)
\Pi]\nonumber \\
&-& {1\over 6}\dot h-{1\over 3}(\dot h+6\dot\eta)
P_2(\mu)\} \nonumber \\
\Delta_P^{S}(k,\tau_0)
 &=& {1\over 4}\int_0^{\tau_0} d\tau e^{i k \mu (\tau -\tau_0)}
e^{-\kappa} \dot\kappa  \Pi P_2^2(\mu).
\label{formal}
\end{eqnarray}

These lead to equations for the multipoles,
\begin{eqnarray}
\Delta^{S}_{Tl}(k,\tau=\tau_0)&=&
\int_0^{\tau_0}S^{S}_{T}(k,\tau)j_l(x)d\tau \nonumber \\
\ _2\Delta^{S}_{Pl}(k,\tau=\tau_0)&=&
\int_0^{\tau_0}S^{S}_{P}(k,\tau){j_l(x) \over x^2}
 d\tau \nonumber \\
x & = &k(\tau_0-\tau)
\end{eqnarray}
where the source functions are,
\begin{eqnarray}
S^{S}_T(k,\tau)&=&g\left(\Delta^{S}_{T0}+2 \dot \alpha
+{\dot{v_b} \over k}+{\Pi \over 4}
+{3\ddot{\Pi}\over 4k^2}\right)\nonumber \\
&+& e^{-\kappa}(\dot \eta+ \ddot \alpha)
+\dot{g}\left({v_b \over k}+\alpha+{3\dot{\Pi}\over 4k^2}\right)
+{3 \ddot{g}\Pi \over
4k^2} \nonumber \\
S^{S}_P(k,\tau)&=&-{3 \over 4} g \Pi. 
\label{finalscalar}
\end{eqnarray}
where $\alpha=(\dot{h}+6\dot{\eta})/2k^2$,
$j_l(x)$ are the spherical Bessel functions and $g$ is the
visibility function $g=\dot{\kappa} \exp(-\kappa)$. 
The source function for the temperature looks rather complicated
because we have done several integrations by parts to express
everything in terms of $j_l(x)$ rather than its derivatives. The polarization
solution is straightforward to check. It is enough to substitute equation
(\ref{finalscalar}) into the hierarchy (\ref{photon}) using the limit
$\lim_{x \rightarrow 0}j_l(x)/x^2 =1/15$ and the recursion,
\begin{equation}
{d \over dx}({j_l(x)\over x^2})=
{1\over (2l+1)}\left[ (l-2){j_{l-1}(x)\over x^2}-(l+3)
{j_{l+1}(x)\over x^2}\right].
\end{equation}
 
Some of the terms in the source function $S^{S}_T(\tau)$ are 
easily recognizable. It may be simpler to express them in the
more familiar conformal Newtonian gauge,
$\Delta_{T0}^{synch}+\dot{\alpha}
= \Delta_{T0}^{conf}+\psi$, $v_b^{synch}+k \dot{\alpha}= v_b^{conf}$
and $\ddot{\alpha}+\dot{\eta}=\dot{\phi}+\dot{\psi}$. Here $\phi$ and
$\psi$ are the conformal gauge gravitational potentials. All moments
higher than $l=1$ are gauge invariant, so in fact the source terms
$S_{(T,P)}$ must be gauge invariant.
The first two contributions in the source $S^{S}_T$ (equation 
\ref{finalscalar})
$\Delta_{T0}+\dot{\alpha}$ are the 
intrinsic temperature and gravitational potential 
contributions from the last-scattering surface, while 
the third contribution
is part of the velocity term, the other part being
$k^{-1}\dot{g}v_b+{\alpha}$  in the next line.
These terms 
make a dominant contribution to the 
anisotropy in the standard recombination models. 
The first term in 
the second line is the so-called integrated
Sachs-Wolfe term and is important after recombination.
It is especially important if matter-radiation equality occurs 
close to the recombination or in $\Omega_{\rm matter}\ne 1$ models. In both
cases the gravitational 
potential decays with time, which leads to an enhancement of 
anisotropies on large angular scales.
Finally we have the terms 
caused by photon polarization and anisotropic
stress, which contribute to $\Pi$.
These terms affect the 
anisotropy spectra at the 10\% level and are important for accurate
model predictions. Moreover, they are the sources 
for photon polarization.

The main advantage of (\ref{finalscalar})
is that it decomposes the anisotropy into a source term 
$S^{S}_{T,P}$, which does 
not depend on the multipole moment $l$ and a geometrical term $j_l$, which 
does not depend on the particular cosmological model. 
The latter only needs to be computed once and can be stored for 
subsequent calculations. 
The source term 
is the same for all multipole moments and only depends on a small number
of contributors in (\ref{finalscalar})
(gravitational potentials, baryon velocity and photon moments
up to $l=2$). By specifying the source term as a function of time one can 
compute the corresponding spectrum of anisotropies. 
Equation (\ref{finalscalar})
is formally an integral system of equations, because 
the first moments appear on both sides of equations. To solve for these
moments it is best to use the equations in their 
differential form (\ref{photon}),
instead of the integral form above.
Once the moments that enter into the source function are computed,
one can solve for the higher moments 
by performing the integration in (\ref{finalscalar}). 

The solution for the tensor modes can similarly be written as an integral 
over the source term,
\begin{eqnarray}
\Delta_{\tilde Tl}^T&=&\sqrt{(l+2)! \over (l-2)!}\int_0^{\tau_0} d\tau\
{1\over 2} [h_t+3 \Pi]\ {j_l(x) \over x^2} \nonumber \\
\Delta_{\tilde El}^T&=&\int_0^{\tau_0} d\tau\
{3 \over 2}\Pi\ \Big[-j_l(x)+j_l''(x)+{2j_l(x) \over x^2}
+{4j_l'(x) \over x}\Big] \nonumber \\
\Delta_{\tilde Bl}^T&=&\int_0^{\tau_0} d\tau\
{3 \over 2} \Pi \  \Big[2j_l'(x)
+{4j_l \over x}\Big]\nonumber \\
x&=&k(\tau_0-\tau)
\label{finaltensor}
\end{eqnarray}

Equations (\ref{finalscalar}) and (\ref{finaltensor}) 
are the main equations of this chapter and form the 
basis of the line of sight integration method
of computing CMB anisotropies. The treatment can be generalized to
arbitrary FRW backgrounds \cite{3.zsb}.
In the next section we will
discuss in more detail the computational advantages of this 
formulation of the Boltzmann equation and its implementation.

\section{Calculational Techniques}

In the previous section we presented the expressions needed 
for the implementation of the line of sight integration
method. 
As shown in  equations (\ref{finalscalar}) and (\ref{finaltensor})
one needs to integrate over
time the source term at time $\tau$ 
multiplied with the sperical Bessel function
evaluated at $x=k(\tau_0-\tau)$. The latter does not depend
on the model and can be precomputed in advance. 
Fast algorithms exist which can compute spherical Bessel functions
on a 
grid in $x$ and $l$ in short amount of time (e.g. \cite{3.Press92}). The
grid is then stored on a disk and used for all the subsequent calculations.
This leaves us with the
task of accurately calculating the source term,
which determines the CMB spectrum for a given model.
Below we discuss some of the calculational techniques needed for
the implementation of the method. 
We especially highlight the differences
between this approach and the standard Boltzmann
integration approach. Our goal is to develop a method which is
accurate to 1\% in $C_l$ up to $l \sim 1000$
over the whole range of cosmological 
parameters of interest. These include models with varying amount of
dark matter, baryonic matter, Hubble constant, vacuum energy, 
neutrino mass, shape of initial spectrum of perturbations, reionization
and tensor modes. The choice of accuracy is based on estimates of
observational accuracies that will be achievable in the next generation 
of experiments and also on the theoretical limitations of model 
predictions (e.g. cosmic variance, second order effects, etc.).
Most of the figures where we discuss the choice
of parameters 
are calculated for the standard CDM model. 
This model is a reasonable choice in the sense that it is a model which 
exhibits most of the physical effects in realistic models, including 
acoustic oscillations, early-time integrated Sachs-Wolfe effect and 
Silk damping. One has to be careful however not to tune the parameters
based on a single model. We compared our results with results from other 
groups \cite{3.BB95,3.sug95} for a number of 
different models. We find a better than 1\% agreement 
with these calculations over most of the parameter space of models.
The computational parameters we recommend below are based on this more detailed
comparison and are typically more stringent than 
what one would find based on the comparison with the standard CDM model only. 

\subsection{Number of coupled differential equations}

\begin{figure*}
\begin{center}
\leavevmode
\end{center}
\caption{CMB spectra produced by 
varying the number of evolved photon multipole moments, together
with the  
relative error (in \%)
compared to the exact case. While using $l_\gamma=5$ produces
up to 2\% error, using $l_\gamma=7$ gives results almost identical to the 
exact case.}

\label{fig1.los}
\end{figure*}

In the standard Boltzmann method the 
photon distribution function is expanded to a high $l_{\rm max}$ 
(\ref{photon}) and typically one has to 
solve a coupled system of several thousand differential equations.
In the integral method one
evaluates the source terms $S(k,\tau)$ as a function 
of time (equations \ref{finalscalar} and  \ref{finaltensor}) and
one only requires the knowledge of photon multipole 
moments up to $l=2$, 
plus the metric perturbations and baryon velocity. 
This greatly reduces 
the number of coupled differential equations that are needed to be solved. 
For an accurate evaluation of the lowest multipoles in the integral
method one has to extend
the hierarchy somewhat beyond $l=2$, because the
lower multipole moments are coupled to the higher multipoles
(\ref{photon}).  
Because power is only being 
transferred from lower to higher $l$  
it suffices to keep a few 
moments to achieve a high numerical accuracy of the first few moments.
One has to be careful however to avoid 
unwanted 
reflections of the power being transferred from low $l$ to high $l$, 
which occur for example if a simple cut-off in the hierarchy is imposed.
This can be achieved by modifying 
the boundary condition for the last term in the 
hierarchy 
using the free streaming approximation
\cite{3.mabert95,3.Hu95}. In the absence of scattering 
(the so-called free streaming regime),
the recurrence relation among the 
photon multipoles in equation (\ref{photon}) becomes the
generator of spherical Bessel functions. 
One can therefore use a different recurrence 
relation among the spherical Bessel functions 
to approximate the last term in the hierarchy without reference to the 
higher terms.
The same approximation can also be used for polarization and
neutrino hierarchies. This type of
closure scheme works extremely well and only a few multipoles
beyond $l=2$ are needed for an accurate calculation of the source 
term. This is shown in 
Figure \ref{fig1.los}, where a relative error in the spectrum is plotted
for several choices of maximal number of photon multipoles.
We choose to end the
photon hierarchy (both anisotropy and polarization) 
at $l_\gamma=8$ and massless neutrino at $l_\nu=7$, which 
results in an error lower than $0.1\%$ compared to the exact case. 
Instead of a few thousand coupled differential equations 
we therefore evolve about 35 equations
and the integration time is correspondingly reduced.

\subsection{Sampling of CMB multipoles}

\begin{figure*}
\begin{center}
\leavevmode
\end{center}
\caption{  
Relative error between the exact and interpolated spectrum,  
where every 20th, 50th or 70th multipole is calculated. The 
maximal error for the three approximations is less
than 0.2, 0.4 and 1.2\%, respectively.  
The rms deviation from the exact spectrum is further improved by
finer sampling, because the
interpolated spectra are exact in the sampled points. For the sampling
in every 50th multipole the rms error is 0.1\%.} 
\label{fig2.los}
\end{figure*}

In the standard Boltzmann integration method one solves for the whole 
photon hierarchy (\ref{photon}) and the 
resultant $\Delta_l$ is automatically obtained for each $l$ up to some 
$l_{\rm max}$. 
The CMB spectra are however
very smooth (see Figure \ref{fig1.los}), except for the lowest $l$ where the
discrete nature of the spectrum becomes important. This means that 
the spectrum need not be sampled for each $l$ and instead  
it suffices to sparsely sample
the spectrum in a number of points and interpolate between them.
Figure \ref{fig2.los} shows the result of such interpolation 
with cubic splines (see e.g. \cite{3.Press92}) when every 
20th, 50th or 70th 
$l$ is sampled beyond $l=100$ with an increasingly denser sampling
towards small $l$, so that each $l$ is sampled below $l=10$. 
While sampling of every 70th $l$ results in maximal error of 
1\%, sampling in 
every 20th or 50th $l$ gives errors below 0.2 and 0.4\%, respectively. 
We choose to compute every 50th $C_l$ beyond $l=100$ in addition to
15 $l$ modes
below $l=100$, so that a total of
45 $l$ modes are calculated up to $l_{\rm max} =1500$. This gives   
a typical (rms) error of 0.1\%, with excursions of up to 0.4\%. 
The number of integrals in equation (\ref{cl}) is thus 
reduced by a factor of 50 at high $l$ and the computational 
time needed for the integrals becomes comparable to or smaller
than the time
needed to solve for the system of differential equations.

\subsection{Integration over time}

For each Fourier mode $k$ the source term is integrated
over time $\tau$ (equation  \ref{finalscalar}). 
The sampling in time need not be 
uniform, because the dominant contribution arises from the epoch
of recombination around $z\sim 1100$, the width of which is 
determined by the visibility function $g$ and is rather narrow in 
look-back time for
standard recombination scenarios. During this epoch the sources
acoustically oscillate on a time scale of $c_sk^{-1}$, 
so that the longest wavelength modes are the slowest to vary.
For short wavelengths the rate of sampling should therefore be higher.
Even for long wavelengths the source function
should still be sampled in several
points across the last-scattering surface. This is because the 
terms in (\ref{finalscalar}) depend on the derivatives
of the visibility function. 
If the visibility function $g$ is narrow then its derivative will 
also be narrow and will sharply 
change sign at the peak of $g$. Its integration 
will lead to numerical roundoff
errors if not properly sampled, even though  
positive and negative contributions nearly cancel out when integrated
over time and make only a small contribution to the integral. 
Figure \ref{fig3} shows 
the error in integration caused by sampling this epoch with 10, 20
or 40 points. Based on comparison with several models
we choose to sample the recombination epoch
with 40 points, which results in very small ($\sim$ 0.1\%) errors. 
After this 
epoch the main contribution to the anisotropies arises from the 
integrated Sachs-Wolfe term. This is typically a slowly changing
function and it is sufficient to sample 
the entire range in time until the present in 40 points.
The exceptions here are models
with reionization, where the visibility function becomes non-negligible
again and a new last-scattering surface is created. In this case a 
more accurate sampling of the source is also needed at lower redshifts.

\begin{figure*}
\begin{center}
\leavevmode
\end{center}
\caption{ Error in the spectrum caused by insufficient
temporal sampling of the source term. 
Inaccurate sampling of the source during recombination leads to 
numerical errors, which can reach the level of 1\% if the source 
is sampled in only 10 points across the recombination epoch. 
Finer sampling in time gives much smaller errors for this model. 
Comparisons with other models indicate that sampling in 40 
points is needed for accurate integration.} 
\label{fig3}
\end{figure*}

\subsection{Integration over wavenumbers}

The main computational cost of standard 
CMB calculations is solving the coupled
system of differential equations.
The number of $k$-modes for which the system is solved is the main 
factor that determines the speed of the method. For results accurate
to $l_{\rm max}$ one has to sample the wavenumbers up to a maximum value
$k_{\rm max}=l_{\rm max}/ \tau_0$.  
In the line of sight integration method 
solving the coupled
system of differential equations still dominates the computational time 
(although for each mode 
the time is significantly shorter than in the 
standard Boltzmann method 
because of a smaller system of equations).
It is therefore instructive to compare the number of $k$ 
evaluations needed in each of the methods to achieve a given
accuracy in the final spectrum.

\begin{figure*}
\begin{center}
\leavevmode
\end{center}
\caption{In (a) $\Delta_{T,150}^{S}(k)$ is plotted as a function of
wavevector 
$k$. In (b) $\Delta_{T,150}^{S}(k)$ is decomposed into the 
source term $S_T^{S}$ integrated over time
and the spherical
Bessel function $j_{150}(k\tau_0)$. The high frequency oscillations 
of $\Delta_{T,150}^{S}(k)$ are caused by oscillations of 
the spherical Bessel function $j_{150}(k\tau_0)$, whereas the source 
term varies much more slowly. This allows one to reduce the number of
$k$ evaluations in the line of sight integration method, because only 
the source term needs to be sampled.  } 
\label{fig4}
\end{figure*}

In the standard Boltzmann method
one solves for $\Delta_{T,l}^{S}(k)$ directly, so this quantity 
must be sampled densely enough for accurate integration.
Figure \ref{fig4}a shows $\Delta_{T,l}^{S}(k)$
for $l=150$ in a standard CDM model. One can see that it is a rapidly 
oscillating function with a frequency $k \sim \tau_0^{-1}$. 
Each oscillation needs to be sampled in at least a few points to assure 
an accurate integration.
To obtain a smooth CMB spectrum one typically 
requires 6 points over one period, implying 
$2l_{\rm max}$ $k$-mode evaluations. 
This number can be reduced somewhat 
by filtering out the sampling noise in the spectrum \cite{3.Hu95}, but even
in this case one requires at least 1-2 points per each period or 
$l_{\rm max}/2$ $k$-mode evaluations. 

To understand the nature of these rapid oscillations in
$\Delta_{T,l}^{S}(k)$ we will consider wavelengths larger than
the width of the last scattering surface.  In this case the
Bessel function in (\ref{finalscalar}) can be pulled out of the integral as
$j_l(k\tau_0)$ because the time at which recombination occurs,
when the dominant contribution to $\Delta_{T,l}^{S}(k)$ is
created, is 
much smaller than $\tau_0$ and $k \Delta\tau\ll 1$ ($\Delta\tau$
is the interval of time for which the visibility function differs
appreciably from zero).
So the final $\Delta_{T,l}^{S}(k)$ is approximately 
the product of $j_l(k\tau_0)$ and $S_T^{S}$ integrated over time,
if the finite width of the last scattering surface and
contributions after recombination can be  ignored.  

Figure \ref{fig4}b shows the 
source term $S_T^{S}$ integrated over time
and the Bessel function $j_l(k\tau_0)$. 
It shows that the high frequency
oscillations in $\Delta_{T,l}^{S}(k)$ seen in Figure \ref{fig4}a
are caused by the oscillation of the spherical
Bessel functions, while the oscillations of the source term have a
much longer period in $k$.
The different periods of the two 
oscillations can be understood
using the tight coupling approximation \cite{3.Hu95,3.Seljak94}.
Prior and during recombination photons are coupled to the baryons and the 
two oscillate together  
with a typical acoustic timescale $\tau_{s}\sim \tau_{\rm rec}/\sqrt{3}
\sim \tau_0/\sqrt{3z_{\rm rec}} \sim \tau_0/50$. The 
frequency of acoustic oscillations $k \sim \tau_{\rm rec}^{-1}$ 
is therefore 50 times higher than
the frequency of oscillations in spherical Bessel functions, which 
oscillate as $\tau_0^{-1}$. 

Because an accurate sampling of the 
source term requires only a few 
points over each acoustic oscillation, the total number of $k$ evaluations
in the integral method 
can be significantly reduced compared to the standard methods. 
Typically a few dozen
evaluations are needed
over the entire range of $k$, compared to about 500 evaluations
in the standard method when a noise filtering technique is 
used and 2000 otherwise 
(for $l_{\rm max} \sim 1000$). Once the source
term is evaluated at these points one can
interpolate it at points with preevaluated 
spherical Bessel functions, which can be much more densely sampled 
at no additional computational cost. 
The end result is the same accuracy as in the standard method,
provided that the source is sampled in sufficient number of points.
Figure \ref{fig5} shows the relative error in the CMB spectrum for the
cases where the 
source term is calculated in 40, 60 and 80 points between 
0 and $k \tau_0 =3000$ (for $l_{\rm max}=1500$). 
While using 40 points results in up to 1\% errors, using 60 points
decreases the maximum error to below 0.2\% for this model. 
In general it suffices to use 
$l_{\rm max}/30$ $k$ modes, which is at least  
an order of magnitude smaller than in 
the standard methods. 
Note that with this method there is no need to 
filter the spectrum to reduce the sampling noise, because  
the latter is mainly caused by insufficient sampling
of the spherical Bessel functions, which
are easy to precompute. The additional operations needed for a higher 
sampling (summation and source interpolation) do not  
significantly affect the overall computational time. Moreover, 
if each $C_l$ is accurately calculated they can be sparsely sampled
and interpolated
(section 3.2). This would not be possible if they had a significant noise
component added to them.  

\begin{figure*}
\begin{center}
\leavevmode
\end{center}
\caption{ Error in the spectrum caused by insufficient 
$k$-mode sampling of the source term. 
Sampling the source with 40 points up to 
$k = 2l_{\rm max}$ leads to 1\% errors, while  
with 60 or 80 points the maximal error decreases to 0.2\%.
Comparisons with other models indicate that sampling in 60 
points is sufficient for accurate integration.} 
\label{fig5}
\end{figure*}

\section{Summary}

In this chapter we presented a new method for accurate calculations of
CMB anisotropy and polarization spectra. 
The method is not based on any approximations and is an
alternative to the standard Boltzmann calculations, which are based 
on solving large numbers of differential equations. The 
approach proposed here uses a hybrid integro-differential 
approach in solving the same system of equations. 
By rewriting the Boltzmann equations in the integral form
the solution for the photon anisotropy 
spectrum can be written as an integral over a
source and a geometrical term. The first is determined by a small number
of contributors to the photon equations of motion and the second is 
given by the radial eigenfunctions, which do not depend on the
particular cosmological model, but only on the geometry of space.

One advantage of the split between geometrical and dynamical 
terms is that
it clarifies their different contributions to
the final spectrum. A good example of this is the temperature anisotropy
in the non-flat universe, which can be be written using a similar 
decomposition, except
that spherical Bessel functions have to be replaced with their 
appropriate generalization \cite{3.abbott86}. 
This is discussed in more detail elsewhere \cite{3.zsb}.
Here we simply remark that replacing radial eigenfunctions
in a non-flat space with their flat space counterpart (keeping comoving
angular distance to the LSS unchanged) is only approximate and
does not become exact even in the large
$l$ (small angle) limit. The geometry of the 
universe leaves its signature in the CMB spectra in a rather 
nontrivial way and does not lead only to a simple rescaling of the 
spectrum by $\Omega_{\rm matter}^{-1/2}$ \cite{3.Jungman95}.

The main advantage of our line of sight integration method is
its speed and accuracy. 
For a given set of parameters it is two orders of magnitude
faster than the standard Boltzmann 
methods, while preserving the same accuracy.
We compared our results with the results by Sugiyama \cite{3.sug95} and
by Bode \& Bertschinger \cite{3.BB95}. In both cases the agreement was 
better than 1\% up to a very high $l$ for all of the models we
compared to.

The method is useful for fast and accurate normalizations
of density power spectra from CMB measurements, 
which for a given model require the CMB anisotropy spectrum and
matter transfer function, both of which are provided by the output
of the method. 
Speed and accuracy are even more important for accurate determination
of cosmological parameters from CMB measurements. In such applications
one wants to perform a search over a large parameter space,
which typically requires calculating the spectra of a
several thousand models (e.g. \cite{3.Jungman95}). 
One feasible way to do so is to use approximation methods 
mentioned in the introduction. These 
can be made extremely fast, but at a cost of sacrificing the 
accuracy. While several percent accuracy is sufficient for analyzing
the present day experiments, it will not satisfy the requirements
for the future all-sky surveys of microwave sky. Provided that 
foreground contributions can be succesfully filtered out one can hope for 
accuracies on the spectrum close to the cosmic variance limit,
which for a broad band averages can indeed reach below 1\% at 
$l>100$. It is at this stage that fast and accurate CMB calculations
such as the one presented in this chapter
will become crucial and might enable one 
to determine many cosmological parameters with an unprecedented
accuracy.

\newpage

\chapter{Information in the CMB Polarization$^1$}\label{chapinfpol}

\setcounter{footnote}{1}
\footnotetext{Based on M. Zaldarriaga, Phys. Rev. D {\bf
55}, 1822 (1997),
U. Seljak \& M. Zaldarriaga, Phys. Rev. Lett. {\bf 78}, 2054 (1997)
and D. N. Spergel \& M. Zaldarriaga, Phys. Rev. Lett. {\bf 79}, 2180
(1997).}
 
\section{Introduction}

It is now well established that temperature 
anisotropies in CMB offer one of the best probes
of early universe, which could potentially lead to a precise
determination of a large number of cosmological parameters 
\cite{4.jungman,4.bet,4.zss}. 
The main advantage of CMB versus more local probes of large-scale 
structure is that the fluctuations were created at an epoch when the
universe was still in a linear regime. While this fact has long been 
emphasized for temperature anisotropies, the same holds also
for polarization in CMB and as such it offers the same advantages
as the temperature anisotropies in the determination of cosmological 
parameters. The main limitation of polarization is that it is 
predicted to be
small: theoretical calculations show that CMB will be polarized 
at the 5-10\% level on small angular scales and much less than that
on large angular scales \cite{4.pol}. However, future
CMB satellite 
missions (MAP and Planck) will be so sensitive that even such small signals
will be measurable.
Even if polarization by itself cannot compete with the
temperature anisotropies, a combination of the two could result in a 
much more accurate determination of certain cosmological parameters,
in particular those that are limited by a finite number of multipoles in 
the sky (i.e. cosmic variance).

\section{The Imprint of an Early Reionization of the Universe}

It has been pointed out that an early reionization of the universe
will greatly enhance polarization \cite{4.be84}. The fact
that in universes that never recombined the polarization would also be
large was noted in many of the above studies. More recently 
Ng \& Ng \cite{4.ng95} discussed the polarization generated
in reionized universes with instantaneous  recombination. 
The Sachs-Wolfe effect was the only source of anisotropies that they
included. They
concluded that reionization at a moderate redshift could boost polarization
to the level of a few percent of the temperature perturbations.
To make detailed predictions
for an experiments (such as that being built at Wisconsin University) 
a realistic
recombination history should be used since polarization is very 
sensitive to the duration of recombination \cite{4.zal95,4.frewin94}. 
Baryons should also be included in the 
calculation as the acoustic oscillation in the photon-baryon
plasma are very important to determine polarization.   

In this section  
we discuss in detail the physics behind the
polarization generated in models where there was an early 
reionization after the usual recombination. 
These models
show very distinct features in the polarization power spectrum 
including 
a new peak at low $l$. This
peak is not present either in the standard recombination scenarios or 
in the cases where the universe never recombined and it is the cause 
for the boost in the polarization.

\subsection{Standard Ionization History}

In this section we review previous results for the CMB polarization
for a standard ionization history in a flat space-time. 
The Boltzmann equations for the perturbations in the scalar case 
are given by equations (\ref{boltzmann1}) and (\ref{eqnv}),
\begin{eqnarray} 
\dot\Delta_T +ik\mu \Delta_T 
&=&-{1\over 6}\dot h-{1\over 3}(\dot h+6\dot\eta)+\dot\kappa\left[-\Delta_T +
\Delta_{T0} +i\mu v_b +{1\over 2}P_2(\mu)\Pi
\right] \nonumber \\   
\dot\Delta_P +ik\mu \Delta_P &=& \dot\kappa \left[-\Delta_P +
{1\over 4} P^2_2(\mu) \Pi\right] 
\label{Boltzmann}
\end{eqnarray}
They can be formally integrated to give (\ref{formal}),
\begin{eqnarray}
\Delta_T(k,\tau_0) &=&\int_0^{\tau_0}d\tau e^{i k \mu (\tau -\tau_0)}
e^{-\kappa} 
\{ \dot\kappa e^{-\kappa} [\Delta_{T0}+i\mu v_b + {1\over 2} P_2(\mu)
\Pi]\nonumber \\
&-& {1\over 6}\dot h-{1\over 3}(\dot h+6\dot\eta)
P_2(\mu)\} \nonumber \\
\Delta_P(k,\tau_0)
 &=& {1\over 4}\int_0^{\tau_0} d\tau e^{i k \mu (\tau -\tau_0)}
e^{-\kappa} \dot\kappa  \Pi P_2^2(\mu).
\label{formal4}
\end{eqnarray}

\begin{figure*}
\begin{center}
\leavevmode
\end{center}
\caption{$l(l+1)C_{l}  / 2\pi$ for both temperature (a) and
polarization (b) for standard CDM and a model where the optical depth to 
recombination is $\kappa_{ri}=1.0$.}
\label{fig1rei}
\end{figure*}
 
Figure \ref{fig1rei} shows the temperature and polarization 
$C_l$ spectra 
for the standard CDM model ($\Omega_0=1$, $H_0=50\, \rm km\,\rm  sec^{-1}$
and $\Omega_{b}=0.05$), normalizing the result to the COBE measurement.
Normalization was carried out using the fits to the shape 
and amplitude of the 4 year COBE data described in \cite{4.bunnwhite}, 
which approximately fixes $C_{T10} \sim 47 \mu K^2$. In Sections
\ref{weaklens} and \ref{corrcdm}
we discussed an approximate solution 
for wavelengths longer than the width of the last scattering surface,
$\Delta\tau_D$ (equation \ref{polapprox}). We showed that 
the polarization perturbation is
\cite{4.zal95}, 
\begin{equation}
\Delta_P =0.51  
(1-\mu^2) e^{i k \mu (\tau_D -\tau_0)} k\Delta\tau_D \Delta_{T1}(\tau_D) 
\label{analitzh}
\end{equation}
where $\tau_D$ is the conformal time of decoupling.
Note that in the tight coupling regime 
\footnote{This velocity is in the longitudinal gauge,
$v_b^{conf}=v_b^{synch}+k\dot{\alpha}$.} $\Delta_{T1}\propto
v_b$. 
The polarization is proportional to the velocity 
difference between places separated by a distance $\Delta\tau_D$, 
the distance photons travel on average during decoupling. 
For the standard adiabatic initial conditions 
$\Delta_{T1}$ and the baryon velocity  vanish as $k \tau \rightarrow 0$
which together with the $k\Delta\tau_D$ factor in the previous expression
explains the dramatic decrease of polarization with decreasing $l$.
For large wavelengths the quadrupole generated in the photon 
distribution as photons travel
between their last scatterings is extremely small  both due to the small
distance they can travel compared to
the wavelength as well as  to the small velocity
differences generated by these small $k$ perturbations. 

\subsection{The Reionized Case}

\begin{figure*}
\begin{center}
\leavevmode
\end{center}
\label{fig2rei}
\caption{Visibility function for standard CDM with reionization
such that the optical depth to 
recombination is $\kappa_{ri}=1.0$.} 
\end{figure*}

In this section we  consider models with early reionization
and try to explain the origin 
of the new features that appear in the polarization power spectrum.
For definiteveness we use a  
standard CDM model with specified optical depth to 
recombination $\kappa_{ri}$. For example if reionization 
occurred at a redshift of around $z_{ri}\sim100$ then
$\kappa_{ri}=1.0$, assuming that all hydrogen atoms are ionized
up to the present epoch ($x_e=1.0$). 
Figure \ref{fig2rei} 
shows the visibility function, $g(\tau)=\dot\kappa \exp{(-\kappa)}$, 
for $\kappa_{ri}=1.0$ . The visibility
function has a very simple interpretation: the probability that a photon 
reaching the observer last scattered between $\tau$ and $\tau + d\tau$
is  $g(\tau) d\tau$. The first peak in Figure \ref{fig2rei}, occurring at 
$\tau \approx 120$ Mpc for sCDM ($h=0.5$)
accounts for the photons that last scattered
at recombination. The area under this peak, the probability
that a photon came directly to us from this epoch, is $\exp(-\kappa_{ri})$.
The area under the second peak gives the probability  that a photon 
scattered after reionization and is equal to $1-\exp(-\kappa_{ri})$. 

Figure \ref{fig1rei} shows the result of numerically integrating the Boltzmann
equations for this reionized case. On small angular scales,
the polarization  acoustic peaks are suppressed, just like in
the temperature case. This is very easy to understand: only
a fraction $\exp(-\kappa_{ri})$ of the photons reaching the observer come from
recombination, so their contribution to the $C_l$ power spectrum is reduced 
by a factor $\exp(-2 \kappa_{ri})$. On large scales new temperature
anisotropies are created that compensate the reduction and leave the
spectrum approximately the same.
New peaks appear in the polarization 
power spectrum while  the temperature anisotropy shows 
no new peaks. These new peaks in the polarization are what boost 
it and may make it detectable.

Let us  try to understand the origin of these peaks. 
Before the decoupling of photons and electrons, for 
low values of $k$ the 
largest perturbation in the photon distribution function is the monopole,
$\Delta_{T0}$
because of 
the tight coupling between photons and electrons before recombination.
Both the dipole and the quadrupole as well as the polarization
perturbations are much smaller. But after photons and electrons decouple, all
the temperature multipoles can grow by free streaming. Power is being
carried from the low multipole 
moments to higher ones. This is a geometrical 
effect.  The
temperature quadrupole is growing by free streaming after recombination and 
by the time the universe  reionizes 
there is an appreciable quadrupole that can
generate polarization. The structure of this quadrupole explains the new 
features in the polarization power spectrum.

The formal line of
sight solution for the polarization perturbation is
\begin{equation}
\Delta_P={1\over 4}\int_0^{\tau_0} d\tau e^{i k \mu (\tau -\tau_0)}
e^{-\kappa} \dot\kappa   
\Pi P_2^2(\mu).
\label{lospol}
\end{equation}
The visibility function has two peaks one
at recombination and one at reionization.  It is convenient
to separate the previous integral in two parts,
\begin{equation}
\Delta_P={1\over 4}P^2_2(\mu)\Big( \int_0^{\tau_{ri}} 
d\tau e^{i k \mu (\tau -\tau_0)} \dot\kappa e^{-\kappa}  \Pi +
\int_{\tau_{ri}}^{\tau_{0}} 
d\tau e^{i k \mu (\tau -\tau_0)} \dot\kappa e^{-\kappa} 
\Pi\Big) 
\label{polsep}
\end{equation}
where $\tau_{ri}$ is the conformal time of the start of reionization.
The first integral  represents the polarization generated at 
recombination and can easily be shown to be 
\begin{equation}
\Delta_P^{(1)}\equiv {1\over 4}P^2_2(\mu) \int_0^{\tau_{ri}} 
d\tau e^{i k \mu (\tau -\tau_0)} \dot\kappa e^{-\kappa} \Pi =
e^{-\kappa_{ri}}\Delta_P^{NR} 
\end{equation}
where $\Delta_P^{NR}$ is the polarization that would be measured
if there was no reionization, as  discussed in the previous section.
This contribution is damped because only a fraction $\exp(-\kappa_{ri})$
of the photons that arrive to the observer came directly from recombination
without scattering again after reionization.

Let us now consider the new contribution arising from reionization.
The polarization source is $\Pi=\Delta_{T2}-6 (_{+2}\Delta_{P2}
+ \ _{-2}\Delta_{P2})$. The temperature quadrupole 
$\Delta_{T2}$ is large and originates  in  the free 
streaming of the monopole at recombination, while the
polarization terms do not grow after decoupling 
and are thus negligible to  first approximation. Equation (\ref{polsep})
shows that the new polarization is basically an average of the value
of the temperature quadrupole during reionization.
This accounts for all the new features in the polarization power spectrum
of Figure \ref{fig1rei}.

To understand the origin of these new  peaks
let us find the amplitude of the temperature quadrupole at the time 
reionization starts $\tau_{ri}$. The monopole at recombination 
is approximately given by \cite{4.hu95c}
\begin{equation}
(\Delta_{T0}+\dot \alpha)(\tau_D)=
{1\over 3}\psi (1+3R) \cos(k c_s \tau_D) -R \psi 
\end{equation}
where $\psi$ is a constant (the value of the conformal gauge gravitational potential), $R=3\rho_b / 4 \rho_{\gamma}|_{\tau_D}
\approx 30 \Omega_b h^2$ and 
$c_s=1 / \sqrt{3(1+R)}$ is the photon-baryon sound speed.
The quadrupole at $\tau_{ri}$ arising from the free streaming 
of this monopole is simply (equation \ref{finalscalar}), 
\begin{equation}
\Delta_{T2}(\tau_{ri})= (\Delta_{T0}+\dot \alpha)(\tau_D) j_2 [k(\tau_{ri}-
\tau_D)]
\label{dt2}
\end{equation}
where $j_2$ is the $l=2$ spherical Bessel function.  

\begin{figure*}
\begin{center}
\leavevmode
\end{center}
\caption{$l(l+1)C_{lP} / 2\pi$ (a) for CDM models with
varying  $\kappa_{ri}=0.5,\ 1.0,\ 1.5$ and 
(b) for models with varying cosmological constant $\Omega_{\Lambda}
=0.3,\ 0.5,\ 0.7$ and a fixed redshift of reionization $z_{ri}=100$. 
Reionized ($\kappa_{ri}=1.0$) CDM models (c) with
varying  $\Omega_b=0.3,\ 0.5,\ 0.8$ and 
(d) with different Hubble constants $H_0
=60,\ 80,\ 100 \ \rm km \; \rm sec^{-1} \rm Mpc^{-1}$. In all cases
reionization was assumed to be complete ($x_e=1$)}
\label{fig3rei}
\end{figure*}

The peaks of $\Delta_{T2}(\tau_{ri})$ as a function of $k$
will show up in the polarization power spectrum. The first 
peak of equation (\ref{dt2}) is approximately at the first peak of
the Bessel function because $ c_s \tau_D \ll (\tau_{ri}-\tau_D)$.
The wavenumber for this 
first peak is approximately given by
$k(\tau_{ri}-\tau_D)\sim 2$, translating
into an  $l$ value as usual according to 
$l\sim k (\tau_0-\tau_{ri})$. The physical size of the wavelength is
translated to an angular size in the sky using the 
distance to reionization, $\tau_0-\tau_{ri}$. In terms of multipoles,
the first peak is at $l \sim k (\tau_0-\tau_{ri}) 
\sim 2 (\tau_0-\tau_{ri}) /(\tau_{ri}-\tau_D)\sim 2  \sqrt{z_{ri}}$.
For the case under consideration, $z\sim 100$ this means 
$l\sim 20$ which agrees very well with the 
the first peak in Figure \ref{fig3rei}. Only the first few 
peaks appear because
the reionization scattering surface is  wide and thus 
for smaller wavelengths the source
in equation (\ref{polsep})
oscillates during its width and cancels out after
integration. 
This cancellation
makes the new polarization small and thus  hidden under the polarization
generated at recombination.
 
The major factor determining the difference in height 
of these new peaks for different models  
is the fraction of photons reaching the observer 
that last scattered after reionization, $1-\exp(-\kappa_{ri})$.
The ratio of the distances
between the observer and reionization to that between the two scattering 
surfaces determines the positions of the peaks, and the optical depth 
determines $\kappa_{ri}$ their heights.
 
To further illustrate these points, Figure \ref{fig3rei}a 
shows the $C_{El}$ spectrum for 
standard CDM models with varying optical depths $\kappa_{ri}$. The peaks not 
only vary in height but also in position, as the redshift of reionization 
increases when $\kappa_{ri}$ does and so the position 
of the peaks gets moved to smaller scales, $l_{peak}\sim 2 \sqrt{z_{ri}}$. 
Figure \ref{fig3rei}b 
on the other hand show how these peaks vary with the cosmological 
constant for a fixed reionization redshift $z_{ri}=100$. 
The positions hardly change as both the distance to reionization
and the distance between the two scattering surfaces
scale approximately in the same way with the matter density (in this 
calculations the matter density is given by $\Omega_0=1-\Omega_{\Lambda}$
where $\Omega_{\Lambda}$ is the energy density due to the cosmological
constant). 
The distance to
a fixed redshift increases with the cosmological constant, the 
optical depth $\kappa_{ri}$ increases, and consequently
the peaks should get higher. The fact that this is not the case
is a consequence of the COBE normalization, because models with larger
values of the cosmological constant have larger additional
contributions to the low $l$ temperature anisotropies from the
ISW effect while polarization is not affected by the late time
variation of the gravitational potential which does not affect
the polarization.
Thus the changes in the normalization to keep the value of $C_{T10}$ fixed  
partially compensate the change in the height
of the new polarization peaks produced by the larger optical depth.   

Figures 3c and 3d explore the dependence of the polarization
power spectrum with the baryon density and the Hubble constant for
a fixed optical depth to decoupling, $\kappa_{ri}=1.0$, keeping
the rest of the
parameters the same as sCDM. The height
of the first peak in the spectrum remains nearly constant as it is determined
by $\kappa_{ri}$ which was kept fixed. The fact that the peaks move is 
due to the fact that the redshift of reionization is changing, it is
given by 
$(1+z_{ri})\approx 100 [\kappa_{ri} (0.5/h)(0.05/\Omega_b) (1/x_e)]^{2/3}$ 
and so $l$ scales
approximately as $l\propto (\kappa_{ri}/ h\Omega_b x_e)^{1/3}$. 

In the sCDM model the reionization must have occurred extremely early
($z_{ri}\approx 100$)
in order to produce an optical depth of unity;
even an optical depth of $\kappa_{ri}=0.5$ 
is only obtained for $z_{ri}\approx 60$. The
situation is different for open models or models with a cosmological
constant. An approximate scaling for the optical depth
valid for $\Omega_0 z_{ri}\gg 1$
is $\kappa_{ri}\propto (h \Omega_b x_e/\Omega_0^{1/2}) (1+z_{ri})^{3/2} $,
so for example reionization starting at $z_{ri}\approx 23$ will produce
an optical depth $\kappa_{ri}\approx 0.5$ in a model with
$\Omega_0=0.2$, $H_0=70 \rm km\ \rm sec^{-1}\ \rm 
Mpc^{-1}$ and $\Omega_b=0.1$.
 
The polarization increase on large scales produced by an early
reionization of the universe can have an important impact on the
accuracy with which future satellite missions will be able to
reconstruct cosmological parameters from CMB. This will be the focus
of chapter \ref{chapconstraints}. 
There will be other polarization experiments from the ground before
the satellites fly, for example POLAR at Wisconsin University 
\cite{4.wisconsin} and 
POLATRON from CalTech \cite{4.polatron}. We illustrate the impact of
the reionization signal in the polarization  by looking at 
the Wisconsin  experiment. It  will try to measure 
both $Q$ and $U$ parameters with an expected sensitivity of  
$1 \mu \rm K$ per pixel.  The instrument will allow measurements with a $7^{o}$
FWHM at an early stage and a $1^{o}$ FWHM afterwards. This corresponds
to a gaussian window function, 
$W_l=\exp[-(l+0.5)^2 \sigma_{\theta}^2],\ \sigma_{\theta} = \theta
/ 2\sqrt{(2\ln2)}$ where $\theta$ is the FWHM of the detector in 
radians. The predicted values for the Stokes parameters were
calculated numerically and the spectra normalized to COBE.
The expected {\it rms} values of $Q$ for standard 
CDM with no reionization are $P(7^{o})=4.8\times 10^{-2} \mu \rm K$ and 
$P(1^{o})=0.77 \mu \rm K$. These values, especially the large angular scale
one, are extremely small and thus very difficult to detect. This is the
reason why the reionized scenarios are the most promising to 
detect polarization.

\begin{figure*}
\begin{center}
\leavevmode
\end{center}
\caption{$l(l+1)C_{l} / 2\pi$ for both temperature (a) and
polarization (b) for standard CDM and a model where the optical depth to 
recombination is $\kappa_{ri}=0.5$ and a spectral index $n=1.2$.}
\label{fig4rei}
\end{figure*}

Reionization will not only change 
the polarization power spectrum but also the temperature one, and
could suppress the acoustic peaks completely. 
There is some degree of degeneracy between the different parameters
determining the CMB spectra, for example a reionization with a moderate
optical depth will decrease the amplitude of the acoustic peaks but 
this effect may be compensated by changing the spectral index
\cite{4.bond94}. In fact only an optical depth in the 
$0.10-0.20$ range seems detectable from temperature maps alone
\cite{4.jungman,4.bet,4.zss}.
Figure \ref{fig4rei} shows both polarization and temperature power 
spectra for standard CDM with a spectral index  $n=1$ and 
a reionized model with $\kappa_{ri}=0.5$ but a spectral index $n=1.2$. 
The differences in the anisotropy power spectra are not large,
while the polarization spectra are
 very different. The {\it rms} $P$ values in 
this reionized case are
$P(7^{o})=1.2 \mu \rm K$ and $P(1^{o})=1.8 \mu \rm K$. 
For the large angular scale experiment the difference with 
standard CDM is more than two orders 
of magnitude and in the one degree case is more than a factor of two.
Thus a polarization measurement would much more easily distinguish
between these two scenarios.

\begin{figure*}
\begin{center}
\leavevmode
\end{center}
\caption{Polarization {\it rms} fluctuations ($\mu K$) as a function of 
the optical depth, $\kappa_{ri}$ for a $7^o$ and $1^o$ FWHM experiments.}
\label{fig5rei}
\end{figure*}

Figure \ref{fig5rei} shows the {\it rms} value of $P$ as a function of $\kappa_{ri}$,
the major parameter determining  the amplitude of the 
polarization perturbation. $P(7^{o})$ only exceeds  $1 \mu \rm K$  
for $\kappa_{ri} \geq 0.5$ but saturates quickly near $1.8 \mu \rm K$.
On the other hand $P(1^{o})$ quickly raises above $1\mu \rm K$ and
reaches  $3.2 \mu \rm K$ for an optical depth of two. This means that even
a negative detection at the  $1 \mu \rm K$ level for the one 
degree experiment 
is enough to rule out some models,
those with optical depth $\kappa_{ri} \ge 0.3$.
Parameters other than $\kappa_{ri}$ do not make much difference in the
height of the peaks. Table \ref{4.table1} explores the dependence
of $P(7^{o})$ and  $P(1^{o})$ with different cosmological parameters
for a fixed $\kappa_{ri}=1.0$. Although the 
height of the peaks remains almost constant in these models, slight shifts 
in their  location  
change the predicted $P$. The $7^o$ $rms$ linear
polarization  is more sensitive to the position of the first peak. 
The $1^{o}$ experiment has the best chance of 
putting interesting constraints on a possible reionization, as the expected
signal is greater because it is sensitive to all the power in the new
peaks of the polarization power spectrum.  A correlation
analysis between the polarization in the 
forty pixels that the experiment will measure
may further improve the above limits.

In summary,
the polarization of the microwave background is very sensitive to the 
ionization history of the universe and an early reionization can 
greatly enhance it.
We have discussed in detail the physics behind the generation
of polarization in reionized scenarios and the appearance of new peaks in
the polarization power spectrum. We have identified the major parameters
determining the location of these peaks, the ratio of distances between 
the observer and the reionization scattering surface to that between 
reionization and recombination. The height of the peaks is mainly function of
$\kappa_{ri}$, the optical depth to recombination.

\vspace{1cm}

\begin{table}
\caption{
Degree of linear polarization in $\mu \rm K$ 
SCDM (first row) and several cosmological models
all with $\kappa_{ri}=1.0$. The value of the
cosmological constant is such that all the above 
models are flat, $\Omega_{total}=1.0$. $H_0$ is the Hubble constant in 
$\rm km \; \rm sec^{-1} \rm Mpc^{-1}$.}
\label{4.table1}
\begin{center}
\begin{tabular}{||l|l|l|l|l||}    \hline
$\Omega_0$&$\Omega_b$&$H_0$&$P(7^o)$&$P(1^o)$ \\ \hline
1.0 & 0.05 & 50 & $4.81\, 10^{-2}$& 0.642 \\ \hline
0.7 & 0.05 & 50 & 1.62 & 2.25 \\ \hline
0.5 & 0.05 & 50 & 1.67 & 2.50 \\ \hline
0.3 & 0.05 & 50 & 1.62 & 2.25 \\ \hline
1.0 & 0.03 & 50 & 1.40 & 2.67 \\ \hline
1.0 & 0.08 & 50 & 1.83 & 2.79 \\ \hline
1.0 & 0.10 & 50 & 1.91 & 2.80 \\ \hline
1.0 & 0.05 & 60 & 1.72 & 2.79 \\ \hline
1.0 & 0.05 & 80 & 1.84 & 2.85 \\ \hline
1.0 & 0.05 &100 & 1.92 & 2.88 \\ \hline
\end{tabular}
\end{center}
\end{table}

\section{The Detection of Non-Scalar Perturbations}\label{secnonscal}
\def\edth{\;\raise1.5pt\hbox{$'$}\hskip-6pt\partial\;}
\def\baredth{\;\overline{\raise1.5pt\hbox{$'$}\hskip-6pt
\partial}\;}
\def\bi#1{\bbox{#1}}
\def\gsim{\raise2.90pt\hbox{$\scriptstyle
>$} \hspace{-6.4pt}
\lower.5pt\hbox{$\scriptscriptstyle
\sim$}\; }
\def\lsim{\raise2.90pt\hbox{$\scriptstyle
<$} \hspace{-6pt}\lower.5pt\hbox{$\scriptscriptstyle\sim$}\; }

\def\bi#1{\hbox{\boldmath{$#1$}}}

Primordial gravity waves produce fluctuations in the tensor 
component of the metric, which could result in a significant 
contribution to the CMB anisotropies on large angular scales. 
Unfortunately, the presence of
scalar modes prevents one from clearly separating one contribution 
from another. If there are only a finite number of multipoles where the
tensor contribution is significant then there is a limit in amplitude 
beyond which 
tensors cannot be distinguished from random fluctuations.  
In a noise free experiment
the tensor to scalar ratio $T/S$ needs to be larger than 
0.15 to be measurable
in temperature maps \cite{4.knox}. 
Independent determination of
the tensor spectral slope $n_T$ is even less accurate and a rejection
of the consistency relation in inflationary models 
$T/S=-7n_T$ is only possible if $|n_T|\gg (T/S)/7$ \cite{4.knox,4.dodelson}.
Polarization produced by tensor
modes has also been studied \cite{4.tenspol}, 
but only in the small scale limit. 
In previous work correlations
between the Stokes parameters $Q$ and $U$ have been used. 
These two variables are not the most suitable for the
analysis as they depend on the orientation of coordinate system.
In Chapter \ref{chapstatcmb} we 
presented a full spherical analysis of
polarization using Newman-Penrose spin-s spherical harmonic decomposition. 
We have shown that there is  
a particular combination of Stokes parameters that vanishes in the case
of scalar modes, which can thus be used as a probe of gravity waves. 
Here we  discuss
the sensitivity needed to detect this signal and compare it to the
expected sensitivities of future CMB satellites 
We use 
the expressions in Chapters \ref{chapstatcmb} and \ref{chaplos}
to evaluate the power spectra
in various theoretical models. 
We use $T/S$ as the parameter
determining the amplitude of tensor polarization.
Figure \ref{fig1tens} shows the predictions for scalar and tensor 
contribution in standard CDM model 
with no reionization 
(a) and in reionized 
universe with optical depth of $\kappa_{ri}=0.2$ (b). The latter value is
typical in standard cosmological models \cite{4.haiman}.
We assumed $T/S=1$ and $n_T=(n_s-1)=-0.15$. In the no-reionization case
both tensor spectra 
peak around $l \sim 100$ and give comparable contributions, although the 
$B$ channel is somewhat smaller. Comparing the scalar and tensor $E$
channels one can see that
scalar polarization dominates for $T/S\lsim 1$. 
Even though 
tensor contribution is larger than scalar at low $l$, the overall power 
there is too small to be measurable. 
Tensor reconstruction in the $E$ channel suffers from similar drawbacks 
as in the case of temperature 
anisotropies: because of large scalar 
contribution, cosmic variance prevents one
from isolating very small tensor contributions
\cite{4.knox}. 
The situation improves if the
epoch of reionization occurred sufficiently early that a moderate
optical depth to Thomson scattering is accumulated (Fig.~\ref{fig1tens}b). 
In this case there is an 
additional peak at low $l$ \cite{4.zalreio} and the 
relative contribution of tensor to
scalar polarization in $E$ channel around $l=10$ 
is higher than around $l=100$. Still, if $T/S \ll 1$ 
cosmic variance again 
limits out ability to extract unambiguously the tensor contribution.
It is in this limit that the importance of the $B$ channel becomes crucial. 
This channel is not contaminated by scalar contributions and is only 
limited by noise, so in principle with sufficient noise sensitivity 
one can detect even very small tensor to scalar ratios. Moreover, a detection
of signal in this channel would be a model independent detection of
non-scalar perturbations. In the following we will discuss sensitivity 
to gravity waves using both only $B$ channel information alone
and all available information.

We can obtain an 
estimate of how well can tensor parameters be reconstructed by 
using only the $B$ channel and assuming that the rest of cosmological
parameters will be accurately determined from the temperature and $E$
polarization measurements. While this test might not be the most 
powerful it is the least model dependent: any detection in $B$ 
channel 
would imply a presence of non-scalar fluctuations and therefore
give a significant constraint on cosmological models.
Because the $B$ channel does not
cross-correlate with either $T$ or $E$ \cite{4.uros,4.kks,4.spinlong} 
only its auto-correlation
needs to be considered. 
A useful method to estimate parameter sensitivity for a given experiment 
is to use the Fisher information matrix introduced in equation 
(\ref{intfisher}) 
\cite{4.jungman,4.uros,4.kks,4.spinlong},
\begin{eqnarray}
\alpha_{ij}&=&\sum_{l=2}^{l_{\rm max}}{(2l+1)f_{\rm sky} \over 2}
\nonumber \\
&\times&\Big[ C_{Bl}+{4\pi\sigma^2 \over N}e^{l(l+1)\sigma_b^2}\Big]^{-2}
\left( {\partial C_{Bl} \over \partial s_i} \right)\left(
{\partial C_{Bl} \over \partial s_j} \right),
\end{eqnarray}
where $f_{\rm sky}$ is the sky coverage.
Receiver noise
can be parametrized by $4 \pi \sigma^2/N$, where $\sigma$ is the noise 
per pixel and $N$ is the number of pixels. Typical values are
$(0.15 \mu{\rm K})^2$ 
for MAP
and $(0.025\mu{\rm K})^2$ for the most
sensitive Planck bolometer channel
in one year of observation.
In our case the parameters $s_i$ can be $T/S$ and $n_T$, so that the matrix is
only 2x2. The error on each parameter 
is given by
$(\alpha^{-1}_{ii})^{1/2}$ 
if the other parameter is assumed to be unknown 
and $(\alpha_{ii})^{-1/2}$ if the other parameter is assumed to be known.
Using this expression we may calculate the experiment
sensitivity to these parameters. 
Current inflationary models and 
limits from large scale structure and COBE 
predict $T/S$ to be less than unity. 
Figure \ref{fig1tens} shows that the expected amplitude in this case 
is below 0.5$\mu$K.
We find that MAP is not sufficiently sensitive in the $B$ channel 
to detect these low levels.
On the other hand, Planck will be much more
sensitive and can detect $T/S >0.2$ if the tensor index $n_T$ is assumed
to be known (for example through the
consistency relation). For the underlying model with 
$T/S=1$ one can determine it
with an error $\Delta( T/S) \sim 0.1$. If the tensor index is not known then 
a combination of the two parameters, which corresponds to the total
power under the $B$ curve in Figure \ref{fig1tens}, can still be 
determined with the same accuracy. 

\begin{figure*}
\begin{center}
\leavevmode
\end{center}
\caption{Multipole moments for the three polarization spectra for 
no-reionization case (a) and reionized case with optical depth of 0.2
(b).
The underlying model is ``standard CDM'' with $T/S=1$.}
\label{fig1tens}
\end{figure*}

Separate determination of the tensor amplitude and slope
from the $B$ channel is only possible 
in reionized models. 
In the no-reionization model the contribution to $B$ is very
narrow in $l$ space and the leverage on $n_T$ independent of $T/S$ is small,
so that 
the correlation coefficient $\alpha_{12}/(\alpha_{11}\alpha_{22})^{1/2}$
is almost always close to unity.   
A modest amount of reionization improves the separation; 
in the reionized models the power spectrum for $B$ is bimodal 
(Figure \ref{fig1tens})
and the overall signal is higher, which gives
a better leverage on $n_T$ independent of $T/S$.
For $\kappa_{ri}=0.2$ the Planck errors are $\Delta (T/S) \sim 0.15$ and
$\Delta n_T \sim 0.1$ for the underlying model with $T/S = 1$. 
These results depend on
the overall amplitude
relative to the noise level. As long as both peaks can be separated 
from the noise one can determine the tensor slope, which allows to 
test the inflationary consistency relation.

Combining temperature, $E$ polarization 
and their cross-correlation
further improves these estimates. In this case other parameters
that affect scalar modes 
such as baryon density, Hubble constant or cosmological constant
enter as well and the results become more
model dependent (Chapter \ref{chapconstraints}). 
The Fisher information matrix has to be generalized to 
include all the parameters that can be degenerate with the tensor 
parameters. 
The results 
depend on the class of models and number of parameters one restricts
to in the analysis, as opposed to the results based on $B$ channel above,
which  depend only on the two main parameters that characterize the
gravity wave production. 
As a typical example, for $T/S=0.1$ and $\kappa_{ri}=0.1$
one can determine $\Delta (T/S)=0.05$ and $\Delta n_T=0.2$ with 
Planck \cite{4.zss}. 
These errors improve 
further if a model with higher $T/S$ or 
$\kappa_{ri}$ is assumed. 
For the same underlying 
model without using polarization the expected errors are 
$\Delta T/S \sim 0.26$ and $\Delta n_T\sim 1$, significantly worse
than with polarization. 
Even for MAP the limits on $T/S$ improve by a factor of 2 when 
polarization information is included.

To summarize the above discussion, 
future CMB missions are likely to reach the sensitivities needed to measure
(or reject) a significant production of primordial gravity waves in 
the early universe through polarization measurements, which will vastly 
improve the limits possible from temperature measurements alone and 
will allow a test of
consistency relation.
The more challenging question is wether
the foreground can be subtracted to  at the required level. 
At low frequencies radio point sources and synchrotron emission 
from our galaxy dominate the foregrounds
and both are polarized at a 10\% level.  
Their contribution 
decreases at higher frequencies and with several frequency 
measurements one can subtract these foregrounds at 
frequencies around 100 GHz
at the required microkelvin level. 
At even higher frequencies dust 
is the dominant foreground, but
is measured to be only a few percent polarized \cite{4.dust}. 

While we discussed only scalar and tensor modes, vector modes, if present
before recombination, will also contribute to both polarization 
channels and so could contaminate the signature of gravity waves.
At present there are no viable cosmological models that would
produce a significant contribution of vector modes without a comparable
amount of tensor modes. In inflationary models vector modes, even if
produced during inflation, decay away and are not significant during 
recombination. In topological 
defect models nonlinear sources continuously 
create both vector and tensor modes and so some of the signal in $B$ 
channel could be caused by vector modes. Even in these models however 
a significant fraction of signal in $B$ will still be generated
by tensor modes and in any case, absence of signal in $B$ channel would
rule out such models. Polarization thus offers a unique way to probe 
cosmological models that is within reach of the next generation of CMB 
experiments.

\smallskip

\section{A Test of the Causal Structure of the Universe}

There are two competing families of models to explain the origin of 
the structure we observe in our universe: defect
models, where a symmetry breaking phase transition generates seeds that
form sub-horizon scale density fluctuations, and inflationary models, where
a period of superluminal expansion turns quantum fluctuations into
super-horizon density perturbations. 
A fundamental difference between these
two mechanisms of structure formation is that only inflation alters the
causal structure of the very early universe and is able to create
correlations on super-horizon scales, while defects are causal and all
correlations vanish for events where both light cones do not overlap.

The COBE satellite observed correlations on angles much larger than that
subtended by the horizon at decoupling $(\theta _h\sim 1.1^o)$ in the CMB
temperature. 
This does not however, imply that there were correlations on super
horizon scales at decoupling because a time dependent gravitational
potential can produce temperature fluctuations at late-times through the 
integrated Sachs Wolfe effect (ISW). For example,  cosmic string or texture
models can reproduce the COBE results despite being causal models by
generating the fluctuations at low redshift.

Measurements of temperature fluctuations at small scales have been suggested
as a potential test of inflation: inflationary models and most 
non-inflationary
ones predict different locations and relative heights for the acoustic
peaks \cite{4.huwhite}. 
Unfortunately, causality alone is insufficient to distinguish inflationary
and non-inflationary temperature power spectra: causal sources that mimic
exactly the inflationary pattern of peaks can be constructed \cite{4.turok}.
While the predicted CMB fluctuations  of the current family of defect models
differ significantly from inflationary predictions \cite{4.urostur}, it is 
useful to have model independent tests of the causal structure of the early
universe.

Polarization fluctuations are produced by Thomson scattering during the
decoupling of matter and radiation. Thus, unlike temperature fluctuations,
they are unaffected by the ISW effect. 
Measurements of the polarization fluctuations are certain to probe the
surface of last scattering. Hence, the detection of correlated polarization
fluctuations on super-horizon scales at last scattering provides a
clear 
signature of the existence of super-horizon scale fluctuations, one of the
distinctive predictions of inflation.\footnote{In this section 
we will consider
the correlation function in real space (ie. as a function of the
separation angle)
rather than the usual power
spectrum. By doing so we can easily express the 
causality constraint,
while it would become a set of integral constraints that the power
spectra have to satisfy in the now more usual treatment in term of $C_l$s.}

We will work in the initially unperturbed synchronous gauge, where the
metric is given by $ds^2=a^2(\tau )[-d\tau ^2+(\delta _{ij}+h_{ij})dx^idx^j]$%
. We will consider only perturbations produced by scalar modes and will
solve the Einstein equations in the presence of sources (e.g., defects)
using the stiff approximation \cite{4.veerasteb}. The sources are
characterized by their covariantly conserved stress energy tensor $\Theta
_{\mu \nu }$. Before recombination, matter and radiation act as a very
tightly coupled fluid, so the evolution of fluctuations can be described by 
\begin{eqnarray}
\ddot \delta _C+{\frac{\dot a}a}\dot \delta _C &=&4\pi
G(\sum_N(1+3c_N^2)\rho _N\delta _N+\Theta _{00}+\Theta )  \nonumber \\
\dot \delta _R &=&{\frac 43}\dot \delta _C-{\frac 43}\nabla \cdot {\bf {v_R}}
\nonumber \\
\dot {{\bf {v_R}}} &=&-(1-3c_S^2){\frac{\dot a}a}{\bf v_R}-{\frac 34}%
c_S^2\nabla \delta _R,  \label{eq1}
\end{eqnarray}
where $\Theta$ is the trace of the spatial part of $\Theta_{\mu\nu}$,
$\Theta /3$ is the source pressure, $\delta _R$ and ${\bf {v_R}}$ describe
the energy density and velocity of the photon-baryon fluid and $\delta _C$
is the energy density of cold dark matter.  In synchronous gauge, the cold
dark matter has zero velocity. The sum over $N$ is carried out over all
species and $c_S$ is the sound speed. Temperature and polarization
anisotropies seen on the sky today depend on $\delta _R$ and ${\bf {v_R}}$
at decoupling. 

Equations (\ref{eq1}) imply that the photon-baryon fluid propagates
information at the speed of sound and thus cannot generate correlations on
scales larger than the sound horizon. Causality on the other hand implies
that the unequal time correlators of the sources $\langle \Theta _{\mu \nu
}(r,\tau )\Theta _{\mu \nu }(0,\tau ^{\prime })\rangle $ vanishes if $r>\tau
+\tau ^{\prime }$. In the absence of initial correlations, these two
conditions together imply that $\langle X|_{\tau _{*}}(\hat {{\bf n}%
}_1)X|_{\tau _{*}}(\hat {{\bf n}}_2)\rangle =0$ if $\theta _{12}>2\theta
_h\sim 2^o$, where $X={\delta }_R,{\bf v_R},\partial _i{\bf v_R}$ and $\tau
_{*}$ is the conformal time of decoupling.

In the thin scattering surface approximation, equations (\ref{eq1}) are
solved up to recombination and then the photons free stream to the observer.
The final temperature anisotropy in direction $\hat {{\bf n}}$ on the
sky is (equation \ref{tempapprox}),
\begin{eqnarray}
T(\hat {{\bf n}})={\frac{\delta _R}4}|_{\tau _{*}}-\hat {{\bf n}}\cdot {\bf {%
v_R}}|_{\tau _{*}}-{\frac 12}\int_{\tau _{*}}^{\tau _0}d\tau \dot h_{ij}\hat
{{\bf n}}^i\hat {{\bf n}}^j.  \label{eqn2.4}
\end{eqnarray}
The first two terms are evaluated at the last scattering surface and the
third term is an integral along the line of sight, the ISW effect. In
non-inflationary models, the first two terms cannot correlate temperature
fluctuations at separations larger than $2\theta _h\sim 2^o$ but because
anisotropies can be created later through the ISW effect these models can
have temperature correlations on larger angular scales. 

For the polarization we had (equation \ref{polapprox}),
\begin{eqnarray}
(Q+iU)(\hat {{\bf n}})\approx 0.17\Delta \tau _{*}{\bf m}^i{\bf m}^j\partial
_iv_j|_{\tau _{*}}.  
\label{eqn3.4}
\end{eqnarray}
This equation shows that the observed
polarization only depends on the state of the fluid at the last scattering
surface. No correlations can be present in the polarization for separations
larger than $\sim 2^o$ in non-inflationary models. An early 
reionization of the universe alters this argument, we will discuss
that later.  

The correlation functions of $Q$ and $U$ (in their  natural
coordinate system) are given by (eq. \ref{QUr}), 
\begin{eqnarray}
C^{(Q,U)}(\theta)&=&\sum_l {\frac{2l+1 }{4\pi}} [C^{(E,B)}_l\ F^1_l(\theta)
- C^{(B,E)}_l\ F^2_l(\theta) ] 
\end{eqnarray}
where $\ _{\pm 2} Y_l^2(\theta,\phi)=\sqrt{(2l+1) / 4\pi}\ [F^1_l(\theta)\pm
F^2_l(\theta)]\ \exp{(2i\phi)}$. Both correlation functions receive
contributions from the $E$ and $B$ channels.
The $E$ channel contains all the cosmological signal if there are no tensor
or vector modes.

We computed both $C^{(Q,U)}(\theta )$ 
for the model proposed by Turok \cite{4.turok} which has a
clever choice of source stress energy tensor that is able to reproduce the
pattern of peaks of inflationary standard CDM (sCDM).
The results are shown in Figure \ref{figdef1}. We see that the inflationary
model is able to produce polarization
correlations on angular scales larger than $\sim 2^o
$, while the other model cannot. On smaller angular scale than shown in
Figure \ref{figdef1}, the two correlation functions coincide. The difference
between the two models is a result of the causal constraints and is
insensitive to  source evolution. It is also worth pointing out that
in inflationary models the large scale polarization is suppressed relative
to the small scale signal, so we are after a small effect.   

Next, we estimate the expected uncertainties in measuring  $C^{(Q,U)}(\theta
)$. Since receiver noise is the likely to be the dominant source of
variance,  we can make a simple estimate of the total noise: it is
proportional to the number of  independent pairs of pixels, $N_p,$ at a
given separation, $\theta $. For an experiment with a beam FWHM
$\theta _{fwhm},$ $N_p=1/2\times (4\pi /\theta
_{fwhm}^2)\times (2\pi \theta /\theta _{fwhm})$.  If $\sigma _{(Q,U)}^2$ is
the noise in the polarization measurement per resolution element, then the
noise in the cross correlation is given by $\Delta C^{(Q,U)}=\sqrt{2/N_p}\
\sigma _{(Q,U)}^2\approx 20\ w_P^{-1}\sqrt{0.2^o/\theta _{fwhm}}\sqrt{%
2^o/\theta }$ where $w_P^{-1}=\sigma _{(Q,U)}^2\ \Omega _{pix}/4\pi $. 

We can make a more accurate determination of the noise 
using the covariance matrix
of the different power spectra (equation \ref{covdiag}): 
\begin{eqnarray}
{\rm Cov }(\hat{C}_{(E,B)l}^2)&=&{2\over 2l+1}(\hat{C}_{(E,B)l}+
w_P^{-1}e^{l^2 \sigma_b^2})^2
\end{eqnarray}
which give the following variances for the correlation functions,
\begin{eqnarray}
(\Delta C^{(Q,U)})^2&=&\sum_l ({2l+1 \over 4\pi})^2 
\{{\rm Cov}( C^2_{(E,B)l})\ [F^{1}_l(\theta)]^2  \nonumber \\ 
&& + {\rm Cov}( C^2_{(B,E)l})\ [F^{2}_l(\theta)]^2\}. 
\label{eqn6}
\end{eqnarray}
Figure \ref{figdef1} shows the  noise in each correlation,
in the limit where
the variances are dominated by receiver noise 
$(\Delta C^{Q})^2 =
(\Delta C^{U})^2$  and
agree perfectly with our previous estimates. 
If either the cosmic 
variance is important or the power spectra of $E$ and $B$ differ,
then the approximate estimate of the previous paragraph is not
accurate and the full calculation should be used to estimate
the noise.

The noise in the correlation functions can be reduced
by focusing on the $E$-like piece of the polarization.  
The noise in the both $C^{(Q,U)}(\theta)$ 
receives contributions from the variances in both 
$E$ and $B$ spectra (equation \ref{eqn6}), 
but by computing both contributions separately we can
show that the  variance in $E (B)$ makes the dominant contribution to 
$\Delta C^Q (\Delta C^U)$. If we  filter the maps to pull out
only the $E$ component, then we remove not only the $B$ signal but also 
some of the noise and  $\Delta C^U$ is reduced
almost by a factor of $\sim 4$.
The assumption that most of the signal is in the $E$ channel can be
checked within the data as both $E$ and $B$ contributions can be
measured separately from the maps.

For the MAP satellite,   without filtering the noise,
$\Delta C^{(Q,U)}\approx 0.36/\sqrt{\theta }\ \mu K^2$, so it will
not be sensitive enough to detect this signal, even if we combine
all of the three highest frequencies.   However,
if we filter the map to extract the E channel signal, then
the noise in the MAP 
experiment drops to 
$\Delta C^U \sim 0.1/\sqrt{\theta }\ \mu K^2$, 
and the $C^U$ signal should be detectable.
The PLANCK satellite, with its very
sensitive bolometers, should be able to achieve $\Delta C^{(Q,U)}\approx
0.003/\sqrt{\theta }\ \mu K^2$ and should easily be able to detect
both $C^U$ and $C^Q$. As cosmic variance is
not the dominant contribution to the noise,  an experiment observing a small
patch of the sky could also potentially detect this signal.


The temperature-polarization cross-correlation \cite{4.coulson} is another
potential test of the origin of fluctuations: 
although ISW effects produce temperature fluctuations
after decoupling,  we still  do not expect correlations
between temperature and polarization on large angular
scales for defect models. 
The  correlations between  temperature and
polarization  fluctuations directions $\hat{\bf n}_1$ and $\hat{\bf
n}_2$ are, 
\begin{eqnarray}
&& \langle Q(\hat{\bf n}_1)T(\hat{\bf n}_2)\rangle = 
\langle Q_1^* T_2^* \rangle - {1\over 2} 
\int_{\tau_*}^{\tau_0} d\tau 
\hat{ \bf n}_1^i \hat{ \bf n}_1^j \langle \dot{h}_{ij}(\tau) Q^*_1 \rangle,
\label{eqn7}
\end{eqnarray}
$T^*$ stands for the first two terms in equation (\ref{eqn2.4}) and 
$Q^*$ is given  by equation
(\ref{eqn3.4}).

In the polarization temperature cross correlation, only the term involving
the line of sight integration could produce correlations on
large angular scales.  This would require 
 correlations between the late time variations of
the metric and the velocity at last scattering. For this
to occur  in defect models, they must be moving very
fast and remain coherent as they evolve from recombination to 
very late times.  As Figure 1(c) shows, even Turok's causal
seed model, which mimics inflation remarkable well in
the temperature correlation does not predict
any temperature-polarization correlation.

If 
gravity waves, rather than scalar modes,
 were the dominant source of the anisotropies, then
they could, in principle, create a cross correlations on large angular scales.
However, if gravity waves were significant enough to create
a large signal, then they would be directly detectable 
in the B channel. 

Figure \ref{figdef1}(c) shows the calculated values of the cross correlation
together with the expected noise. The signal is well
above the noise for MAP and Planck. The detection of a large angular scale cross correlation with no 
appreciable signal in the polarization $B$ channel would put very
stringent limits on the physics of models trying to mimic inflation.


There is one caveat to our argument: reionization. 
If the universe reionized very
early, a significant fraction of the  observed polarization will come
from the rescattering of photons at late times. 
In most scenarios, the 
fraction of rescattered photons 
is thought to be less than $\sim 20\%$ \cite{4.haiman}. 
Reionization has  two
effects on our argument. First, it reduces the amplitude of the 
correlation function by a factor $\exp{(-2\kappa_{ri})}$,
where  $\kappa_{ri}$ is the optical depth to decoupling
($\kappa_{ri} \leq 0.2$). Second, it creates further structure in the correlation
function on large angular scales. Fortunately, the effect of reionization 
can  be separated from that of the primordial anisotropies: 
it leaves a very specific signature in the power spectrum, a peak 
at very low $l$ 
that is easily distinguish from the $l^6$ dependence
expected from causality constraints alone
\cite{4.zalreio,4.Huwhite97,4.Battye}.
Because of
the form of $F_l^1(\theta)$ and $F_l^2(\theta)$, this peak produces an
almost constant positive offset in 
$C^Q$ and $C^U$ for angles $\theta\sim 2^o$. Because the offset 
in $C^U$ is positive $(F^2_l(\theta) < 0$ for $\theta \sim 2^o$ and
$l < 70$) reionization at a relatively recent epoch  can never
create the negative peak at $\theta\sim 2^o$  
predicted by inflationary models. 

There is a precise signature in $C^{(Q,U)}(\theta)$
on $\sim 2^o$ scales that would 
allow  an unambiguous test of inflation.
The signal is small, but within 
reach for the new generation of experiments. The cross correlation 
between temperature and polarization is also expected to provide 
strong constraints that could distinguish inflation
from non-inflationary models: this signal is
much larger and will be well above the noise for MAP.
The next generation of satellites or even polarization measurements
from the ground could provide a definitive test of the inflationary
paradigm in the relatively near future.

\begin{figure*}
\begin{center}
\leavevmode
\end{center}
\caption{Correlation functions for  Q (a)  and  U (b) Stokes parameters
for sCDM and the causal seed model discussed in the text. The noise in
their determination is shown for both MAP and Planck.   Panel (b)
shows the expected noise for MAP if the CMB maps are flltered
to include only the $E$ channel signal.  Panel (c) shows
the cross correlation between temperature and polarization and the
noise for MAP, the expected variance for Planck is even smaller.
Each resolution element in the correlation function should
be considered independent.}
\label{figdef1}
\end{figure*}

\def\edth{\;\raise1.0pt\hbox{$'$}\hskip-6pt\partial\;}
\def\baredth{\;\overline{\raise1.0pt\hbox{$'$}\hskip-6pt
\partial}\;}
\def\bi#1{\hbox{\boldmath{$#1$}}}
\def\gsim{\raise2.90pt\hbox{$\scriptstyle
>$} \hspace{-6.4pt}
\lower.5pt\hbox{$\scriptscriptstyle
\sim$}\; }
\def\lsim{\raise2.90pt\hbox{$\scriptstyle
<$} \hspace{-6pt}\lower.5pt\hbox{$\scriptscriptstyle\sim$}\; }

\chapter{Predictions for Future Experiments$^1$}\label{chapconstraints}

\setcounter{footnote}{1}
\footnotetext{Based on M. Zaldarriaga,
D. N. Spergel \& U. Seljak, Astrophys. J. {\bf 488}, 1 (1997)}

It has long been recognized that the 
microwave sky is sensitive to 
many cosmological parameters, so that a high resolution
map may lead to their accurate determination 
\cite{5.bond94,5.knox,5.jungman,5.bet}. 
The properties of the microwave background fluctuations are sensitive to 
the geometry of the universe, the baryon-to-photon ratio,
the matter density, 
the Hubble constant, the cosmological constant, and the optical
depth due to reionization in the universe. 
A stochastic background of gravitational waves also leaves an imprint on 
the CMB and their amplitude and slope
may be extracted from the observations.
In addition,  
massive neutrinos and a change in the slope 
of the primordial spectrum also lead to potentially observable features.

Previous calculations trying to determine how well the various
parameters could be constrained were based on approximate
methods for computing the CMB spectra \cite{5.jungman}. 
These approximations have an accuracy of several percent, which
suffices for the analysis of the present-day data. 
However, 
the precision of the future missions will be so high that the use  of such
approximations will not be sufficient for an accurate determination of
the parameters. 
Although high accuracy calculations are not needed at present to 
analyze the observations, they are needed to determine how 
accurately cosmological parameters can be extracted from a given 
experiment.
This is important not only for illustrative purposes but may also help
to guide the experimentalists in the design of the detectors. 
One may for example address the question of how much improvement one 
can expect by increasing the angular resolution of an experiment (and 
by doing so increasing the risk of systematic errors) 
to decide whether this 
improvement is worth the additional risk. Another question of current 
interest is whether it is worth sacrificing some sensitivity in the
temperature maps to gain additional information from the polarization of
the microwave background.
When addressing these questions,
the shortcomings of approximations become particularly problematic. 
The sensitivity to a certain parameter depends on  
the shape of the likelihood function around the maximum, which 
in the simplest approach used so far 
is calculated by 
differentiating the spectrum with respect
to the relevant parameter. This differentiation strongly amplifies  
any numerical inaccuracies and 
 almost always leads to an unphysical breaking of
degeneracies among  parameters and  misleadingly 
optimistic results. 

Previous analysis of CMB sensitivity to cosmological parameters used
only temperature information. However, CMB experiments can measure not
just the temperature fluctuations, but also even weaker variations in 
the polarization of the microwave
sky. Instead of one power spectrum,  
one can measure up to four and so increase 
the amount of information in the two-point correlators 
\cite{5.uros,5.spinlong,5.kks}.
Polarization can provide particularly 
useful information regarding the ionization history
of the universe \cite{5.zal} and the presence of a tensor contribution
\cite{5.letter,5.kks}. Because these parameters 
are partially degenerate with others,
any improvement in their determination leads to a better reconstruction
of other parameters as well.
The two proposed satellite missions
are currently investigating the possibility
of adding or improving their ability to measure polarization, so it is
particularly interesting to address the question of improvement in the 
parameter estimation that results from polarization.

The purpose of this Chapter is to re-examine the determination of
cosmological parameters by CMB experiments 
in light of the issues raised above. It
is particularly timely to perform such an analysis now, when the 
satellite mission parameters are roughly defined. We use
the best current mission parameters in the calculations and
hope that our study provides a useful guide for mission
optimization. 
As in previous work \cite{5.jungman}, 
we use the Fisher information matrix 
to answer the question of how accurately parameters
can be extracted from the CMB data. This approach 
requires a fast and accurate method for calculating the spectra and
we use the CMBFAST
code developed in Chapter \ref{chaplos} \cite{5.sz96a} 
with an accuracy of about 1\%.  
We test the Fisher information  method by 
performing a more general exploration of the shape of the 
likelihood function around its maximum and find that this 
method is sufficiently accurate for the present purpose. 

The outline of this chapter is the following: in Section 
\ref{methods}, we present the methods
used, reviewing the calculation of theoretical spectra 
and the statistical methods to address the question of sensitivity 
to cosmological parameters. In Section \ref{tempdata}, we investigate the  
parameter sensitivity that could be obtained 
using temperature information only and in Section \ref{poldata},
we repeat this analysis using both temperature and polarization information. 
In Section \ref{shape}, 
we explore the accuracy of the Fisher method by performing a more 
general type of analysis and investigate the 
effects of prior information in the accuracy of the reconstruction.
We present our conclusions in Section \ref{concl}.

\section{Methods}\label{methods}

In this section, we review the methods used  to calculate
the constraints on different cosmological parameters 
that could be obtained by the future CMB satellite experiments. 
We discuss the Fisher information matrix approach,
as well as the more general 
method of 
exploring the shape of the likelihood function around the minimum.

\subsection{The Fisher information matrix}

The Fisher information matrix is a measure of the width and shape of the
likelihood function around its maximum. Its elements are
defined as expectation values of the
second derivative of a logarithm of the likelihood function with 
respect to the corresponding pair of parameters.
It can be used to estimate the accuracy
with  which the parameters in the cosmological model could be reconstructed
using the CMB data \cite{5.jungman,5.tegmark96}. 
If only temperature information is
given then for each $l$ a derivative of the temperature
spectrum $C_{Tl}$ 
with respect to the parameter under consideration is computed
and this information is then summed over all $l$ weighted  
by ${\rm Cov }^{-1}(\hat{C}_{Tl}^2)$. 
In the more general case implemented here, 
we have a vector of four derivatives and the
weighting is given by the inverse of the covariance matrix,
\begin{equation}
\alpha_{ij}=\sum_l \sum_{X,Y}{\partial C_{Xl} \over \partial s_i}
{\rm Cov}^{-1}(\hat{C}_{Xl},\hat{C}_{Yl})
{\partial C_{Yl} \over \partial s_j}.
\end{equation}
Here $\alpha_{ij}$ is the Fisher information 
matrix, ${\rm Cov}^{-1}$ is the inverse of the covariance matrix,
$s_i$ are the cosmological parameters one would like to 
estimate and $X,Y$ stands for $T,E,B,C$. For each $l$, one has to
invert the covariance matrix and sum over $X$ and $Y$. The derivatives
were calculated by finite differences and the step was usually
taken to be about $5 \%$ of the value of each parameter. We explored the
dependence of our results on this choice and found that the dependence
is less than  $10 \%$. This indicates that
the likelihood surface is  approximately  Gaussian, an assumption of
the Fisher matrix method that only consideres the curvature around the
maximum of the likelihood. Further tests of this
assumption are discussed in Section  \ref{shape}.

The full covariance matrix between the power spectrum estimators  
was presented in \cite{5.uros,5.spinlong,5.kks}.
The diagonal terms are given by
\begin{eqnarray}
{\rm Cov }(\hat{C}_{Tl}^2)&=&{2\over (2l+1)f_{sky}}({C}_{Tl}+
w_T^{-1}B_l^{-2})^2
\nonumber \\
{\rm Cov }(\hat{C}_{El}^2)&=&{2\over (2l+1)f_{sky}}({C}_{El}+
w_P^{-1}B_l^{-2})^2
\nonumber \\
{\rm Cov }(\hat{C}_{Bl}^2)&=&{2\over (2l+1)f_{sky}}({C}_{Bl}+
w_P^{-1}B_l^{-2})^2
\nonumber \\
{\rm Cov }(\hat{C}_{Cl}^2)&=&{1\over (2l+1)f_{sky}}\left[{C}_{Cl}^2+
({C}_{Tl}+w_T^{-1}B_l^{-2})
({C}_{El}+w_P^{-1}B_l^{-2})\right],
\label{5.eqn7}
\end{eqnarray}
while the non-zero off diagonal terms are
\begin{eqnarray}
{\rm Cov }(\hat{C}_{Tl}\hat{C}_{El})&=&{2\over (2l+1)f_{sky}}{C}_{Cl}^2
\nonumber \\
{\rm Cov }(\hat{C}_{Tl}\hat{C}_{Cl})&=&{2\over (2l+1)f_{sky}}{C}_{Cl}
({C}_{Tl}+w_T^{-1}B_l^{-2})
\nonumber \\
{\rm Cov }(\hat{C}_{El}\hat{C}_{Cl})&=&{2\over (2l+1)f_{sky}}{C}_{Cl}
({C}_{El}+w_P^{-1}B_l^{-2}).
\label{eqn8}
\end{eqnarray}
We have defined $w_{(T,P)}^{-1}=4\pi \sigma_{(T,P)}^2 / N_{pix}$ where
$\sigma_T$ and $\sigma_P$ are
noise per pixel in the temperature 
and either $Q$ or $U$ polarization measurements (they are assumed
equal) and $N_{pix}$ is the number of pixels.
We will also assume that noise is uncorrelated between different pixels
and between different polarization components
$Q$ and $U$. This is only
the simplest possible choice and more complicated noise correlations
arise if all the components are obtained from a single set of
observations. If both temperature and polarization are
obtained from the same experiment by adding and differentiating
the two polarization states, then $\sigma_T^2=\sigma_P^2/2$ and
noise in the temperature is uncorrelated with the noise in
polarization components.
The window function $B_l^{-2}$ accounts
for the beam smearing and in the Gaussian approximation is given by
$B_l^{-2}=\exp{l^2 \sigma_b^2}$, with $\sigma_b$ measuring the width of
the beam.
We introduced $f_{sky}$ as the
fraction of the sky that can be 
used in the analysis. In this chapter we assume $f_{sky}=0.8$.
It should be noted that equations (\ref{5.eqn7}) and (\ref{eqn8}) are
valid only in the limit of uniform sky coverage. 

Both satellite missions will measure in several frequency
channels with different angular resolutions: 
we combine them using $w_{(T,P)}=\sum w_{(T,P)}^c$, where
subscript $c$ refers to 
each channel component. 
For the MAP mission we adopt a noise level $w_T^{-1}=(0.11 \mu K)^2$
and $w_P^{-1}=(0.15 \mu K)^2$ for 
the combined noise of the three highest frequency channels, 
with  
conservatively updated MAP beam sizes:
$0.53^\circ$, $0.35^\circ$ and $0.25^\circ$. 
These beam sizes are smaller than
those in the MAP proposal and represent improved estimates
of MAP's resolution. 
The most recent estimates
of MAP's beam sizes are even smaller\footnote{See the MAP homepage at
http://map.gsfc.nasa.gov.}:
$0.47^\circ$, $0.35^\circ$ and $0.21^\circ$.
For Planck, we assume $w_T^{-1}=(0.011 \mu K)^2$ and 
$w_P^{-1}=(0.025 \mu K)^2$, and combining 
$140$ GHz and $210$ GHz 
bolometer channels.  
For Planck's polarization sensitivity,
we assume a proposed design in which eight out of 
twelve receivers in each channel 
have polarizers. The angular 
resolution at these frequencies is $0.16^\circ$ and 
$0.12^\circ$ FWHM.  We
also  explore the possible science return from an enhanced
bolometer system that achieves polarization sensitivity of $w_P^{-1} =
(0.015 \mu K)^2$.

In our analysis, we are assuming that foregrounds can be subtracted
from the data to the required accuracy.
Previous studies of temperature anisotropies have shown that this 
is not an overly optimistic assumption at least on large 
angular scales (e.g. \cite{5.tegef96}). 
On smaller scales, point source removal as well as secondary processes
may make extracting the signal more problematic. This would mostly 
affect our results on Planck, which 
has enough angular resolution to measure features in the spectrum to 
$l \sim 3000$. For this reason, we compared the results by 
changing the maximum $l$ from $3000$ to $1500$. We find that they 
change by less than $30 \%$, so that the conclusions we find
should be quite robust.  
Foregrounds for
polarization have not been studied in detail yet. Given that there
are fewer foreground sources of polarization and that polarization 
fractions in CMB and foregrounds are comparable, we will make the
optimistic assumption that 
the foregrounds can be subtracted
from the polarization data with sufficient accuracy as well. 
However, as we will show, most of the additional information from 
polarization comes from very large angular scales, where the 
predicted signal is very small. Thus, one should
take our results on polarization as preliminary,  
until a careful analysis of foreground subtraction in polarization
shows at what level can polarization signal be extracted.

The inverse of the Fisher matrix, $\alpha^{-1}$,
is an estimate of the
covariance matrix between  parameters and $\sqrt{(\alpha^{-1})_{ii}}$
approximates 
the standard error in the estimate of the parameter $s_i$.
This is the lower limit because 
Cram\' er-Rao inequality guarantees that for an unbiased estimator 
the variance on the $i$-th parameter has to be equal to or 
larger than $\sqrt{(\alpha^{-1})_{ii}}$.   
In addition to the diagonal elements of $\alpha^{-1}$, we will also
use $2\times 2$ submatrices of $\alpha^{-1}$ to analyze 
the covariance between various
pairs of parameters. {\it The Fisher matrix 
depends not only on the experiment
under consideration, but also on the assumed family of models 
and on the number of parameters that are being extracted from 
the data.}
To highlight this dependence
and 
to assess how the errors on the parameters depend on these choices 
we will vary their number and
consider
several different underlying models.

\subsection{Minimization}

The Fisher information matrix approach assumes
that the shape of the likelihood function around the maximum can be
approximated by a Gaussian. In this section, we drop this assumption and
explore directly the shape of the likelihood function.
We use the
PORT optimization routines \cite{5.gay90} to explore one direction
in parameter space at a time by fixing one parameter 
to a given value 
and allowing the minimization routine to explore the
rest of parameter space to find the minimum of 
$\chi^2=\sum_l \sum_{X,Y}(C_{Xl}-C_{Xl}^*)
{\rm Cov}^{-1}(\hat{C}_{Xl}\hat{C}_{Yl})
(C_{Yl}-{C}_{Yl}^*)$, where $C_{Xl}^*$ denotes the underlying spectrum. 
The value of
$\chi^2$ as a function of this parameter can be compared directly
with the Fisher matrix prediction, $\Delta\chi^2=(s_i-s_i^*)^2/
(\alpha^{-1})_{ii}$, where $s_i^*$ is the value of the 
parameter in the 
underlying 
model. This comparison  tests not only 
the shape of the likelihood function around the maximum but also the
numerical inaccuracies resulting from differentiating the spectrum 
with respect to the relevant parameter. 
The minimization method is also useful for finding explicit
examples of degenerate models, models with 
different underlying parameters but almost indistinguishable spectra.

The additional advantage of the minimization approach is that one can 
easily impose 
various prior information on the data in the form of constraints
or inequalities. Some of these priors  
may reflect theoretical prejudice on the part of the person performing 
the analysis, while others are likely to
be less controversial, such as the requirement  that 
matter density, baryon density and optical depth are all positive.
One might also
be interested in incorporating priors into the estimation 
to take other 
astrophysical information into account, e.g. the limits on the 
Hubble constant or $\Omega_m$ from the local measurements. 
Such additional information can 
help to break some of the degeneracies present in the CMB data, as
discussed in Section 3. 
Note that prior information on the parameters 
can also be incorporated into the Fisher matrix analysis, but 
in its simplest formulation
only in the form of gaussian constraints and not in the 
form of inequalities.  

The main disadvantage of this more general analysis is the
computational cost. At each step the minimization routine has to 
compute  derivatives with respect to all the parameters to find
the direction in parameter space towards the minimum.
If the initial model is sufficiently close to the minimum, then 
the code
typically requires 5-10 steps to find it and to sample the 
likelihood shape this has to be repeated for several values of the 
parameter in question (and also for several parameters). This computational
cost is significantly higher than in the Fisher matrix approach, where 
the derivatives with respect to each parameter 
need to be computed only once. 

\section{Constraints from temperature data}\label{tempdata}

In this section, we investigate how measurements of the CMB temperature 
anisotropies alone 
can constrain different cosmological parameters. 
The 
models studied here are approximately normalized to COBE, which sets
the level of signal to noise for a given experiment.
We will start with models in a 
six-dimensional parameter space 
$\bi {s}_6=(C_{2}^{(S)},h,\Omega_\Lambda,\Omega_b,\kappa_{ri},n_s)$, 
where the parameters are respectively, 
the amplitude of the power spectrum for scalar perturbations at
$l=2$ in units of $\mu K^2$, the Hubble 
constant in units of 100 km/s/Mpc, the  
cosmological constant and baryon density in units of 
critical density, the reionization optical depth and the slope of 
primordial density spectrum.
In models with a non-zero optical depth, we assume that the
universe is instantaneously and fully reionized, so that the ionization 
fraction is 0 before the 
redshift of reionization $z_{ri}$ and 1 afterwards. We limit to this 
simple case because only the total optical
depth $\kappa_{ri}$ can be usefully constrained 
without polarization information. We discuss the more general case
when we discuss polarization below.

The underlying model is standard CDM with 
$\bi{s}_6=(796,0.5,0,0.05,0.05,1.0)$. 
Our base model has an optical depth 
of $0.05$, corresponding to the epoch of reionization at
$z_{ri}\approx 13$. Models include  gravity waves, fixing 
the tensor amplitude using the consistency 
relation predicted by inflation $T/S=-7n_T$ and assuming 
a relation between the scalar and tensor spectral slopes
$n_T=n_s-1$ for $n_s<1$ and $n_T=0$ otherwise, 
which is predicted by the simplest models of inflation. 
The results for MAP 
are summarized in Table \ref{5.table1}.   
It is important to keep in mind that the parameters are highly 
correlated. 
By investigating confidence contour plots in 
planes across the parameter space, one can identify combinations 
of parameters that can be more accurately determined. Previous analytical 
work \cite{5.uros94,5.husugi} showed  that the physics of
the acoustic oscillations is mainly determined by two parameters,
$\Omega_bh^2$ and $\Omega_mh^2$, where $\Omega_m$ is the 
density of matter in units of critical density. 
There is an
approximately  flat direction in the three-dimensional space
of $h$, $\Omega_b$ and $\Omega_m$: for example,
one can change 
$\Omega_m$ and adjust $h$ and $\Omega_b$ to keep
$\Omega_bh^2$ and $\Omega_mh^2$ constant, which will not change the 
pattern of acoustic oscillations.
This degeneracy can be broken in two ways. 
On large
scales, the decay of the gravitational potential 
at late times in $\Omega_m \ne 1$ models 
(the so called late time integrated Sachs-Wolfe or ISW term) 
produces an additional component in the  microwave 
anisotropy power spectrum, which depends only on $\Omega_\Lambda$ 
\cite{5.kofman}. Because the cosmic variance (finite 
number of independent multipole moments) is large for small $l$,
this effect cannot completely break the degeneracy. 
The second way is through the change in the angular 
size of the acoustic 
horizon at recombination, which shifts all the features in the 
spectrum by a multiplicative factor.
Around $\Omega_m=1$, this shift is a rather weak function of 
$\Omega_m$ and scales approximately as $\Omega_m^{-0.1}$, leading 
to almost no effect at low $l$, but is increasingly more important
towards higher $l$. MAP is  sensitive to multipole moments
up to $l \sim 800$, where this effect is small.
Consequently MAP's ability to determine the cosmological constant
will mostly come from large scales and thus will be
limited by  the large cosmic variance. 
Planck has a higher angular resolution and significantly 
lower noise, so it is 
sensitive to the change in the angular size of the horizon. 
Because of this Planck can break the parameter degeneracy 
and determine the cosmological constant to a high precision, as 
shown in Table \ref{5.table2}.

\begin{figure*}
\begin{center}
\leavevmode
\end{center}
\caption{MAP
confidence contours  
(68\% and 95\%)
for models in the six parameter space (a) and 
seven parameter space 
with $T/S$ added as a free parameter (b). Parameters are
normalized to their value in the underlying model denoted with
an asterisk.} 
\label{5.fig1}
\end{figure*}

Figure \ref{5.fig1}a shows 
the confidence contours in the $\Omega_m - h$ and $\Omega_b - h$
planes. The error ellipses are significantly elongated along the lines
$\Delta\Omega_b/\Omega_b + 2.1 \Delta h /h=0$  and
$\Delta\Omega_m/\Omega_m + 3.0 \Delta h /h=0$. 
The combinations $\Omega_bh^{2.1}$ and $\Omega_mh^3$ 
are thus better determined than the parameters 
$\Omega_m$, $\Omega_b$ and $h$ 
themselves, both to about 3\% for MAP. 
It is interesting to note that it is $\Omega_mh^3$ rather 
than $\Omega_mh^2$ that is most accurately determined, which 
reflects the fact that ISW tends to break the 
degeneracy discussed above. However, because the ISW
effect  itself can be mimicked by 
a tilt in the spectral index the degeneracy remains, 
but is shifted to 
a different combination of parameters. One sigma standard errors
on the two physically motivated parameters 
are $\Delta (\Omega_bh^2) /\Omega_bh^2\approx 3 \%$ 
and $\Delta(\Omega_mh^2)/\Omega_mh^2\approx 5 \%$.
The fact that there is a certain degree of degeneracy between
the parameters has already been noted in previous work (e.g.  
\cite{5.bond94}).

Another approximate 
degeneracy present in the temperature spectra is between 
the reionization optical depth $\kappa_{ri}$ and amplitude $C_2^{(S)}$. 
Reionization uniformly suppresses the  
anisotropies from 
recombination by $e^{-\kappa_{ri}}$. On large angular scales, new
anisotropies are generated during reionization 
by the modes that have not yet entered the horizon. 
The new anisotropies compensate the $e^{-\kappa_{ri}}$ suppression, 
so that there is no suppression of anisotropies on COBE scales.
On small scales, the
modes that have entered the horizon have wavelengths 
small compared to the width of the new visibility function and so 
are suppressed because of cancellations between positive and
negative contributions along the line of sight and become
negligible.
The net result is that on small scales the spectrum is suppressed 
by $e^{-2\kappa_{ri}}$ compared to the large scales.
To break the
degeneracy between $C_2^{(S)}$ 
and $\kappa_{ri}$ one has to be 
able to measure the amplitude of the anisotropies 
on both large and small scales and
this is again limited on large scales by cosmic variance.
Hence one cannot accurately determine the two parameters separately, while
their combination $C_2e^{-2\kappa_{ri}}$ 
is much better constrained. Figure \ref{5.fig3} shows that indeed the
error ellipsoid is very elongated in the direction $\Delta C_2/C_2
-0.1\Delta \kappa_{ri}/\kappa_{ri}=0$, which corresponds to 
the above combination  for $\kappa_{ri}=0.05$.

\begin{figure*}
\begin{center}
\leavevmode
\end{center}
\caption{Confidence contours $(68\%\  \& \ 95\%)$ in the
$C_2^{(S)}-\kappa_{ri}$ plane
for models in the six parameter space described in the text
with (dotted lines) 
or without (solid lines) polarization information. } 
\label{5.fig3}
\end{figure*}

We now allow for one more free parameter, the ratio of the tensor
to scalar quadrupole anisotropy
$T/S$, fixing the tensor spectral index $n_T$ using the consistency 
relation predicted by inflation $T/S=-7n_T$ but not assuming a relation
between $n_T$ and $n_s$. 
The variances 
for  MAP are again summarized in Table \ref{5.table1}. 
A comparison with the previous case shows that most variances
have increased.
The error bars
for $h$ and  $\Omega_\Lambda$  are approximately five times
larger than before 
while that for $n_s$ has increased by a factor of six and that for
$\Omega_bh^2$ by almost four. On the other
hand, the error bar for $\kappa_{ri}$ remains unchanged.
It is instructive to look again at the contour plots in the 
$\Omega_m - h$ and $\Omega_b - h$ planes shown in
Figure \ref{5.fig1}b. The degeneracy on individual parameters
is significantly worse because the large angular scale amplitude 
can now be adjusted freely with the new extra degree
of freedom, the tensor to scalar ratio $T/S$. This can therefore
compensate any large scale ISW term and so the degeneracy between 
$h$, $\Omega_\Lambda$ and $\Omega_b$  cannot be broken as easily.
However, a combination of the two parameters is still 
well constrained as shown in Figure \ref{5.fig1}. 
The degenerate lines are now given by
$\Delta\Omega_b/\Omega_b + 1.66 \Delta h /h=0$  and
$\Delta\Omega_m/\Omega_m + 3.0 \Delta h /h=0$, with 
relative errors
$\Delta (\Omega_bh^{1.66}) /\Omega_bh^{1.66}\approx 4 \%$ 
and $\Delta(\Omega_mh^3)/\Omega_mh^3\approx 4 \%$, almost unchanged 
from the 6-parameter case. 
On the other hand for the physically relevant parameters
$\Omega_bh^2$ and $\Omega_mh^2$
we now have 
$\Delta (\Omega_bh^2) /\Omega_bh^2\approx 10 \%$ 
and $\Delta(\Omega_mh^2)/\Omega_mh^2\approx 25 \%$,
which is worse than before. This example indicates 
how the errors on individual parameters can change dramatically 
as we add more parameters while certain combinations of them 
remain almost unaffected.

\begin{figure*}
\begin{center}
\leavevmode
\end{center}
\caption{Power spectra of (a) temperature and (b) polarization 
for two  
models that will be degenerate for MAP 
if only temperature information is used. 
The model with $\Omega_\Lambda=0.6$ is
the result of the minimization relative to the
sCDM for models constrained to have $\Omega_\Lambda=0.6$.
Polarization helps to break this degeneracy.} 
\label{5.fig2}
\end{figure*}

The output of a minimization run trying to fit sCDM temperature
power spectra with models constrained to have 
$\Omega_{\Lambda}=0.6$ shows how different parameters 
can be adjusted in order to keep the power spectrum nearly the same.
The minimization program found the
model $\bi{s}_7=(610,0.67,0.6,0.03,0.09,1.1,0.68)$
where the last number now corresponds to the $T/S$ ratio, as
a model almost indistinguishable from the underlying one. 
The two models differ by $\Delta\chi^2=1.8$ and are
shown in Figure \ref{5.fig2}.

It is interesting to analyze how  different parameters are adjusted
to reproduce the underlying model.
By adding gravity waves 
and increasing both the spectral index and the 
optical depth, the    
ISW effect from the cosmological constant can be compensated so that 
it is only noticeable for the first couple of $C_l$'s. 
The relatively high amount of tensors ($T/S \sim 0.7$)
lowers the scalar normalization
and thus the height
of the  acoustic peaks, which is compensated by the increase in the
spectral 
index to $n_s=1.1$
and the decrease of $\Omega_m h^2$ from $0.25$ to 
$0.18$. The latter moves 
the matter radiation equality closer to recombination
increasing the height of the peaks. 
This is the reason why the degeneracy line is not that of 
constant
$\Omega_m h^2$ as Figure \ref{5.fig1} shows. Changes in
$\Omega_m h^2$ change the structure of the peaks
and this can 
be compensated by changing other parameters 
like the optical depth 
or the slope of the primordial
spectrum. 
This cannot be achieved across all the spectrum
so one can expect that the degeneracy  
will be lifted as one increases the angular resolution,
which is what happens if Planck specifications are used 
(Table \ref{5.table1}).

Note that the amount of gravity waves
introduced to find the best fit does not follow the relation 
between $n_T$ and $n_s$ 
predicted by the simplest inflationary models 
discussed previously:
for $n_s=1.1$ no gravity waves are predicted.  
This explains why the addition of $T/S$ as a free parameter
increases the sizes of most error bars compared to the 6-parameter case. 

While the two models shown in Figure \ref{5.fig2} have very similar
temperature anisotropy spectra, they make very different 
astronomical predictions.  Figure \ref{5.fig4} shows the matter
power spectra of the two models.  An interesting effect is
that the two models are nearly identical on the scale
of $k = 0.1 h $ Mpc$^{-1}$, which corresponds to
$l \sim k\tau_0 \sim 600$, the $l$ range where MAP is very
sensitive and  gravity waves are unimportant.  However, the two
models differ significantly on the $0.01 h $ Mpc$^{-1}$
scale and the power spectrum 
shape 
is very different.
The current observational situation is still controversial 
(e.g. \cite{5.peacock}), but
measurements of
the spectrum by the Sloan Digital Sky Survey (SDSS)
should significantly improve
the power spectrum determination.
The models also
make different predictions for cluster abundances:
the matter dominated model has $\sigma_8 \Omega_m^{0.6} = 1.2$,
while the vacuum dominated model has $\sigma_8 \Omega_m^{0.6}
= 0.8$. Analysis of cluster X-ray temperature and luminosity
functions suggests 
$\sigma_8  \Omega_m^{0.6}
= 0.5 \pm 0.1$ \cite{5.eke}, inconsistent with both
of the models in the figure. 
These kind of measurements can break some of the 
degeneracies in the CMB data.

\begin{figure*}
\begin{center}
\leavevmode
\end{center}
\caption{Hubble diagram for Type Ia supernovae (a) and 
CDM linear power spectra (b) for sCDM and the $\Omega_\Lambda=0.6$ model
described in the text.} 
\label{5.fig4}
\end{figure*}

Observations of 
Type Ia supernovae at redshifts $z\sim 0.3-0.6$ is another
very promising way of measuring cosmological parameters. 
This test complements the CMB constraints
because the combination of $\Omega_m$ and $\Omega_\Lambda$ that leaves
the luminosity distance to a redshift $z\sim 0.3-0.6$ unchanged
differs from the one that leaves the position of the  acoustic peaks
unchanged. Roughly speaking, the SNe observations are sensitive to
$q_0 \simeq \Omega_m/2 - \Omega_\Lambda$, while the CMB observations are
sensitive to the luminosity distance which depends on a roughly
orthogonal combination, $\Omega_m + \Omega_\Lambda$.
Figure \ref{5.fig4}a 
shows the  apparent magnitude vs. redshift plot for
supernovae in the two models of Figure \ref{5.fig3}.
The analysis 
in \cite{5.perlmutter} of the first seven supernovae already excludes the
$\Omega_\Lambda=0.6$ model with a high confidence.  However,
it remains to be seen however whether this test will be free 
of systematics such as 
evolutionary effects
that have plagued other classical cosmological tests
based on the luminosity-redshift relation.

Finally, we may also relax the relation between tensor spectral index
and its amplitude, thereby testing the consistency relation of inflation.
For MAP, we studied two sCDM models, one
with $T/S=0.28$ and one with $T/S=0.1$, but with
$\kappa_{ri}=0.1$, 
for Planck we only used the latter model.
Table \ref{5.table1} summarizes
the obtained one sigma limits.
A comparison between the $T/S=0.28$ model and previous results
for sCDM with seven parameters shows that the addition of $n_T$ as a new
parameter does not significantly change the expected sensitivities to
most parameters.
The largest change, as expected, 
is for the tensor to scalar ratio. We now find
$\Delta T/S\sim 0.7$ which means that the consistency 
relation will only be poorly  tested from the temperature 
measurements.  
If $T/S=0.1 and \kappa_{ri}=0.1$ most error bars
are smaller than if $T/S=0.28 and \kappa_{ri}=0.05$ case. The
reason for this is that the higher 
value of the optical depth in the underlying model makes its detection
easier and this translates to smaller error bars on the other parameters. 
The only exception is $n_T$, which has a significantly higher error
if $T/S=0.1$ 
than if $T/S=0.28$ as expected on the basis  
of the smaller contribution of tensor modes to the total anisotropies.
A comparison between the expected MAP and Planck performances
for the $T/S=0.1$ model shows that Planck error bars 
are significantly smaller.
For $h$, $\Omega_bh^2$ and $\Omega_\Lambda$ the improvement is by a factor
of $10-20$, while for $T/S$ and $C_2^{(S)}$ by 
a factor of $2-3$. The limits on $\kappa_{ri}$ and $n_T$ remain 
nearly unchanged, reflecting the fact that these parameters are 
mostly constrained on large angular scales which are cosmic variance
and not noise/resolution limited. It is for these parameters that 
polarization information helps significantly, as discussed in the 
next section.

The accuracy with which certain parameters can be determined 
depends not only on the number of parameters but
also on their ``true'' value. We tested
the sensitivity of the results by repeating the analysis around a cosmological 
constant model $\bi{s}_8=(922,0.65,0.7,0.06,0.1,1.0,0.1,0.0)$, where 
the last number corresponds to the tensor spectral index $n_T$.
Results for MAP specifications are given in Table \ref{5.table1}.  
The most dramatic change is for the cosmological constant, which 
is a factor of ten better constrained in this case. This is 
because the underlying model has a large ISW effect which increases
the anisotropies at small $l$. 
This cannot be mimicked by adjusting the tensors,
optical depth and scalar slope as can be done if
the slope of the underlying model is flat, such as for sCDM model
in Figure \ref{5.fig2}.
Because of the degeneracy between $\Omega_\Lambda$ and $h$ ,
a better constraint on the former will also improve the latter, as
shown in Table \ref{5.table1}. Similarly, because a change 
in $\Omega_\Lambda$ affects $T/S$, 
$\kappa_{ri}$ and $n_s$ on large scales, 
the limits on these parameters will
also change. 
On the other hand, errors on $\Omega_bh^2$, $C_2^{(S)}$ and $n_T$ 
do not significantly change. 
This example clearly shows that the effects of the 
underlying model can be rather significant for certain parameters, so 
one has to be careful in quoting the numbers without specifying the 
``true'' parameters of the underlying model as well.

So far we only discussed flat cosmological models. CMBFAST can 
compute open cosmological models as well,
and we will now address the question 
of how well can curvature be determined using temperature data. 
We consider models in a six parameter space 
$\bi {s}_6=(C_{2}^{(S)},h,\Omega,\Omega_b, \kappa_{ri},n_s)$, 
with no gravity waves and where
$\Omega_\Lambda=0$, so that $\Omega=\Omega_m$.
We will consider as the underlying
model $\bi{s}_6=(1122,0.65,0.4,0.06,0.05,1.0)$. 
Fisher matrix results are displayed in Table \ref{5.table1}. Within this family
of models $\Omega$ can be determined very precisely by both MAP
and Planck due to its effect on the position of the  acoustic peaks.
This conclusion changes drastically if we also allow cosmological 
constant, in which case $\Omega=\Omega_m+\Omega_\Lambda$.
Both $\Omega$ and $\Omega_\Lambda$ change 
the angular size of the sound horizon at recombination 
so it is possible to change 
the two parameters without changing  
the angular size, hence the two parameters will be nearly
degenerate in general. 
We will discuss this degeneracy in greater detail in the next section, 
but we can already say  that including both parameters in the analysis
increases the error bar on the curvature dramatically.

To summarize our results so far, keeping in mind that 
the precise numbers depend on the underlying 
model and the number of parameters being extracted, we may 
reasonably expect that using temperature information only
MAP (Planck) will be able to achieve accuracies of
$\Delta C_2^{(S)}/C_2^{(S)} \sim 0.5 (0.1)$, 
$\Delta h \sim 0.1 (0.006)$, $\Delta \Omega_{\Lambda} \sim 0.6 (0.03)$,
$\Delta (\Omega_bh^2)/\Omega_bh^2 \sim 0.1(0.008) $,
$\Delta\kappa_{ri} \sim 0.1 (0.1)$,
$\Delta n_s\sim 0.07 (0.006)$, $\Delta (T/S) \sim 0.7 (0.3)$ 
and $\Delta n_T \sim 1 (1)$. 
It is also worth 
emphasizing that there are combinations of the parameters that are 
very well constrained, 
e.g. $\Delta(\Omega_mh^3)/\Omega_mh^3\sim 0.04$ and 
$\Delta(C_2^{(S)})/C_2^{(S)}- 2\Delta{\kappa_{ri}}\sim 0.05$. 
For the family of models with curvature but no cosmological 
constant, MAP (Planck) will be able to achieve 
$\Delta \Omega \sim 0.007 (0.0006)$, determining the curvature of 
the universe with an impressive accuracy.
 
These results agree qualitatively, but not 
quantitatively, with those in \cite{5.jungman}. The discrepancy
is most significant for 
$C_2^{(S)}$, $h$ and $\Omega_\Lambda$, for which 
the error bars obtained here
are significantly larger.  
The limit we obtain for  $\Omega_bh^2$ is 
several times smaller than that in \cite{5.jungman},  
while for the rest of the
parameters the results agree. The use of different codes
for computing model predictions is probably the main cause of 
discrepancies and emphasizes
the need to use high 
accuracy computational codes when performing 
this type of analysis. 

\section{Constraints from temperature and polarization data}\label{poldata}

In this section, we consider the constraints on cosmological parameters
that could be obtained
when both temperature and polarization data are used.
To generate
polarization, two conditions have  to be satisfied: photons need to 
scatter (Thomson scattering has a polarization dependent
scattering cross-section) and the angular distribution of
the photon temperature must have a non-zero quadrupole moment. 
Tight
coupling between photons and electrons prior to recombination makes
the photon temperature distribution nearly isotropic and the
generated polarization very small, specially on scales larger than 
the width of the last scattering surface. For this reason
polarization 
has not been considered previously as being important for the determination 
of cosmological parameters. However,
early reionization increases the polarization amplitude
on large angular scales  in a way which 
cannot be mimicked with variations in other parameters
\cite{5.zal}.
The reason for this is that after recombination 
the quadrupole moment 
starts to grow due to the photon free streaming. If there
is an early reionization with sufficient optical depth, then 
the new scatterings can transform this
angular anisotropy into polarization. This effect 
 dominates on the angular scale of  the horizon when reionization occurs.
It will produce a peak in the polarization power
spectrum with an amplitude proportional to the optical depth,
$\kappa_{ri}$,
and a position
$l\sim 2\sqrt{z_{ri}}$, where $z_{ri}$ is the redshift of
reionization.

We first consider
the six parameter space described in the previous section.
Table \ref{5.table2} 
contains the one sigma errors on the parameters for MAP 
specifications. 
Compared to the temperature case,
the errors improve particularly 
on the amplitude, the reionization optical depth  
and the spectral index $n_s$. 
Figure \ref{5.fig3} shows the 
confidence contours for $C_2^{(S)}$ and $\kappa_{ri}$ with and without 
polarization. One can see from this figure how
the information in the polarization breaks the degeneracy between the
two parameters by reducing the error on $\kappa_{ri}$, 
but does not really improve their non-degenerate
combination, which is well determined from the temperature data alone.

We now allow for one more parameter,
$T/S$. Again, polarization improves the errors 
on most of the parameters 
by a factor of two compared to the no-polarization case,
as summarized in Table \ref{5.table2}.
The optical depth and the amplitude are better constrained for the same 
reason as for the six parameter model discussed above. 
Without polarization,
the extra freedom allowed by the gravity waves made it
possible to compensate the changes on large angular 
scales caused by the ISW, 
while the amplitude of small scale fluctuations could be adjusted
by changing the optical depth and the spectral index.
Changing $n_s$ also changes the slope on 
large angular scales, 
compensating for the change caused by the ISW. 
When polarization is included, a
change on the optical depth produces a large effect in the spectrum:
see the model  with $\Omega_\Lambda=0.6$ 
in Figure \ref{5.fig3}, which has $\kappa_{ri}=0.1$. 
The difference in $\chi^2$ between the two models in Figure 
\ref{5.fig3} becomes 10 instead of 1.8 (for MAP). 

\begin{figure*}
\begin{center}
\leavevmode
\end{center}
\caption{Confidence contours $(68\%\  \& \ 95\%)$ in the
(a) $\Omega_b-h$ plane and (b) $\Omega_m-h$ plane
for models in the seven parameter space described in the text
with or without using polarization information. } 
\label{5.fig5}
\end{figure*}

Figure \ref{5.fig5} shows how the confidence contours
in the $\Omega_m-h$ and $\Omega_b-h$ planes are improved by including
polarization. The $95\%$ confidence contour  
corresponds roughly to the $68\%$ confidence contour that could be
obtained from temperature 
information alone, while the orientation of the 
error ellipsoids does not change. 
As before the well determined combination 
is  constrained from the temperature data alone.
The constraints on tensor parameters also improve
when polarization is included. Again, this results from 
the better sensitivity to the ionization history, which is 
partially degenerate with the tensor contribution, as discussed 
in the previous section.  
The $B$ channel to which only gravity waves contribute
is not providing additional information in the model
with $T/S=0.28$ for MAP noise levels. Even in a model
with $T/S=1$ the $B$ channel does not provide additional 
information in the case of MAP.

With its very sensitive bolometers, Planck
has the potential to detect the $B$ channel
polarization produced by tensor modes: the 
$B$ channel provides a signature free of scalar mode contribution
\cite{5.letter,5.kks}. However, 
it is important to realize that even though for a model
with $T/S=1$ only $20\%$ of the sensitivity of Planck  to tensor
modes is
coming from the $B$ channel. Planck can detected
primordial gravity waves in models 
with $T/S \sim 0.3$ 
in the $B$ channel alone.  If the bolometer sensitivities are improved so that  $w_P^{-1} =
(0.015 \mu K)^2$, then Planck can detect gravity waves
in the $B$ channel even if  $T/S\sim 0.1$. 
We also analyzed the 8-parameter models presented in Table \ref{5.table2}.
For the models with $T/S=0.28$ and $T/S=0.1$,
MAP will not have sufficient sensitivity to test 
the inflationary consistency relation $T/S=-7n_T$. Planck
should have sufficient sensitivity to determine $n_T$ with an error 
of 0.2 if $T/S \sim 0.1$, which would  allow a reasonable test 
of the consistency relation. 

Polarization is helping to constrain most of the 
parameters mainly by better constraining  $\kappa_{ri}$
and thus removing some of its degeneracies with 
other parameters. Planck will be able to determine not only
the total optical depth through the amplitude of the reionization peak but 
also the ionization fraction, $x_e$,
through its position. To investigate this, we assumed that the 
universe reionized 
instantaneously at $z_{ri}$ and that $x_e$ remains constant but different 
from 1 for
$z<z_{ri}$).
The results given in Table \ref{5.table2} indicate 
that $x_e$ can be determined with an accuracy of 15\%. This 
together with the optical depth will be an important test of
galaxy formation models which 
at the moment are consistent with wildly different ionization 
histories and cannot be probed otherwise \cite{5.haiman,5.gnedin}.
We also investigated the modified Planck design, where both 
polarization states in bolometers are measured.
An improved polarization noise of $w_P^{-1} =
(0.015 \mu K)^2$ for Planck will shrink the error bars presented
in Table \ref{5.table2} by an additional $6-20\, \%$. Error bars on
$\Omega_\Lambda$ and $\Omega_bh^2$ are reduced by $20\%$, those
in $h$, $\kappa_{ri}$ and $x_e$  $10-15\, \%$ and for $T/S$, $n_T$
and $n_s$ the improvement is  approximately $6\%$.

We can examine in more detail how polarization helps to constrain 
different cosmological parameters by investigating the angular scales 
in the polarization power spectra that contribute the 
most information. To do
so, we will consider the $T/S=0.1$ model and
perform a Fisher matrix analysis that includes
all the temperature information,  
but polarization information only up to  maximum $l$. 
Figures \ref{5.fig6} (MAP) and \ref{5.fig7} (Planck)
show the increase in accuracy as a function of maximum $l$
for various parameters. 
In the case of Planck, we added the ionization fraction after reionization 
as another  parameter.
Most of the increase in information is coming from the low $l$ portion 
of polarization spectrum, primarily from
the peak produced by reionization around $l\sim 10$. 
The first  acoustic peak in 
the polarization
spectra at $l\sim 100$ explains the second increase in information
in the MAP case. The better noise properties and resolution 
of Planck  
help to reach the higher $l$ polarization
acoustic peaks, which
add additional information for constraining
$h$, $\Omega_bh^2$ and $\Omega_\Lambda$. For Planck, on the other hand,
some of the degeneracies will already be lifted
in the temperature data alone
and so less is gained when polarization data is used
to constrain the ionization history. 

\begin{figure*}
\begin{center}
\leavevmode
\end{center}
\caption{Relative improvement in the parameter estimation 
as a function of the maximum $l$ up to which polarization information is
used for the MAP mission.}
\label{5.fig6} 
\end{figure*}

\begin{figure*}
\begin{center}
\leavevmode
\end{center}
\caption{Same as figure 6 but for Planck mission 
parameters.}
\label{5.fig7} 
\end{figure*}

An interesting question that we can address with 
the methods developed here
is to what extent is one willing 
to sacrifice the sensitivity in temperature to gain sensitivity 
in polarization. A specific example is the 140 GHz channel in Planck, 
where the current proposal is to have
four bolometers with no polarization sensitivity and eight
bolometers which are polarization sensitive so that they transmit only one
polarization state while the other is being thrown away. One 
can compare the
results of the Fisher information matrix analysis for this case 
with the one where
all twelve detectors have only temperature sensitivity, but 
with better overall
noise because no photons are being thrown away. The results 
in this case for
the 8 parameter model with $T/S=0.1$ are 10-20\% better 
than the results given in the fifth column of 
Table \ref{5.table1}. These results should be compared to the 
same case with polarization in Table \ref{5.table2}. The latter case is clearly
better for all the parameters, especially for those that are 
degenerate with 
reionization parameters, where the improvement can be quite dramatic.  
Based on this example it seems clear that it is worth 
including polarization sensitivity in the 
bolometer detectors, even at the expense of some sensitivity
in the temperature. However, it remains to be seen whether 
such small levels of polarization can be separated from the 
foregrounds.

The Fisher matrix results for the six parameter open models 
are presented in Table \ref{5.table2}. 
As expected polarization improved the constraints on 
$C_{2}^{(S)}$ and $\kappa_{ri}$ the most. So far we have explicitly 
left $\Omega_\Lambda$ out of the analysis; 
as discussed in Section 3 the 
positions of the peaks depends on both 
$\Omega$ and $\Omega_\Lambda$
and it is possible to change 
the two parameters without changing  
the spectrum.
For any given value of $\Omega_m$ we may adjust 
$h$ and $\Omega_b$ to keep  $\Omega_bh^2$ and $\Omega_m h^2$
constant, so that acoustic oscillations will not change. 
If we then in addition adjust also $\Omega_\Lambda$ to match the 
angular size of the acoustic features, then 
the power spectra for two models with different underlying 
parameters remain almost unchanged. 
As mentioned in previous section the effect of 
$\Omega_\Lambda$ on the positions of the peaks 
is rather weak around $\Omega_m=1$ and the peak positions
are mostly sensitive to the curvature $\Omega$. 
The lines of constant $l_{peak}$, the inverse of the angular 
size of acoustic horizon, roughly coincide with those of constant 
$\Omega$ near flat  
models, making it possible to weigh the universe using the
position of the peaks. In the more general case, it is not 
$\Omega$ that can be determined from the CMB observations.
but a particular combination of $\Omega_m$ and $\Omega_\Lambda$
that leaves $l_{peak}$ unchanged.
Figure \ref{5.fig9} shows  confidence contours in the 
$\Omega_m-\Omega_\Lambda$ plane. The 
contours approximately agree with the constant $l_{peak}$ (dotted) line,
which around $\Omega_m=1$ coincides with the 
constant $\Omega$  line (dashed) 
but not around $\Omega_m=0.4$. 
The squares and triangles correspond to the minima found by the
minimization routine when constrained to move in subspaces of 
constant $\Omega$ and agree with the ellipsoids
from the Fisher matrix approach.   

\begin{figure*}
\begin{center}
\leavevmode
\end{center}
\caption{Confidence contours $(68\%\  \& \ 95\%)$ in the
$\Omega_\Lambda-\Omega_m$ plane 
for open models in the seven parameter space described in the text.
The dots show the positions of the $\chi^2$ minima found by the
minimization routine when constrained to a subspace of 
constant $\Omega$.}
\label{5.fig9}
\end{figure*}

Figure \ref{5.fig10} shows the temperature and polarization
spectra for the basis model and one found
by the minimization routine with  
$\bi{s}_o=(1495,.87,0.6,0.033,0.051,1.0,0.39)$ where the last number is
now $\Omega_\Lambda$. This model differs
from the basis model by a $\chi^2=2$ 
and so is practically indistinguishable
from it. Only on 
large angular scales do the two models differ somewhat, but 
cosmic variance prevents  an accurate separation between the two.
In this case, polarization does not help to break the degeneracy.
The agreement on the large angular scales is better for polarization 
than for temperature 
because the former does not have a contribution from  
the ISW effect, which is the only effect that can break this degeneracy.
When both $\Omega$ and $\Omega_\Lambda$ are included in the analysis
the $1\sigma$ error bars for both MAP and Planck increase. The greatest
change is for the error bars on the 
curvature that now becomes $\Delta\Omega=0.1$ for
both MAP and Planck.
Note that improving 
the angular resolution does not help to break the degeneracy, which is
why MAP and Planck results are similar. 
If one is willing to allow for both cosmological 
constant and curvature then there is a genuine degeneracy present in
the microwave data and constraints from other cosmological probes 
will be needed to break this degeneracy. 

\begin{figure*}
\begin{center}
\leavevmode
\end{center}
\caption{Power spectra for 
(a) temperature and 
(b) polarization. The model with $\Omega = 0.6, \Omega_\Lambda=0.4$ is
the output of the minimization code when made to fit the
$\Omega=0.4, \Omega_\Lambda = 0$ model. 
Temperature and polarization data were used for this fit. 
the two models differ in $\chi^2$ by 2.} 
\label{5.fig10}
\end{figure*}

\section{Shape of the likelihood function, 
priors and gravitational lensing}\label{shape}

As mentioned in Section 2, the Fisher information matrix approach used so
far assumes that the likelihood function is Gaussian around the maximum.
In previous work \cite{5.jungman}, this assumption was tested
by calculating the likelihood along several directions in parameter space.
This approach could miss potential problems in other directions, particularly
when there are degeneracies between parameters.
We will further test the Gaussian  assumption by 
investigating the shape of the likelihood function varying 
one parameter at a time but marginalizing over the others. 
We fix the relevant parameter and  
let the minimization routine vary all the others in its search for the 
smallest $\chi^2$.  We then repeat the procedure for a different value
of this parameter, mapping the shape of the likelihood function
around the minimum.
The minimization routine is  exploring  parameter space
in all but one direction. These results may be 
compared with the prediction of the
Fisher matrix which follow a parabola in the parameter versus 
log-likelihood plot. 
This comparison tests the gaussianity of the likelihood function
in one direction of parameter space at a time.

The panels in Figure \ref{5.fig12} show two examples of the results 
of this procedure.
In most cases, the agreement between the
Fisher matrix results and those of the minimization code is very good, 
especially very near the minimum (i.e., $\Delta \chi^2 \lsim 2$). 
As illustrated in the $\Omega_b$ panel, there are cases when 
$\chi^2$ increased more rapidly than predicted by the Fisher matrix. 
This 
is caused by the requirements that $\kappa_{ri}$, $\Omega$, $\Omega_\Lambda$
and $\Omega_b$ are all positive, which 
can be enforced easily in the minimization code. 
Of course, such priors are most relevant if the underlying model is
very close to the boundary enforced by the prior and are only important 
on one side of the parameter space.
The importance of this effect therefore depends on the underlying
model. If the amount of information in the CMB data on a given  
parameter is sufficiently high,
then the prior will have only a small effect 
near the maximum of the likelihood function. 

\begin{figure*}
\begin{center}
\leavevmode
\end{center}
\caption{Comparison between the Fisher matrix expansion of the likelihood
around the minimum (solid lines) and direct minimization for two
different cosmological parameters. In most cases the agreement near 
the minimum is good. In the upper panel 
full triangles (crosses) 
correspond to fits of sCDM within the six parameter 
family described in the text, including (not including) the effects of
gravitational lensing. The lower curve belongs to the $T/S=0.28$
model in the eight parameter space.
In the lower panel, the $\Omega_\Lambda>0$ prior is reached for 
sCDM when $\Omega_b >0.05$, which is why the minimization results 
differ from the Fisher matrix results.} 
\label{5.fig12}
\end{figure*}

We also investigated the effect of gravitational
lensing on the parameter reconstruction. As shown in \cite{5.uros96a},
gravitational lensing smears somewhat 
the acoustic oscillations but leaves
the overall shape of the power spectrum unchanged. The amplitude of
the effect depends on the power spectrum of density fluctuations. 
Because the CMBFAST output consists of both CMB and density power spectra
one can use them as an input for the 
calculation of the weak lensing effect following the method in 
\cite{5.uros96a}. The gravitational lensing effect is treated 
self-consistently by normalizing the power spectrum for each model 
to COBE. 
We find that the addition of gravitational
lensing to the calculation does not appreciably 
change the expected sensitivity 
to different parameters that will be attained
with the future CMB experiments. This conclusion again depends somewhat
on the underlying model, but even for sCDM where  COBE
normalization predicts two times larger small scale normalization 
than required by the cluster abundance data, the lensing effect 
is barely noticeable in the error contours for various parameters.
 
\section{Conclusions}\label{concl}

In this chapter,
we have analyzed how accurately cosmological parameters can
be extracted from the CMB measurements
by two future satellite missions. Our work differs from previous
studies on this subject  in that we
use a more accurate 
computational code for calculating the theoretical spectra and we 
include the additional information that is present in the polarization 
of the microwave background. We also investigate how the results 
change if we vary the number of parameters to be modeled 
or the underlying model around which the parameters are 
estimated. Both of these variations can have a large effect 
on the claimed accuracies of certain parameters, so the numbers presented
here should not be used as firm numbers but more as
typical values. Of course, once the underlying model is revealed
to us by the observations then
these estimates can be made more accurate.
The issue of variation of the errors on the number of parameters however 
remains, and  results will always 
depend to some extent on the prior belief. If, for example,
one believes that gravity waves are not generated in inflationary
models (e.g. \cite{5.lyth}) or that they are related to the scalar 
perturbations through a simple relation (e.g. \cite{5.turner}),
then the MAP errors on most parameters shrink by
a factor of 2. Similarly, one may decide that models with 
both curvature and cosmological constant are not likely, which removes
the only inherent degeneracy present in the CMB data.

Using temperature data alone, MAP should be able to make
accurate determinations (better than 10\%) of the scalar
amplitude ($\sigma_8$), the baryon/photon ratio
($\Omega_b h^2$), the matter content ($\Omega_m h^2)$,
the power spectrum slope ($n_s$) and the angular
diameter distance to the surface of last scattering (a combination
of $\Omega$ and $\Omega_\Lambda$).  If we restrict ourselves
to models with no gravity wave content, then MAP should also be able
to make accurate determinations of $\Omega_\Lambda$,
the Hubble constant and the optical depth, $\kappa_{ri}$.  However,
in more general models that include the gravity wave amplitude and
spectral 
slope as additional parameters, the degeneracies between these parameters
are large and they cannot be accurately determined. 

Several other 
measurements of the CMB anisotropies from the ground and from 
balloons are now in progress and accurate results are likely to
be available by the time MAP flies. This additional information 
will help constrain the models further, especially determinations of 
the power spectrum at the smaller angular scales.   

Astronomical data can significantly reduce these degeneracies.
The two nearly degenerate models, sCDM and a tilted
vacuum dominated model ($\Omega_\Lambda = 0.6$) shown in Figure \ref{5.fig2}
can be distinguished already by current 
determinations of $\sigma_8 \Omega^{0.6}$, or by measurements of
the shape of the galaxy power spectrum,
or by measurements of the distance-magnitude relationship with
SNeIa's.

MAP's measurements of polarization will significantly enhance
its scientific return.  These measurements will accurately 
determine the optical depth between the present and the surface of last
scatter.  This will not only probe star formation during the
``dark ages'' ($5 < z < 1300$), but will also enable accurate determination
of the Hubble constant and help place interesting constraints
on $\Omega_\Lambda$ in models with tensor slope and amplitude
as free parameters.

Planck's higher sensitivity and smaller angular resolution
will enable further improvements in the parameter determination.
Particularly noteworthy is its ability to constrain $\Omega_\Lambda$
to better than 5\%  and the Hubble constant
to better than 1\% even in the most general
model considered here.  
The proposed addition of polarization-sensitive bolometer channels
to  Planck significantly enhances its science return. Planck 
should be able to measure the ionization
history of the early universe, thus studying primordial star formation.
Sensitive polarization measurements should enable Planck to determine
the amplitude and slope of the gravity wave spectrum. This is particularly
exciting as it directly tests
the predicted tensor/scalar relations in the inflationary theory
and is a probe of Planck scale physics.  
The primordial gravity wave contribution can at present only be measured
through the CMB observations. One may therefore ask  how
well Planck can determine $T/S$ assuming that other cosmological parameters 
are perfectly known by combining CMB and other astronomical data. The
answer sensitively depends on reionization optical depth. Without 
reionization, $T/S \sim 0.1$ can be detected, while with $\kappa_{ri}=0.1$
this number drops down to $T/S \sim 0.02$. The equivalent number without 
polarization information is $0.2$,
regardless of optical depth.
Improvements in 
sensitivity will further improve these numbers, particularly 
in the $B$ polarization channel which is not cosmic variance
limited in the sense that tensors cannot be confused with scalars.
A detection of a $B$ component
would mean a model independent detection of a stochastic
background of gravitational waves or vector modes
\cite{5.letter}.

The most exciting science return from polarization measurements
come  from measurements at large angular scales (see Figures
\ref{5.fig6} and  \ref{5.fig7}).  
These measurements can only be made
from satellites as systematic effects will swamp balloons
and ground based experiments on these scales.  The low $l$ measurements
enable determinations of the optical depth and the ionization history
of the universe and may lead to the detection of gravity
waves from the early universe. Both foregrounds and systematic
effects may swamp the weak polarization signal, even in
space missions, thus it is important that the satellite experiment
teams adopt scan strategies and frequency coverages 
that can minimize systematics and foregrounds at large
angles.

We explored the question of how priors such as positivity of 
certain parameters or constraints
from other cosmological probes
help reduce the uncertainties from the CMB data alone.
For this purpose, we compared the predictions from the Fisher information 
matrix  with those of the brute-force minimization which allows
the  easy incorporation of inequality priors. As expected, 
we find that positivity changes the 
error estimates only on the parameters that are not well constrained 
by the CMB data. On the other hand, using some additional constraints
such as the limits on the Hubble constant, age of the universe, 
dark matter power spectrum or $q_0$ measurements from 
type Ia supernovae can significantly reduce the error estimates
because the degeneracies present in these cosmological tests are typically
different from those present in the CMB data.
The minimization approach also allows testing the 
assumption that the log-likelihood is well described by a quadratic 
around the minimum, which is implicit 
in the Fisher matrix approach.
We find that this
 is a good approximation close to the minimum, with no 
nearby secondary minima that could be confused with the global one. 
Finally, we also tested the effect of gravitational lensing on the 
reconstruction of parameters and found that its effect on the 
shape of the likelihood function can be neglected. 

In summary, future CMB data will provide us with an unprecedented
amount of information in the form of temperature and polarization 
power spectra. Provided that the true cosmological model belongs
to the class of models studied here these data will enable us to constrain 
several combinations of cosmological parameters with an exquisite
accuracy. While some degeneracies between the cosmological
parameters do exist, and in principle do not allow some of them to be 
accurately determined individually, these can be removed by including 
other cosmological constraints. Some of 
these degeneracies
belong to contrived cosmological models, which may not survive 
when other considerations are included. 
The microwave background is at present
our best hope for an accurate determination of classical cosmological 
parameters.

\newpage
\begin{table}
\caption{Fisher matrix one-sigma error bars for different
cosmological parameters when only temperature is included. 
Table 5.3 gives the cosmological parameters for each of the models.
Columns with $+$ correspond to MAP and those
with $\times$ to Planck.}
\begin{center}
\scriptsize
\begin{tabular}{|l|l|l|l|l|l|l|l|l|} \hline
$Param.$&${\rm sCDM}^+$&${\rm sCDM}^+$&${T \over S}=0.28^+$&
${T \over S}=0.1^+$ 
&${T \over S}=0.1^\times$&$\Omega_\Lambda=0.7^+$
& Open$^+$ & Open$^\times$\\ \hline
$\Delta \ln C_2^{(S)}$ & $2.1\, 10^{-1}$ & $4.2\, 10^{-1}$& 
$4.8\, 10^{-1}$ 
& $4.7 \, 10^{-1}$ & $7.4\, 10^{-2}$&$ 4.1\, 10^{-1}$
& $1.2 \, 10^{-1}$ & $4.7 \, 10^{-2} $ \\ \hline
$\Delta h$& $1.7\, 10^{-2}$  & $9.2\, 10^{-2}$& 
$1.1\, 10^{-1}$ & $1.0 \, 10^{-1}$ 
& $5.1\, 10^{-3}$ &$ 4.1\, 10^{-2}$ & $2.0 \, 10^{-2}$ & 
$1.1 \, 10^{-3}$ \\ \hline
$\Delta \Omega_\Lambda$& $9.8 \, 10^{-2}$& $5.3 \, 10^{-1}$ 
& $6.1 \, 10^{-1}$ & $5.8 \, 10^{-1}$& $2.9 \, 10^{-2}$ 
& $5.0 \, 10^{-2}$ & - & -\\ \hline
$\Delta \Omega_b h^2$& $3.0\, 10^{-4}$ & $1.0\, 10^{-3}$ & 
$9.8\, 10^{-4}$ & $1.2\, 10^{-3}$ 
& $1.2\, 10^{-4}$ & $9.7\, 10^{-4}$
& $1.1 \, 10^{-3}$ & $1.3 \, 10^{-4}$ \\ \hline
$\Delta \kappa_{ri}$& $1.2\, 10^{-1}$ & $1.3\, 10^{-1}$ & 
$1.4\, 10^{-1}$& $1.1\, 10^{-1}$&$8.2 \, 10^{-2}$ &$1.9 \, 10^{-1}$
& $7.2 \, 10^{-2}$ & $3.3 \, 10^{-2}$ \\ \hline
$\Delta n_s$& $9.8\, 10^{-3}$ & $5.9\, 10^{-2}$ & 
$6.7\, 10^{-2}$ & $6.4\, 10^{-2}$ & $5.9\, 10^{-3}$ & $2.9\, 10^{-2}$
& $2.4\, 10^{-2}$ & $3.3 \, 10^{-3}$\\ \hline
$\Delta {T \over S}$& - & $3.9\, 10^{-1}$ & $6.8\, 10^{-1}$ & 
$5.3 \, 10^{-1}$ & $2.5 \, 10^{-1}$ & $3.2 \, 10^{-1}$& - & -\\ \hline
$\Delta n_T$& - & - & $3.9\, 10^{-1}$ & $9.1\, 10^{-1}$ &
$9.4\, 10^{-1}$ & $9.9 \, 10^{-1}$& - & -\\ \hline
$\Delta \Omega$& - & - & - 
&- & - &- & $6.6\, 10^{-3}$& $5.2\, 10^{-4}$ \\ \hline
\label{5.table1}
\end{tabular}
\normalsize
\end{center}
\end{table}

\vspace{1cm}

\begin{table}
\caption{Fisher matrix one-sigma error bars for different
cosmological parameters when both temperature and polarization is included. 
Table 5.3 gives the cosmological parameters for each of the models.
Columns with $+$ correspond to MAP and those
with $\times$ to Planck.}
\begin{center}
\scriptsize
\begin{tabular}{|l|l|l|l|l|l|l|l|l|} \hline
$Param.$&${\rm sCDM}^+$&${\rm sCDM}^+$&${T \over S}=0.28^+$&${T \over S}=0.1^+$ 
&${T \over S}=0.1^\times$&$\Omega_\Lambda=0.7^+$
& Open$^+$ & Open$^\times$\\ \hline
$\Delta \ln C_2^{(S)}$ & $4.8\, 10^{-2}$ & $2.4\, 10^{-1}$& 
$2.8\, 10^{-1}$ & $2.4 \, 10^{-1}$ & $1.0\, 10^{-2}$& $8.3\, 10^{-2}$
&$6.5\, 10^{-2}$ & $1.2\, 10^{-2}$\\ \hline
$\Delta h$& $1.6\, 10^{-2}$  & $5.1\, 10^{-2}$& 
$5.8\, 10^{-2}$ & $5.0\, 10^{-2} $ & $3.0\, 10^{-3}$& $3.8\, 10^{-2}$
&$1.9\, 10^{-2}$ & 
$1.0\, 10^{-3}$\\ \hline
$\Delta \Omega_\Lambda$& $9.3 \, 10^{-2}$& $2.9 \, 10^{-1}$ 
& $3.3 \, 10^{-1}$ & $2.9 \, 10^{-1}$& $1.7 \, 10^{-2}$ & $4.6\, 10^{-2}$
& - & - \\ \hline
$\Delta \Omega_b h^2$
& $2.8\, 10^{-4}$ & $6.1\, 10^{-4}$ & 
$7.1\, 10^{-4}$ & $6.2\, 10^{-4}\ $ & $5.7\, 10^{-5}\ $ & $8.9\, 10^{-4}$&
$9.5\, 10^{-4}$ & $1.1\, 10^{-4}$\\ \hline
$\Delta \kappa_{ri}$& $2.1\, 10^{-2}$ & $2.1\, 10^{-2}$ & 
$2.0\, 10^{-2}$& $2.0\, 10^{-2}$&$5.5 \, 10^{-3}$
& $2.0\, 10^{-2}$&$3.2 \, 10^{-2}$ & $ 3.5\, 10^{-3}$\\ \hline
$\Delta n_s$& $4.8\, 10^{-3}$ & $3.1\, 10^{-2}$ & 
$3.5\, 10^{-1}$ & $3.0\, 10^{-2}$ & $3.0\, 10^{-3}$
& $2.6\, 10^{-2}$&$1.7 \, 10^{-2}$ & $2.6\,  10^{-3}$\\ \hline
$\Delta {T \over S}$& - & $2.2\, 10^{-1}$ & $4.3\, 10^{-1}
$ & 
$3.0 \, 10^{-1}$ & $4.5 \, 10^{-2}$& $2.1\, 10^{-1}$& - & -\\ \hline
$\Delta n_T$& - & - & $3.9\, 10^{-1}$ & $8.1\, 10^{-1}$ &
$1.7\, 10^{-1}$ & $7.8\, 10^{-1}$& - & -\\ \hline
$\Delta x_e$& - & - & -& - & $1.4\, 10^{-1}$& - & - &-\\ \hline
$\Delta \Omega$& - & - & - 
&- & - &- & $6.1\, 10^{-3}$& $4.1\, 10^{-4}$ \\ \hline
\label{5.table2}
\end{tabular}
\normalsize
\end{center}
\end{table}
\vspace{0.5cm}

\vspace{1cm}

\begin{table}
\caption{Cosmological parameters for the models we studied.
All models were normalized to COBE.}
\begin{center}
\scriptsize
\begin{tabular}{|l|l|l|l|l|l|} \hline
$Param.$&${\rm sCDM}$&${T \over S}=0.28$&${T \over S}=0.1$ 
&$\Omega_\Lambda=0.7$
& Open \\ \hline
$h$& $0.5$  & $0.5$& 
$0.5$ & $.65$ & $.65$\\ \hline
$\Omega_\Lambda$& $0.0$& $0.0$ 
& $0.0$ & $0.7$& $0.0$ \\ \hline
$\Omega_b $
& $0.05$ & $0.05$ & 
$0.05$ & $0.06$ & $0.06$\\ \hline
$\kappa_{ri}$& $0.05$ & 
$0.05$& $0.1$&$0.1$
& $0.05$\\ \hline
$n_s$& $1.0$ & $0.96$ & 
$0.99$ & $1.0$ & $1.0$\\ \hline
${T \over S}$& $0.0$ & $0.28$& $0.1$ & 
$0.0$ & $0.0$\\ \hline
$n_T$& - & $0.04$ & $0.01$ & - & -\\ \hline
$\Omega$& $1.0$ & $1.0$ & $1.0$ & $1.0$ & $0.4$ \\ \hline
\label{5.table3}
\end{tabular}
\normalsize
\end{center}
\end{table}

\appendix
\chapter{Spin-weighted functions} \label{appa}

In this Appendix we review the theory of spin-weighted functions
and their expansion in spin-$s$ spherical harmonics. This was used
in the main text to make an all-sky expansion of the  $Q$ 
and $U$ Stokes parameters  
(Chapter \ref{chapstatcmb}). The main application of these functions
in the past was in the theory of gravitational wave radiation (see e.g. 
\cite{appa.thorne}).
Our discussion follows closely that of Goldberg et al. \cite{appa.goldberg},
which is based on the work by Newman and Penrose \cite{appa.np}.
We refer to these references for a more detailed discussion.

For any given direction on the sphere
specified by the angles $(\theta,\phi)$, one can define three
orthogonal vectors, one radial and two tangential to
the sphere. Let us denote the radial 
direction vector with ${\bi n}$
and the tangential with $\hat{{\bi e}}_1$,  $\hat{{\bi e}}_2$.
The latter two are only defined up to 
a rotation around ${\bi n}$.

A function $\;_sf(\theta,\phi)$
defined on the sphere is said to have spin-$s$ if under
a right-handed rotation of ($\hat{{\bi e}}_1$,$\hat{{\bi e}}_2$)
by an  angle $\psi$  it transforms as 
$\;_s f^{\prime}(\theta,\phi)=e^{-is\psi}\;_sf(\theta,\phi)$.
For example, given an arbitrary vector ${\bf a}$ on the sphere the
quantities ${\bf a}\cdot \hat{{\bi e}}_1+i{\bf a}\cdot\hat{{\bi e}}_2$, 
${\bf n} \cdot {\bf a}$ and ${\bf a}\cdot 
\hat{{\bi e}}_1-i{\bf a}\cdot\hat{{\bi e}}_2$ have spin
$1$,$0$ and $-1$ respectively.  
Note that we use a different convention for
rotation than Goldberg et al.
\cite{appa.goldberg} to agree with the previous literature on 
CMB polarization.

A scalar field on the sphere can be expanded in spherical 
harmonics, $Y_{lm}(\theta,\phi)$, which form  
a complete and orthonormal basis. These functions
are not appropriate to expand spin weighted functions with $s\neq 0$.
There exist analogous sets of functions  
that can be used to expand spin-$s$ functions, the so called spin-$s$ 
spherical harmonics $\;_sY_{lm}(\theta,\phi)$. These sets of functions
(one set for each particular spin) satisfy the same completness and 
orthogonality relations,
\begin{eqnarray}
\int_0^{2\pi} d\phi \int_{-1}^1 d\cos \theta \;_sY_{l^{\prime}m^{\prime}}^*(\theta,\phi)
\;_sY_{lm}(\theta,\phi)&=&\delta_{l^{\prime}l}\delta_{m^{\prime}m} 
\nonumber \\
\sum_{lm} \;_sY_{lm}^*(\theta,\phi)
\;_sY_{lm}(\theta^{\prime},\phi^{\prime})
&=&\delta(\phi-\phi^{\prime})\delta(\cos\theta-\cos\theta^{\prime}).
\end{eqnarray}   

An important property of spin-$s$ functions is that there exists
a spin raising (lowering) 
operator $\edth$ ($\baredth$) with the property of 
raising (lowering) the spin-weight of a function,
$(\edth \,_sf)^{\prime}=e^{-i(s+1)\psi}
\edth \,_sf$, $(\baredth \,_sf)^{\prime}=e^{-i(s-1)\psi}
\baredth \,_sf$. Their explicit expression is given by
\begin{eqnarray}
\edth \;_sf(\theta,\phi)&=&-\sin^{s}(\theta)
\left[{\partial \over \partial\theta}+i\csc(\theta)
{\partial \over \partial\phi} \right]\sin^{-s}(\theta)\;_sf(\theta,\phi) 
\nonumber \\
\baredth \;_sf(\theta,\phi)&=&-\sin^{-s}(\theta)
\left[{\partial \over \partial\theta}-i\csc(\theta)
{\partial \over \partial\phi} \right]\sin^{s}(\theta)\;_sf(\theta,\phi) 
\label{edth}
\end{eqnarray}

In this paper we are interested in polarization, which is a quantity 
of spin $\pm 2$. 
The $\baredth$ and $\edth$ 
operators acting twice on a function $\;_{\pm 2}f(\mu,\phi)$ 
that satisfies $\partial_{\phi}\;_sf=im\;_sf$ can be expressed as
\begin{eqnarray} 
\baredth^2 \;_2f(\mu,\phi)&=&
\left(-\partial \mu + {m \over 1-\mu^2}\right)^2 \left[
(1-\mu^2) \;_2f(\mu,\phi)\right]  \nonumber\\ 
\edth^2 \;_{-2}f(\mu,\phi)&=&
\left(-\partial \mu - {m \over 1-\mu^2}\right)^2 \left[(1-\mu^2) \;_{-2}
f(\mu,\phi)\right], 
\label{operators1}
\end{eqnarray}
where $\mu=\cos(\theta)$.
With the aid of these operators one can express  
$_sY_{lm}$ in terms of the spin zero spherical harmonics $Y_{lm}$,
which are the usual spherical harmonics,
\begin{eqnarray}
_sY_{lm}&=&
 \left[{(l-s)!\over (l+s)!}\right]^{1\over 2}\edth^s Y_{lm}
\;\;\; , (0 \leq s \leq l) 
\nonumber \\ 
_sY_{lm}&=&
 \left[{(l+s)!\over (l-s)!}\right]^{1\over 2}(-1)^s 
\baredth^{-s} Y_{lm}
\;\;\; ,(-l \leq s \leq 0). 
\end{eqnarray}

The following properties of spin-weighted harmonics are also useful
\begin{eqnarray}
\:_sY^*_{lm}&=&(-1)^{s}{}_{-s}Y_{l-m} \nonumber \\
\edth \:_sY_{lm}&=&\left[(l-s)(l+s+1)\right]^{1\over 2}
\:_{s+1}Y_{lm} \nonumber \\
\baredth \:_sY_{lm}&=&-\left[(l+s)(l-s+1)\right]^{1\over 2}
\:_{s-1}Y_{lm} \nonumber \\
\baredth\edth \:_sY_{lm}&=&-(l-s)(l+s+1)
\:_{s}Y_{lm} 
\label{propYs}
\end{eqnarray}
Finally, to construct a map of polarization one needs an explicit 
expression for the spin weighted functions,
\begin{eqnarray}
&{}&_sY_{lm}(\hat{n})=e^{im\phi}\Big[{(l+m)!(l-m)!\over (l+s)!(l-s)!}
{2l+1\over 4\pi}\Big]^{1/2}\sin^{2l}(\theta/2) \nonumber \\
&\times& \sum_r {l-s \choose r}{l+s \choose r+s-m}
(-1)^{l-r-s+m}{\rm cot}^{2r+s-m}(\theta/2).
\label{expl}
\end{eqnarray}
\smallskip

For the special case $|s|=2$ a more useful expression
(\ref{spinuseful}) is provided in
Chapter \ref{chapstatcmb}.

\end{document}